\documentclass[11pt]{article}
\pdfoutput=1

\usepackage[utf8]{inputenc}
\usepackage{multirow}
\usepackage{amsmath, amsfonts, amssymb}
\usepackage{comment}
\usepackage{graphicx}
\usepackage{psfrag}
\usepackage{amsthm}

\allowdisplaybreaks
\usepackage[dvipsnames,svgnames,table]{xcolor}
\usepackage{enumerate}
\usepackage{arydshln}
\usepackage{MorrisIn,lettrine,Romantik}

\usepackage{soul}
\usepackage{array}
 \usepackage{slashed}
 \usepackage{nicefrac}
 \usepackage[most]{tcolorbox}
\usepackage[font=small,labelfont=bf]{caption} 
\usepackage{subcaption}
\usepackage{mathtools}
\usepackage{colortbl}
\usepackage{mathrsfs}
 \usepackage{a4wide}
  \usepackage{tikz}
  \usepackage{tikz-cd}
  \usetikzlibrary{shapes.geometric}
  \usepackage{tcolorbox}
\usepackage{diagbox}

  \usepackage{color}
  \definecolor{dark-gray}{gray}{0.20}
  \definecolor{gray}{gray}{0.30}
  \definecolor{light-gray}{gray}{0.80}
  \definecolor{dark-red}{rgb}{0.7,0,0}
  \definecolor{dark-green}{rgb}{0.1,0.4,0}
  \definecolor{dark-blue}{rgb}{0.3,0.3,0.7}
  \definecolor{light-blue}{rgb}{0.8,0.8,1}
      \definecolor{swamp}{RGB}{240, 199, 197}
       \definecolor{landscape}{RGB}{180, 250, 199}
          \definecolor{undecided}{RGB}{252, 252, 197}
\definecolor{myRED}{HTML}{FF0000}
\definecolor{myGREEN}{HTML}{00AA00}
\definecolor{myBLUE}{HTML}{0055D4}

  \usepackage{pifont}

\usepackage{newunicodechar} 
\usepackage{setspace}

\newcommand{\beq}{\begin{equation}}  \newcommand{\eeq}{\end{equation}}
\newcommand{\bal}{\begin{aligned}}   \newcommand{\eal}{\end{aligned}}
\newcommand{\be}{\begin{equation}}
\newcommand{\ee}{\end{equation}}

\def\be{\begin{equation}}
\def\ee{\end{equation}}
\def\bea{\begin{eqnarray}}
\def\eea{\end{eqnarray}}

\newcommand{\dd}{\mathrm{d}}

\def\simleq{\; \raise0.3ex\hbox{$<$\kern-0.75em
      \raise-1.1ex\hbox{$\sim$}}\; }
   \def\simgeq{\; \raise0.3ex\hbox{$>$\kern-0.75em
      \raise-1.1ex\hbox{$\sim$}}\; }

\numberwithin{equation}{section}

\usepackage{jheppub}
\usepackage{hyperref}

\hypersetup{
	colorlinks=true,
	linkcolor=dark-blue,
	citecolor=dark-red,
	urlcolor=dark-green,
	linktoc=page
}

\theoremstyle{remark}

\renewcommand{\d}{\partial}
\newenvironment{eqn}
    {\begin{equation}
    \begin{aligned}
    }
    { 
    \end{aligned}
    \end{equation}
    \ignorespacesafterend
    }

\title{\centering Alice in Warpland: KK modes, \\Warped Compactifications and the Swampland}

\author{Salvatore Raucci$^{1,2}$, Ignacio Ruiz$^{3}$}\affiliation{$^{1}$Instituto de F\'{i}sica Te\'{o}rica IFT-UAM/CSIC,
C/ Nicol\'{a}s Cabrera 13-15, Campus de Cantoblanco, 28049 Madrid, Spain}
\affiliation{$^2$Departamento de F\'{i}sica Te\'{o}rica, Universidad Aut\'{o}noma de Madrid, Cantoblanco, 28049 Madrid, Spain}
\author{and Irene Valenzuela$^{1,2,3}$}\affiliation{$^3$CERN, Theoretical Physics Department, 1211 Meyrin, Switzerland}
\emailAdd{salvatore.raucci@uam.es , ignacio.ruiz.garcia@cern.ch, irene.valenzuela@cern.ch}
\preprint{IFT-UAM/CSIC-26-22\\ \vspace*{-0.1cm} 
\hfill CERN-TH-2026-040}

\abstract{We investigate the asymptotic behavior of Kaluza–Klein (KK) towers in warped compactifications to Minkowski space. Focusing on the overall decompactification limit, we derive the  scaling of KK masses at large KK momentum for scalar fluctuations in lower-dimensional Planck units. In codimension-one warped backgrounds sourced by a higher-dimensional exponential potential, we solve explicitly for the internal profiles and obtain a closed expression for the exponential mass decay rate $\lambda_{\rm KK}$ of the tower in terms of the moduli space distance. We find that warping reduces $\lambda_{\rm KK}$ relative to the unwarped case, in such a way that sufficiently strong warping could in principle violate the Sharpened Distance Conjecture bound. Remarkably, this sharpened bound is still satisfied precisely when the higher-dimensional potential obeys the condition forbidding asymptotic accelerated expansion, establishing a direct link between the Sharpened Distance Conjecture and the Strong de Sitter condition in one higher dimension. We also argue that for higher-codimension warped backgrounds the asymptotic KK scaling remains unmodified.
}

\begin{document}

\hypersetup{pageanchor=false}
\makeatletter
\let\old@fpheader\@fpheader

\makeatother

\maketitle

\hypersetup{
    pdftitle={Warped KK decompactification},
    pdfauthor={Salvatore Raucci, Ignacio Ruiz, Irene Valenzuela},
    pdfsubject={Warped (de)compactifications and scaling of KK modes}
}

\newcommand{\remove}[1]{\textcolor{red}{\sout{#1}}}

\section{Introduction}

Determining the spectrum of Kaluza–Klein (KK) modes remains a long-standing challenge in a theory with extra dimensions, like string theory. The problem typically involves two steps: identifying the appropriate Laplacian operators governing the fluctuations and then solving for their eigenvalue spectra.

In direct product compactifications, this program is in principle tractable. The Laplacians factorize, mixing between different excitations is mild, and the task reduces to studying differential operators on the internal manifold. Exact spectra can be obtained for sufficiently simple spaces where the internal metric is known such as tori, spheres, or coset manifolds. While the detailed spectrum is highly case-dependent, the asymptotic growth of eigenvalues for large KK momenta is universally governed by Weyl’s law \cite{Weyl1911}, and various bounds on spectral gaps are known (see e.g.~\cite{PMIHES_1961__10__5_0,Alday:2019qrf,Bonifacio:2019ioc,Collins:2022nux,Bonifacio:2023ban}).

The situation becomes substantially more complicated in warped compactifications, which are the focus of this work. In this setting, the warp factor breaks the product structure of spacetime, and may induce non-trivial mixing between fields of different spins. As a result, even identifying the correct KK operators is technically non-trivial. When these operators can be written down, they typically take the form of weighted Laplacians, for which analytic spectra are generally out of reach except in very simple models. Remarkably, however, Weyl’s law continues to hold even in the warped case and some bounds on spectral gaps are known~\cite{setti1998eigenvalue,hassannezhad2013eigenvalues,charalambous2015eigenvalue,DeLuca:2021mcj,DeLuca:2021ojx,DeLuca:2024fbc,DeLuca:2025klz}. See \cite{Randall:1999ee,Randall:1999vf,Davoudiasl:1999jd,Douglas:2009zn,Boos:2012zz,Blumenhagen:2019qcg,Blumenhagen:2022dbo,Blumenhagen:2022zzw,ValeixoBento:2022qca,Etheredge:2023odp,Lust:2025vyz,Reig:2025dpz} for recent works on the KK mass scaling in warped compactifications of string theory.

In this work, we take a step forward in this long-standing challenge by identifying universal features of the KK spectrum in warped compactifications. We study general warped compactifications to Minkowski space and derive the asymptotic behavior of KK masses for scalar fluctuations in the limit where the overall internal space decompactifies. This provides analytic control over the scaling of the KK tower at large momentum, even in the presence of strong warping effects. In particular, for the case of one-dimensional warped compactifications, we explicitly solve for the internal field profiles and obtain the resulting KK mass scaling in lower-dimensional Planck units in setups where the warping is sourced by a higher-dimensional exponential potential.\\

Even if we expect our results to be useful for a wide range of applications involving warped compactifications, our primary motivation for addressing this technical challenge comes from the Distance Conjecture \cite{Ooguri:2006in} in the Swampland program \cite{Vafa:2005ui,Brennan:2017rbf,Palti:2019pca,vanBeest:2021lhn,Grana:2021zvf,Harlow:2022ich,Agmon:2022thq}. This conjecture implies the existence of infinite towers of states becoming exponentially light in terms of the field space distance at every infinite-distance limit in moduli space. This is a remarkably universal feature observed across the string landscape, and it is expected to be satisfied in any effective field theory that admits a UV completion in quantum gravity.

In recent years, there has been substantial activity aimed at quantifying this statement \cite{Baume:2016psm,Klaewer:2016kiy, Blumenhagen:2017cxt, Grimm:2018ohb,Heidenreich:2018kpg, Blumenhagen:2018nts, Grimm:2018cpv, Buratti:2018xjt, Corvilain:2018lgw, Joshi:2019nzi,  Erkinger:2019umg, Marchesano:2019ifh, Font:2019cxq,Andriot:2020lea,Gendler:2020dfp, Lanza:2020qmt,Bedroya:2020rmd,Klaewer:2020lfg, Lee:2021qkx,Lee:2021usk,Rudelius:2022gbz,Alvarez-Garcia:2023gdd,Alvarez-Garcia:2023qqj,Rudelius:2023mjy,Aoufia:2024awo,Hassfeld:2025uoy,Monnee:2025ynn,Hattab:2025aok,Monnee:2025msf}. Recent progress in string theory has revealed strikingly universal patterns \cite{Lee:2019wij,Etheredge:2022opl,Castellano:2023stg,Castellano:2023jjt,Etheredge:2024tok}: not only does such a tower of states always appear, but its microscopic nature and the exponential rate at which it becomes light are highly constrained across the landscape. In particular, there seems to always exist a duality frame in which the tower can be interpreted either as a Kaluza–Klein tower—signaling the decompactification of extra dimensions—or as a tower of string oscillator modes—signaling a perturbative string limit. This expectation is known as the Emergent String Conjecture \cite{Lee:2019wij}, also extensively tested, see e.g.~\cite{Lee:2018urn, Lee:2019xtm,Baume:2019sry, Xu:2020nlh, Baume:2020dqd, Perlmutter:2020buo, Lanza:2021udy, Baume:2023msm, Rudelius:2023odg, Calderon-Infante:2024oed, Grieco:2025bjy}. 

Related to this, a lower bound was proposed for the exponential rate governing the mass behavior of the leading tower of states,
\begin{equation}\label{eq. sharpened DC}
m\sim M_{{\rm Pl},d}\, e^{-\lambda \Delta\varphi} \, , \qquad \lambda\geq \frac1{\sqrt{d-2}} \, ,
\end{equation}
where $d$ is the spacetime dimension, $M_{{\rm Pl},d}$ the associated Planck mass and $\Delta\varphi$ is the geodesic distance in field space. This is often referred to as the Sharpened Distance Conjecture \cite{Etheredge:2022opl}. The bound is saturated by the string tower of the fundamental string propagating in flat space, where $\lambda_{\rm str}=\frac{1}{\sqrt{d-2}}$.

Kaluza–Klein towers arising from unwarped compactifications to flat space exhibit an exponential rate given by
\begin{equation}\label{eq. KK rate}
\lambda_{\rm KK}=\sqrt{\frac{d+n-2}{n(d-2)}} \,,
\end{equation}
where $n$ is the number of extra dimensions decompactifying. Since $\lambda_{\rm KK}>\lambda_{\rm str}$, the sharpened bound \eqref{eq. sharpened DC} is satisfied in unwarped decompactification limits. However, it is known that the KK spectrum is modified in the presence of warping. In fact, in \cite{Etheredge:2023odp} the KK spectrum was explicitly computed for the Polchinski–Witten warped solution of Type I$'$ \cite{Polchinski:1995df}, and it was found that $\lambda_{\rm KK}$ is reduced with respect to \eqref{eq. KK rate}, although it remains larger than \eqref{eq. sharpened DC}.\\

This naturally raises an important question: could warping effects in more general backgrounds decrease $\lambda_{\rm KK}$ so drastically that the sharpened bound \eqref{eq. sharpened DC} is violated? To what extent is this bound truly universal, rather than a lamppost effect of unwarped compactifications?
After all, just as Alice learned, rules can bend in Warpland.

In this work, we address this question by explicitly computing the asymptotic form of KK masses at large KK momentum in warped compactifications to Minkowski space. The warping will be sourced by a higher-dimensional scalar potential together with suitable localized sources. We focus on decompactification limits of the overall volume modulus and assume that the higher-dimensional potential decreases exponentially with field distance, with an arbitrary exponential rate.

Our main result is the asymptotic exponential rate of KK towers in codimension-one backgrounds, namely when the warp factors depend only on a single internal dimension:
\begin{equation}\label{eq. warped intro}
\lambda_{\rm KK}=\sqrt{\frac{d-1}{d-2}}\left[1+\frac{4(d-2)}{(d-1)\gamma^2}\right]^{-1/2}\,,
\end{equation}
which is determined in terms of the exponential rate $\gamma$ of the higher-dimensional potential, i.e. $V\sim V_0 e^{-\gamma \Delta\varphi}$. In particular, $\lambda_{\rm KK}$ decreases as $\gamma$ decreases, reflecting the fact that smaller $\gamma$ corresponds to stronger warping effects (the unwarped result is recovered in the limit $\gamma\to \infty$). This implies that for sufficiently small $\gamma$, it is in principle possible to violate \eqref{eq. sharpened DC}. However, this requires a very slowly varying higher-dimensional potential.
Indeed, we find that the sharpened bound is satisfied as long as
\begin{equation}
\gamma>\frac2{\sqrt{D-2}} \, ,
\end{equation}
with $D=d+1$, which is precisely the condition for the absence of asymptotic accelerated expansion in the higher-dimensional theory. This bound plays a key role in the Swampland literature and is often referred to as the Strong de Sitter Conjecture \cite{Obied:2018sgi,Agrawal:2018rcg,Rudelius:2021azq}, which is expected to hold at least in asymptotic regimes of moduli space \cite{Hertzberg:2007wc,Andriot:2019wrs,Andriot:2020lea,Calderon-Infante:2022nxb,Shiu:2023fhb,Shiu:2023nph,Cremonini:2023suw,Hebecker:2023qke,VanRiet:2023cca,Seo:2024fki,Shiu:2024sbe}. Hence, our results reveal an intriguing correlation between satisfying the Strong de Sitter Conjecture and the Sharpened Distance Conjecture upon compactification.

We also study how the taxonomy rules proposed in \cite{Etheredge:2024tok} are modified in the presence of warping, and we provide arguments for how the exponential rate of a string tower is expected to change. In particular, we find that it becomes larger than the KK rate whenever $\gamma<\frac2{\sqrt{D-2}}$. We further explore higher-dimensional warped backgrounds, arguing that in those cases the warping can never become strong enough—due to the higher dimension of the internal space —to modify the asymptotic scaling of KK towers.\\

The journey through Warpland unfolds as follows.
After a brief review of the usual KK dimensional reduction, in Section \ref{sec. comp} we step through the looking glass into the realm of warped compactifications. We perform, for general warped Minkowski backgrounds, the dimensional reduction of the Einstein–scalar action and the solution of the corresponding Laplacian of scalar KK modes. This yields the general expression for KK masses in lower-dimensional Planck units as well as the moduli space metric. In Section \ref{sec:warped_codim_1}, we specialize to one-dimensional warped compactifications and derive the general scaling result \eqref{eq. warped intro}. Readers primarily interested in the main lessons of our story may follow the white rabbit directly to Section \ref{ss. results and bounds}, where we summarize our results and compare them with the Swampland bounds. Section \ref{sec:higher_cod} extends the analysis to higher-dimensional warped backgrounds, where we argue that warping effects are diluted in the decompactification limit, so that the asymptotic KK scaling remains unmodified. We conclude in Section \ref{sec. conc}, followed by several appendices providing technical guidance for those who might otherwise lose their way in the Warpland.

\section{Compactifications and scaling of KK masses\label{sec. comp}}

As we have explained in the Introduction, our focus is on extracting the asymptotic behavior of KK masses in decompactification limits with warping. We start by reviewing unwarped, homogeneous compactifications in Section \ref{ssec:review_unwarped} to set our notation and highlight some subtleties, and move to the warped case in Section~\ref{ssec:warped}. The analysis without warping can also be found e.g.~in \cite{Roest:2004aqa,Ortin:2015hya,VanRiet:2023pnx,Castellano:2023jjt}, with different levels of detail.

    \subsection{Review of KK modes in unwarped compactifications}\label{ssec:review_unwarped}
    
    Consider a $D$-dimensional action for a scalar field $\hat\varphi$ coupled to Einstein gravity, which in the Einstein frame\footnote{See~\cite{ValeixoBento:2025emh} for a detailed discussion on frames in string theory compactifications.} is
    \begin{equation}\label{eq:scalar_fiel_norm}
        S_D=\frac{1}{2\kappa_D^2}\int \dd^Dx\sqrt{-G_D}\left\{\mathcal{R}_D-(\partial\hat\varphi)^2\right\}\,.
    \end{equation}
    We refer to scalars $\hat{\varphi}$ normalized as in~\eqref{eq:scalar_fiel_norm} as \emph{canonically normalized} scalar fields.
    We compactify on an $n$-manifold $X_n$ to $d=(D-n)$ dimensions and study the decompactification limit in which $X_n$ grows homogeneously. To this end, we consider the following metric on $X_n$:
    \begin{equation}\label{eq:unwarped_internal_metric}
        \dd s^2_n=G_{ij}\dd y^i\dd y^j=e^{2 \sigma}\mathsf{M}_{ij}\dd y^i\dd y^j\,,\quad \text{with}\quad \int_{X_n}\dd^ny\sqrt{\mathsf{M}}=V_{X,0}\ell_D^n\,,
    \end{equation}
    where $\mathsf{M}_{ij}$ is a metric on $X_n$ with fixed volume $V_{X,0}$ in $D$-dimensional Planck units, and where we use $M_{{\rm Pl},D}=\kappa_D^{-\frac{2}{D-2}}$ and $\ell_D=M_{{\rm Pl},D}^{-1}$. In~\eqref{eq:unwarped_internal_metric}, $\sigma$ is the overall volume modulus, defined such that the physical volume $V_X$ of $X_n$ (with respect to the metric $G_{ij}$) is
    \begin{equation}
        V_X=M_{{\rm Pl},D}^n\int_{X_n}\dd^n y\sqrt{G_n}=V_{X,0}e^{n\sigma}\,.
    \end{equation}
 After the compactification, $\sigma$ produces a scalar field in the $d$-dimensional effective action. This can be viewed as a perturbation $\delta\sigma(x)$ of the metric field $\sigma$ that appears in~\eqref{eq:unwarped_internal_metric}. To simplify the notation, we simply use $\sigma$ to denote the sum of the background field $\sigma$---the volume---and the lower-dimensional scalar field: $\sigma\to\sigma+\sigma(x)$. We adopt this same convention for all scalar fields across the paper. 
    Given this setup, we expand the $D$-dimensional metric $G_D$ as
    \begin{align}
        \dd s^2_D=G_{MN}\dd x^M\dd x^N&=e^{2\alpha\sigma}g_{\mu\nu}\dd x^\mu\dd x^\nu+e^{2\sigma}\mathsf{M}_{ij}\dd y^i\dd y^j\notag\\&=e^{2\sigma}\left[e^{2(\alpha-1)\sigma}g_{\mu\nu}\dd x^\mu\dd x^\nu+\mathsf{M}_{ij}\dd y^i\dd y^j\right]\,,
    \end{align}
    where $\alpha$ is a numerical constant and $g_{\mu\nu}$ is the Einstein-frame metric of the $d$-dimensional action. A double conformal rescaling of the Ricci scalar $\mathcal{R}_{D}$ leads to
    \begin{align}\label{eq.Ricci red unwarp}
        \mathcal{R}_D&=e^{-2\alpha\sigma}\left\{\mathcal{R}_g+e^{2(\alpha-1)\sigma}\mathcal{R}_\mathsf{M}-2[(d-1)(\alpha-1)+(d+n-1)]\Delta_g\sigma\right.\notag\\
        &\quad\left.-\left[\alpha ^2 (d-2) (d-1)+2 \alpha    (d-2) n+ n (n+1)\right](\partial\sigma)^2\right\}\,,
    \end{align}
    together with
    \begin{equation}
        \sqrt{-G_D}=\sqrt{-g}e^{(d\alpha+n)\sigma}\sqrt{\mathsf{M}}\,,\qquad G^{MN}\partial_M\hat\varphi\partial_N\hat\varphi=e^{-2\alpha\sigma}g^{\mu\nu}\partial_\mu\hat\varphi\partial_\nu\hat\varphi\,,
    \end{equation}
    where we have further imposed that $\hat\varphi$ depends only on the external dimensions. In order for $g_{\mu\nu}$ to be the lower-dimensional Einstein-frame metric, the value of the numerical constant $\alpha$ must be set to
    \begin{equation}
        \alpha=-\frac{n}{d-2}\,,
    \end{equation}
    and the term $\Delta_g\sigma$ in \eqref{eq.Ricci red unwarp} becomes a total derivative in the lower-dimensional Einstein frame. This results in the following $d$-dimensional effective action,
    \begin{equation}\label{eq. eff act d unwarped}
        S_d\supset \frac{1}{2\kappa_d^2}\int \dd^dx\sqrt{-g_d}\left\{\mathcal{R}_d-\frac{n(d+n-2)}{d-2}(\partial\sigma)^2-(\partial\hat\varphi)^2-2\kappa_d^2 V_{\mathcal{R}_{\mathsf{M}}}(\sigma)\right\}\,,
    \end{equation}
    where
    \begin{equation}\label{eq. Mpl pot}
        \frac{1}{\kappa_d^2}=\frac{V_{X,0}\ell_D^n}{\kappa_D^2}\,,\qquad \frac{V_{\mathcal{R}_{\mathsf{M}}}(\sigma)}{M_{{\rm Pl},d}^2}=-\frac{1}{2}e^{-\frac{2(d+n-2)}{d-2}\sigma}\int_{X_n}\dd^n\sqrt{\mathsf{M}}\mathcal{R}_{\mathsf{M}}\,,
    \end{equation}
    are the relations between the higher- and lower-dimensional gravitational couplings and the on-shell scalar potential induced by the internal curvature of the compact manifold. 
    
    From the effective action \eqref{eq. eff act d unwarped}, we can define a canonically normalized radion modulus,
    \begin{equation}\label{eq:sigma_hat}
        \hat{\sigma}=\sqrt{\frac{n(d+n-2)}{d-2}}\sigma=\sqrt{\frac{d+n-2}{n(d-2)}}\log\frac{V_X}{V_{X,0}}\,,
    \end{equation}
    such that $\mathsf{G}_{\hat{\sigma}\hat{\sigma}}=1$.
    The main interest of this work is the scaling of the Kaluza--Klein modes as they become light when $X_n$ decompactifies, i.e., $V_X \to\infty$.  To this end, we study the spectrum of the scalar Laplacian operator associated with the internal manifold,
        \begin{equation}            \Delta_{G_n}\Psi+\lambda^2\Psi=0\Longrightarrow \Delta_{\mathsf{M}}\Psi+\left(\lambda e^{\sigma}\right)^2\Psi=0\,,
        \end{equation}
        where we have used the fact that $\sigma$ is constant in the internal dimensions. For any internal topology and suitable boundary conditions of $X_n$, the behavior of the eigenvalues for large values of Kaluza--Klein momenta $\mathbf{k}$ follows Weyl's law \cite{Weyl1911}:\footnote{Given a compact $n$-dimensional Riemannian manifold $(X,g)$, the $k$-th eigenvalue (including multiplicity) of the Laplace operator $\Delta_g$ scales, for large $k$, as
        \begin{equation}\label{eq. WEYLS LAW}
            \lambda_k^2\sim\frac{4\pi^2}{\omega_n^{2/n}}\left(\frac{k}{{\rm vol}_X}\right)^{2/n}\,,\quad\text{where }\ \omega_n=\frac{\pi^{n/2}}{\Gamma(1+\frac{n}{2})}\,,\quad {\rm vol}_X=\int_X\dd^nx\sqrt{g}\,.
        \end{equation}
        See \cite{schnirelman,Verdiere:1985,Zelditch:1987,DeLuca:2024fbc} for more details and for the analysis on the validity of this result, including cases of manifolds with singularities.}
        \begin{equation}
           \lambda_{\mathbf{k}}=\frac{m_{{\rm KK},\mathbf{k}}}{M_{{\rm Pl},D}}\sim  e^{-\sigma}\sim{V_X^{-\frac 1n}}\qquad\text{for }\;|\mathbf{k}|\to\infty.
        \end{equation}
        The above result is given in $M_{{\rm Pl},D}$ units. To obtain the expression in lower-dimensional Planck units, we see that from \eqref{eq. Mpl pot} and \eqref{eq:sigma_hat},
        \begin{equation}\label{eq. Mpl unwarped}
            M_{{\rm Pl},d}=M_{{\rm Pl},D}V_X^{\frac{1}{d-2}}\,,\quad \text{so that}\quad \frac{M_{{\rm Pl},D}}{M_{{\rm Pl},d}}=V_X^{-\frac{1}{d-2}}=\exp\left\{-\sqrt{\frac{n}{(d-2)(d+n-2)}}\hat{\sigma}\right\}\,.
        \end{equation}
        We can use the above expression to obtain 
        \begin{equation}\label{eq. KK unwarped}
            \frac{m_{\rm KK}}{M_{{\rm Pl},d}}\sim V_X^{-\frac{d+n-2}{n(d-2)}}\sim \exp\left\{-\sqrt{\frac{d+n-2}{n(d-2)}}\hat{\sigma}\right\}\,.
        \end{equation}
        This recovers the well-known scaling of the KK modes with the canonically normalized volume in homogeneous decompactification limits~\cite{Etheredge:2022opl,Castellano:2023jjt,Etheredge:2024tok}. 
        
       We can also derive the scaling of higher-dimensional massive states upon compactification. Consider a tower of states whose mass scales exponentially with $\hat \varphi$ as  $\hat\varphi\to\infty$ in the higher-dimensional theory: 
        \begin{equation}
                \label{eq. additional unwarped-1}
            \frac{m_{\hat{\varphi}}}{M_{{\rm Pl},D}}\sim e^{-\lambda_D\hat{\varphi}}\ .
            \end{equation}
        Upon compactification, this becomes
            \begin{equation}
            \label{eq. additional unwarped}
            \frac{m_{\hat{\varphi}}}{M_{{\rm Pl},d}}\sim\exp\left\{-\lambda_D\hat{\varphi}-\sqrt{\frac{n}{(d+n-2)(d-2)}}\hat{\sigma}\right\}\,,
        \end{equation}
        thus picking an additional dependence on the new modulus $\hat \sigma$.
        
        The scalings \eqref{eq. KK unwarped} and \eqref{eq. additional unwarped} are at the core of the \emph{Taxonomy Rules} for asymptotic towers derived in \cite{Etheredge:2024tok}. In the context of moduli spaces with several compact scalars, where the exponential rate of the light tower depends on the trajectory, it is useful to introduce the \emph{scalar charge-to-mass ratio} vectors, also known as $\zeta$-vectors or \emph{scaling vectors} \cite{Calderon-Infante:2020dhm} (see also \cite{Etheredge:2022opl, Etheredge:2023odp, Calderon-Infante:2023ler,Etheredge:2023usk,Etheredge:2025ahf}),
        \begin{equation}
        \label{scalingvector}
            \vec\zeta=-\vec{\nabla}\log\frac{m(\varphi)}{M_{{\rm Pl,}d}}\,,
        \end{equation}
        where the gradient is taken with respect to the moduli. In this way, given a trajectory with an asymptotic unit tangent vector $\hat{T}$, the exponential mass decay rate is given by
        \begin{equation}
        \label{exprate}
            \lambda=\hat{T}\cdot\vec\zeta\,,
        \end{equation}
        where the product is taken with the moduli space metric. For one-dimensional (submanifolds of) moduli spaces parameterized by $\varphi$, with metric $\mathsf{G}_{\varphi\varphi}$, we simply have
        \begin{equation}\label{eq. 1 mod lambda}
            \lambda=-\mathsf{G}_{\varphi\varphi}^{-\frac{1}{2}}\partial_\varphi\log\frac{m(\varphi)}{M_{{\rm Pl,}d}}
        \end{equation}
        along the $\varphi\to\infty$ limit.
        
    \subsection{Warped compactifications}\label{ssec:warped}
    
    We now turn to the generalization of \eqref{eq. KK unwarped} to warped compactifications.    
    In this section, we determine the general expression for the moduli space metric and the scaling of KK modes, leaving explicit examples for later sections. As we have explained in the Introduction, this is a challenging problem when addressed in full generality. We thus proceed with some additional assumptions, which are not too restrictive for the cases that we will analyze in detail. 

    The main assumption concerns mixing. We assume that in the limit of large KK momentum, the mixing of different fields has no effect on the asymptotic scaling of the KK masses as one takes the limit in which the overall volume of the compactification becomes large. 
    In general, we take this as an assumption, partly inspired by Weyl's law and supported by the physical intuition that large-momentum KK modes are only sensitive to local physics. In fact, we only consider decompactification limits that preserve the structure of the compactification---for example, if the compact space is an Einstein manifold, we only deform its overall volume while keeping fixed the Einstein structure---and under these conditions, it is reasonable to assume that mixing, among fields with the same or different spin, does not change our results in the large KK momentum regime. For the codimension-one backgrounds of Section~\ref{sec:warped_codim_1}, the analysis of perturbations in~\cite[Sec.~5]{Basile:2022ypo} and the arguments of~\cite{Etheredge:2023odp} show that the Laplacian operator that we will use is the correct one in the limit of large KK momentum for spin 0 fields, so that mixing is indeed subleading and can be safely ignored.
    
    Given these premises, we consider a set of $D$-dimensional scalars $\hat{\varphi}^{\hat{a}}$ coupled to Einstein gravity, with the action 
    \begin{equation}\label{eq:warped_action}
        \int\dd^Dx\sqrt{-G_D}\left\{\frac{1}{2\kappa_D^2}\left[\mathcal{R}_D-\hat{\mathsf{G}}_{\hat{a}\hat{b}}\partial_M\hat{\varphi}^{\hat{a}}\partial^M\hat{\varphi}^{\hat{b}}\right]-\hat{V}(\hat{\varphi})\right\}+S_{\rm loc.}\,,
    \end{equation}
    where $S_{\rm loc.}$ refers to the contribution from localized sources. We include a higher-dimensional bulk potential that is the source of the warping. 
    
    We compactify on a warped product of a $d$-dimensional Minkowski space and an $n$-dimensional internal manifold $X_n$, with metric
    \begin{equation}\label{eq:warped_cpt}
        \dd s_D^2=G_{MN}\dd x^M\dd x^N=e^{2\rho(y)}\eta_{\mu\nu}\dd x^\mu \dd x^\nu+e^{2\sigma(y)}\mathsf{M}_{ij}\dd y^i \dd y^j\,.
    \end{equation}
    From now on, we assume that it is always possible to include some suitable localized sources to cancel the contribution from the scalar potential and yield a lower-dimensional Minkowski solution, as discussed in detail in Appendix~\ref{app:lower_potential}. For simplicity, we only discuss the dimensional reduction of the bulk contribution in the main text, as it is enough to obtain the asymptotic scaling of the KK towers, leaving further details to Appendix~\ref{app:lower_potential}.
    
    Unlike in the unwarped case, $\sigma$ and $\rho$ in \eqref{eq:warped_cpt} explicitly depend on the internal coordinates $y^i$. We introduce the parameter $\theta$, which allows us to work in the $d$-dimensional Einstein-frame metric,
    \begin{equation}\label{eq:theta}
        g_{\mu\nu}=e^{2\theta}\eta_{\mu\nu}\,, \quad {\text{with}}\quad 
        e^{(d-2)\theta}=\int_{X_n}\dd^n y \sqrt{\mathsf{M}}e^{(d-2)\rho+n\sigma} \ell^{-n}\,.
    \end{equation}
    Here, $\ell$ is an arbitrary length scale; for instance, we can choose the $D$-dimensional Planck length $\ell_D$. We fix the volume of the internal space, measured by the metric $\mathsf{M}_{ij}$, as in~\eqref{eq:unwarped_internal_metric}. Without loss of generality, we take $X_n$ to have unit volume in higher-dimensional Planck units, i.e., $V_{X,0}=1$.

    From now on, $\rho$, $\sigma$, and the scalars $\hat{\varphi}^{\hat{a}}$ will be promoted to functions of both $y^i$ and $x^\mu$, the former encoding the background profiles and the latter the dependence on the lower-dimensional moduli. This is in a similar spirit to what we did for $\sigma$ in Section~\ref{ssec:review_unwarped}.

    Our goal is to compute the rate at which Kaluza--Klein modes become light in terms of the scalar field distance as one moves in the moduli space. The first step is then to compute the kinetic terms for the lower-dimensional moduli, from which one can extract the natural metric on the moduli space. 
    Since we are interested in decompactification limits, we focus only on the moduli coming from the higher-dimensional scalars and the overall volume. We expect that other scalar perturbations from the shape of $X_n$ do not change the asymptotic behavior of the KK modes associated with the overall volume. In practice, this is the assumption that mixing with these modes is subleading in the regime of large KK momentum.

    In the $D$-dimensional action of~\eqref{eq:warped_action}, we must expand the Ricci scalar and the kinetic term of the scalar fields. The former is
    \begin{align}\label{eq. exp warp ricci}
        \mathcal{R}_D   =&  e^{-2(\rho-\theta)}\left[\mathcal{R}_g - 2(d-1)\Delta_g(\rho - \theta) - 2n\Delta_g\sigma - (d-1)(d-2)\d_\mu(\rho - \theta)\d^\mu(\rho - \theta)\right. \notag\\
            &\left. - n(n+1)\d_\mu \sigma\d^\mu\sigma - 2n(d-2)\d_\mu(\rho - \theta)\d^\mu\sigma\right] + e^{-2\sigma}\left[\mathcal{R}_{\mathsf{M}} - 2(n-1)\Delta_{\mathsf{M}}\sigma \right.\notag\\
                & \left.- 2d\Delta_{\mathsf{M}}\rho - (n-1)(n-2)\d_i\sigma\d^i\sigma - d(d+1)\d_i\rho\d^i\rho - 2d(n-2)\d_i\sigma\d^i\rho \right]\,,
    \end{align}
    where all quantities refer to the two metrics $g_{\mu\nu}$ and $\mathsf{M}_{ij}$. For example, $\Delta_{g}$ and $\Delta_{\mathsf{M}}$ are the two Laplace operators associated with $g_{\mu\nu}$ and $\mathsf{M}_{ij}$, $\mathcal{R}_g$ and $\mathcal{R}_{\mathsf{M}}$ are their Ricci scalars, and the indices are contracted with the two unwarped sub-metrics.
    Using~\eqref{eq. exp warp ricci} in~\eqref{eq:warped_action}, integrating by parts, and recalling the definition of $\theta$---which is now a function of $x$---we can isolate the terms with two spacetime derivatives, which lead to the desired metric for the $d$-dimensional scalars $\varphi^a$:
    \begin{align}\label{eq:moduli_space_metric}
        \mathsf{G}_{ab}\d_\mu\varphi^a \d^\mu\varphi^b  =& (d-1)(d-2)(\d\theta)^2 -e^{-(d-2)\theta}\int_{X_n}\dd^n y \sqrt{\mathsf{M}}\,e^{(d-2)\rho+n\sigma}\left[(d-1)(d-2)(\d\rho)^2\right. \notag\\
            &\left.+n(n-1)(\d\sigma)^2+2n(d-1)\d_\mu\rho\d^\mu\sigma-\hat{\mathsf{G}}_{\hat{a}\hat{b}}\d_\mu\hat{\varphi}^{\hat{a}}\d^\mu\hat{\varphi}^{\hat{b}}\right]\,,
    \end{align}
    where $\d$ denotes a spacetime derivative. This expression gives the moduli space metric provided that one knows the dependence of $\rho$, $\sigma$, and $\hat{\varphi}^{\hat{a}}$ on the lower-dimensional moduli through the internal profile along $X_n$. The remaining terms from~\eqref{eq. exp warp ricci}---those with internal derivatives---and the scalar potential from~\eqref{eq:warped_action} are discussed in Appendix~\ref{app:lower_potential}. 
    Before proceeding, we should clarify an important aspect of our derivation of~\eqref{eq:moduli_space_metric}. It is known that to compute the kinetic metric, one should include compensator fields because of subtleties with the gauge fixing: see, e.g., \cite{Douglas:2008jx}.\footnote{Note that we cannot compare the results of this paper with ours because in \cite{Douglas:2008jx}, the reduction is not performed down to the lower-dimensional Einstein frame.} It turns out that our expression is the correct one when an additional condition holds for the background and moduli fields, as derived in Appendix~\ref{app:compensators}; see~\eqref{eq:condition_no_compensator}. In the following, we will assume that such condition holds, and we will verify this for all the explicit solutions that we analyze.

    In addition to the metric of~\eqref{eq:moduli_space_metric}, we need the dependence of the KK masses on the moduli. To extract this, consider a $D$-dimensional scalar field $\hat{\psi}$ and expand it with a complete basis of functions $f_\mathbf{n}$ that we will fix momentarily:
    \begin{equation}\label{eq:KK_expasion_scalar}
        \hat{\psi}=\psi(x)+\sum_{\mathbf{n}}\psi_\mathbf{n}(x)f_\mathbf{n}(y)\,.
    \end{equation}
    The leading contributions for large KK momentum come from the higher-dimensional kinetic term, effectively resulting in a WKB-like approximation. Given the expansion in~\eqref{eq:KK_expasion_scalar}, the relevant terms in the scalar field action are then
    \begin{align}\label{eq:kk_mass_derivation_first_step}
    -\int\dd^D x\sqrt{-G_D}(\partial\hat{\psi})^2&=
        -\int \dd^D x \sqrt{-g}\sqrt{\mathsf{M}} \, \sum_{\mathbf{k},\mathbf{l}}\Big[{e^{(d-2)(\rho-\theta)+n\sigma}}g^{\mu\nu}\d_\mu\psi_\mathbf{k} \d_\nu\psi_\mathbf{l} f_\mathbf{k} f_\mathbf{l} \notag\\&~~~~~~~~~~~~+ e^{d(\rho-\theta)+(n-2)\sigma}\psi_\mathbf{k}\psi_\mathbf{l} \mathsf{M}^{ij}\d_i f_\mathbf{k} \d_j f_\mathbf{l}\Big]\,.
    \end{align}
    We now fix the basis $\{f_\mathbf{k}\}_\mathbf{k}$ as eigenfunctions of a Laplacian operator on the internal $X_n$ space, requiring that the lower-dimensional scalars $\{\psi_\mathbf{k}\}_\mathbf{k}$ be canonically normalized. To do this, we define the weighted Laplace operator
    \begin{equation}\label{eq:laplacian_weighted}
        \frac{1}{\mu(y)}\d_i\left[\mu(y)\mathsf{M}^{ij}\d_j \right]\,,
    \end{equation}
    where the measure $\mu(y)$ is
    \begin{equation}
        \mu(y)=\sqrt{\mathsf{M}}e^{(d-2)(\rho-\theta)+n\sigma}\,.
    \end{equation}
    In these expressions, $\rho$, $\sigma$, and $\theta$ are the background metric fields (without their fluctuations as in~\eqref{eq:moduli_space_metric}). Note that the weighted Laplace operator in~\eqref{eq:laplacian_weighted} is self-adjoint with respect to the canonical inner product on the weak $L^{2}(X_n,\mu)$ space. Note also that in the unwarped limit, $\mu(y)$ reduces to $\sqrt{\mathsf{M}}$, see~\eqref{eq:theta}, and that \eqref{eq:laplacian_weighted} becomes the scalar Laplace operator on $X_n$ with the metric $\mathsf{M}_{ij}$. We then take $\{f_\mathbf{k}\}_\mathbf{k}$ to be the eigenfunctions of the operator in~\eqref{eq:laplacian_weighted}:
    \begin{equation}\label{eq:eigenfunctions-laplacian}
        \frac{1}{\mu(y)}\d_i\left[\mu(y)\mathsf{M}^{ij}\d_j f_\mathbf{k}\right]=-\lambda_\mathbf{k}^2 f_\mathbf{k}\,,
    \end{equation}
    normalized so that 
    \begin{equation}\label{eq.norm}
        \int_{X_n}\dd^n y \, \mu(y)f_\mathbf{n} f_\mathbf{m}=\delta_{\mathbf{mn}}\,.
    \end{equation}
    Using \eqref{eq.norm} in \eqref{eq:kk_mass_derivation_first_step} and integrating by parts in the second term, we obtain
    \begin{align}\label{eq:kk_mass_warped_derivation}
        & \int \dd^d x \sqrt{-g}\left[-\sum_{\mathbf{l}}(\d_\mu\psi_\mathbf{l})^2+\sum_{\mathbf{k},\mathbf{l}}\psi_\mathbf{k}\psi_\mathbf{l} \int_{X_n} \dd^n y f_\mathbf{k} \d_i\left(\sqrt{\mathsf{M}} \, e^{d(\rho-\theta)+(n-2)\sigma}\mathsf{M}^ {ij}\d_j f_\mathbf{l}\right) \right]\notag\\
            =&\int \dd^d x \sqrt{-g}\left[-\sum_{\mathbf{l}}(\d_\mu\psi_\mathbf{l})^2+\sum_{\mathbf{k},\mathbf{l}}\psi_\mathbf{k}\psi_\mathbf{l} \int_{X_n} \dd^n y \, \mu(y) f_\mathbf{k} \frac{1}{\mu(y)} \d_i\left(\mu(y) e^{2(\rho-\theta-\sigma)}\mathsf{M}^ {ij}\d_j f_\mathbf{l}\right) \right]\,.
    \end{align}
    Since we can decompose
    \begin{equation}\label{eq. op exp}
        \frac{1}{\mu(y)} \d_i\left(\mu(y) e^{2(\rho-\theta-\sigma)}\mathsf{M}^ {ij}\d_j f_\mathbf{l}\right)= e^{2(\rho-\theta-\sigma)}\Bigg[\underbrace{\frac{1}{\mu(y)} \d_i\left(\mu(y)\mathsf{M}^ {ij}\d_j f_\mathbf{l}\right)}_{\mathcal{O}(\lambda_\mathbf{l}^2)}+\underbrace{2\mathsf{M}^{ij}\partial_i(\rho-\sigma)\partial_jf_\mathbf{l}}_{\mathcal{O}(\lambda_\mathbf{l})}\Bigg]\,,
    \end{equation}
    the second term is subleading for large values of KK momenta and can be neglected when studying the scaling of KK masses. Note that this does not necessarily hold in strongly curved regions, where the variation of the warp factors can be arbitrarily large; see~\cite{DeLuca:2024fbc}. With this proviso, \eqref{eq:kk_mass_warped_derivation} becomes
    \begin{equation}\label{eq:kk_mass_warped_derivation_2}
        \int \dd^d x \sqrt{-g}\left[-\sum_{\mathbf{l}}(\d_\mu\psi_\mathbf{l})^2-\sum_{\mathbf{k},\mathbf{l}}\psi_\mathbf{k}\psi_\mathbf{l} \lambda_\mathbf{l}^2 \int_{X_n} \dd^n y \, \mu(y) e^{2(\rho-\theta-\sigma)}f_\mathbf{k} f_\mathbf{l} \right]\,.
    \end{equation}
    In the limit of large KK momenta, we can further simplify the mass term using results from (weighted) quantum ergodicity~\cite{schnirelman,Verdiere:1985,Zelditch:1987,DeLuca:2024fbc}:\footnote{Note that (weighted) quantum ergodicity is conjectured to hold for manifold with not too many isometries, see \cite[Appendix A]{DeLuca:2024fbc}. However, this will not be a problem for the specific spaces considered in this paper.}
    \begin{equation}\label{eq. q erg}
        \int_{X_n} \dd^n y \, \mu(y)  e^{2(\rho-\theta-\sigma)} f_\mathbf{k} f_\mathbf{l} \to \frac{ \int_{X_n} \dd^n y \, \mu(y) e^{2(\rho-\theta-\sigma)} }{\int_{X_n} \dd^n y \, \mu(y)} \delta_{\mathbf{kl}}\quad\text{for large }\mathbf{k}\\,\mathbf{l}\,.
    \end{equation}
    Therefore, \eqref{eq:kk_mass_warped_derivation_2} takes the form
    \begin{equation}
        \int \dd^d x \sqrt{-g}\sum_{\mathbf{l}}\left[-(\d_\mu\psi_\mathbf{l})^2-\lambda_\mathbf{l}^2\psi_\mathbf{l}^2 \int_{X_n} \dd^n y\sqrt{\mathsf{M}} \, e^{d(\rho-\theta)+(n-2)\sigma} \right]\,.
    \end{equation}
    For large $\mathbf{k}$, $|\lambda_\mathbf{k}|\sim\mathcal{O}(\mathbf{k}^{1/n})$ in~\eqref{eq:eigenfunctions-laplacian}, see \eqref{eq. WEYLS LAW}, and it is independent of $\mu(y)$. In fact, in this limit, the terms that contain derivatives of the warp factors are subleading, and the result follows from the normalization of the volume in the metric $\mathsf{M}_{ij}$.
    Then, in the large-momentum limit, the KK masses scale as
    \begin{equation}\label{eq:KK_mass_estimate}
        \frac{m_{\rm KK}}{M_{{\rm Pl},d}}\sim \left[\int_{X_n} \dd^n y\sqrt{\mathsf{M}}\,  e^{d(\rho-\theta)+(n-2)\sigma} \right]^{\frac{1}{2}}\,.
    \end{equation}
    Note that in~\eqref{eq:KK_mass_estimate}, the KK scale is written in terms of the $d$-dimensional Planck mass. In the unwarped result where both $\rho$ and $\sigma$ do not have dependence on the internal coordinates, one has $\theta=\rho+\frac{n}{d-2}\sigma$, see \eqref{eq:theta}, so that \eqref{eq:KK_mass_estimate} reduces to    
    \begin{equation}
        \left.\frac{m_{\rm KK}}{M_{{\rm Pl},d}}\right|_{\rm unwarp.}\sim \exp\left\{\frac{1}{2}\big[d(\rho-\theta)+(n-2)\sigma\big]\right\}=\exp\left\{-\sqrt{\frac{d+n-2}{n(d-2)}}\hat{\sigma}\right\}\,,
    \end{equation}
    thus recovering \eqref{eq. KK unwarped}.

   Consider now a higher-dimensional tower of states that becomes exponentially light when moving towards an infinite-distance limit $\hat{\Delta}$ along the higher-dimensional moduli space, parameterized by $\{\hat{\varphi}^{\hat{a}}\}_{\hat a}$ (not necessarily canonically normalized).\footnote{Given a geodesic trajectory $\gamma:[0,\infty)\to\hat{\mathcal{M}}$ on the higher-dimensional moduli space, 
   with metric $\hat{\mathsf{G}}_{\hat a\hat b}$, the traveled geodesic distance is straightforwardly computed as
   \begin{equation}
       \hat{\Delta}(\sigma)=\int_0^{\sigma}\dd s\,\sqrt{\hat{\mathsf{G}}_{\hat a\hat b}(\hat{\varphi}(s))\partial_s\hat{\varphi}^{\hat a}(s)\partial_s\hat{\varphi}^{\hat b}(s)}\,.
   \end{equation}
   For the case where we have a single scalar $\hat\varphi$, one can always perform a field redefinition such that $\mathsf{G}_{\hat{\varphi}\hat{\varphi}}=1$, which allows the identification $\hat{\Delta}\equiv\hat{\varphi}$.} To express the scaling of this tower in terms of all moduli upon compactification, we need the relation between higher- and lower-dimensional Planck units. To this end, take one of the massive fields in the $D$-dimensional theory (for simplicity, we take the zero mode of such field, and assume that the $y$-dependence is mild enough so that it can be safely ignored for this computation) and reduce:
   \begin{align}
       S_D&\supset-\frac{1}{2\kappa_D^2}\int\dd^Dx\sqrt{-G_D}\Big[(\partial\chi)^2+m_\chi(\hat\varphi)^2\chi^2\Big]\notag\\&=-\int\dd^d x\sqrt{-g}\Bigg\{\frac{1}{2\kappa_d^2}(\partial\chi)^2-\chi^2\frac{1}{2\kappa_D^2}\int_{X_n}\dd^ny\sqrt{\mathsf{M}}\, e^{d(\rho-\theta)+n\sigma}m_\chi(\hat\varphi)^2\Bigg\}\,,
   \end{align}
   where the scalars $\{\hat{\varphi}^{\hat{a}}\}_{\hat a}$ that govern the higher-dimensional mass scale can also depend on the internal coordinates due to the warping. We then obtain
    \begin{equation}\label{eqref. red tow warp}
        \frac{m_{{\chi}}}{M_{{\rm Pl},d}}=\left[\int_{X_n}\dd^n y\sqrt{\mathsf{M}}\,e^{d(\rho-\theta)+n\sigma}\left(\frac{m_\chi({\hat{\varphi})}}{M_{{\rm Pl},D}}\right)^2\right]^{\frac{1}{2}}\,.
    \end{equation}
    As a concrete application, we take our higher-dimensional modulus $\hat{\varphi}$ to be canonically normalized, with the characteristic mass of the tower of states scaling as in~\eqref{eq. additional unwarped-1}, $m_{\chi} =M_{{\rm Pl},D}e^{-\lambda_D\hat{\varphi}}$ for some $\lambda_D=\mathcal{O}(1)$. Then, the above expression can be rewritten as 
    \begin{equation}\label{eqref. red tow warp 1}
        \frac{m_{{\chi}}}{M_{{\rm Pl},d}}=\left[\int_{X_n}\dd^n y\sqrt{\mathsf{M}}\,e^{d(\rho-\theta)+n\sigma-2\lambda_D\hat{\varphi}}\right]^{\frac{1}{2}}\,.
    \end{equation}
    
    An immediate consequence of \eqref{eqref. red tow warp} is the scaling of the higher-dimensional Planck mass. Taking $m_{\hat{\varphi}}=M_{{\rm Pl},D}$, one recovers
    \begin{equation}\label{eq. warped MplD}
        \frac{M_{{\rm Pl},D}}{M_{{\rm Pl},d}}=\left[\int_{X_n}\dd^ny\sqrt{\mathsf{M}}e^{d(\rho-\theta)+n\sigma}\right]^{\frac{1}{2}}\,,
    \end{equation}
    which again reduces to \eqref{eq. Mpl unwarped} in the unwarped limit.

    Recently, \cite{DeLuca:2024fbc} also analyzed the behavior of the KK scale in warped compactifications, focusing on KK copies of spin 2 fields. The result was given, though, in higher-dimensional Planck units, so the comparison is not immediate. If one is interested in the result in lower-dimensional Planck units, one should do the dimensional reduction in one-go as we do in this Section. Using \eqref{eq. warped MplD}, the rough scaling of the masses at first glance seems to be compatible with our results, but it would be nice to perform a more thorough comparison.

\subsection{Highly warped limits}\label{ssec:highly_warped_limit}

The above expressions are valid for any point in the moduli space, provided the KK momentum and the overall volume are large enough, and, as we have seen, recover the unwarped result \eqref{eq. KK unwarped} when both $\rho$ and $\sigma$ are constants. For most trajectories in the moduli space that approach the decompactification limit, the warping gets diluted in the infinite-distance limit, and the asymptotic scaling of the KK mass is well approximated by the unwarped result. However, in this paper, we are interested in those particular trajectories where the warping does not get diluted asymptotically, thus modifying the KK scaling. We denote these trajectories as ``\emph{highly warped limits}''. They correspond to those asymptotic geodesic trajectories approaching an infinite-distance point for which the warping profiles do not go to a constant, i.e., 
\begin{equation}
    \lim_{\Delta\varphi\to\infty}\vec{\nabla}_y\rho(y)\neq 0\quad\text{and/or}\quad \lim_{\Delta\varphi\to\infty}\vec{\nabla}_y\sigma(y)\neq 0\,,
\end{equation}
where the gradient $\vec{\nabla}_y$ is taken with respect to the internal coordinates of $X_n$. Through the equations of motion (see \eqref{eq. gen EOM} and \eqref{e.eom} in Appendix \ref{app:lower_potential}), this implies that the higher-dimensional scalars $\hat{\varphi}^{\hat a}$ can also have non-trivial internal profiles. As we move towards the decompactification limit and the compact manifold $X_n$ grows, the warping does not ``dilute'' or ``go away'', and the resulting higher-dimensional spacetime is not simply Minkowski space; it can contain some defects backreacting on the non-compact space or correspond to some running time-dependent solution.

In the different examples that will be considered in this paper, these \emph{highly warped limits} will correspond to a set of trajectories in the lower-dimensional moduli space with \emph{measure zero} in the space of asymptotic geodesics, see \cite[Appendix C]{Castellano:2023jjt}. This can be explicitly checked for our highly warped limits decompactifying one dimension, as the moduli that we leave fixed as we decompactify act as \emph{impact parameters} that are completely subleading in the moduli space metric, and thus do not affect the asymptotic direction, see Appendix \ref{app:codim_1_general} and \cite[Appendix C]{Etheredge:2023odp}.

As we will see in the following, the behavior of the KK mass in terms of the moduli space distance along these highly warped limits can be drastically different from the unwarped expectations.

\section{Warped codimension-one backgrounds}\label{sec:warped_codim_1}

We now apply the results of Section~\ref{ssec:warped} to warped codimension-one backgrounds. 
These correspond to taking a one-dimensional internal manifold $X_n$, so that the metric ansatz \eqref{eq:warped_cpt} becomes
\begin{equation}\label{eq.met1d}
    \dd s^2=e^{2\rho(y)}\eta_{\mu\nu}\dd x^\mu\dd x^\nu+ e^{2\sigma(y)}\dd y^2\,,
\end{equation}
where $y$ is the coordinate parameterizing the compact space. Equivalently, the localized defects that source the warping have codimension one. The results obtained in this Section can be easily generalized to the case in which we have additional unwarped compact dimensions.

In the following, we focus on a single higher-dimensional scalar $\hat{\varphi}$ with a canonically normalized kinetic term, $\hat{\mathsf{G}}_{\hat\varphi\hat\varphi}=1$, and an exponential potential 
\begin{equation}
\label{V1d}
    \hat{V}(\hat{\varphi})=\kappa_D^{-2}V_0e^{\gamma\hat{\varphi}}\,.
\end{equation}
We make no assumption about the sign of $V_0$, but since the action is invariant under $(\hat{\varphi},\gamma)\to(-\hat{\varphi},-\gamma)$, we lose nothing by taking $\gamma\geq0$.
The bulk equations of motion (see \eqref{e.eom}) for the metric fields and the scalar $\hat{\varphi}$ are\footnote{These equations we describe here, together with their solutions, have been extensively described before in the literature; see, e.g., \cite{Poletti:1994ff,Cvetic:1996vr,DeWolfe:1999cp,Sonner:2005sj}.}
\begin{subequations}\label{e.eom1d}
    \begin{align}\label{e.eom rhosigma}
        \hat{\varphi}''+ \hat{\varphi}'(d\rho'-\sigma')-\gamma V_ 0e^{2\sigma+\gamma \hat{\varphi}}&=0\,,\\
        \rho''+\rho'(d\rho'-\sigma')+\frac{2V_0}{d-1}e^{2\sigma+\gamma \hat{\varphi}}&=0\,,\\
        \rho''-\rho'\sigma'+\frac{1}{d-1}( \hat{\varphi}')^2&=0\,,
    \end{align}
\end{subequations}
where all derivatives are with respect to the variable $y$. For our purposes, we only need to consider the bulk part of the equations. See Appendix \ref{app:lower_potential} for the full discussion involving localized sources.

\begin{figure}[hbt]
    \centering
    \includegraphics[width=0.4\linewidth]{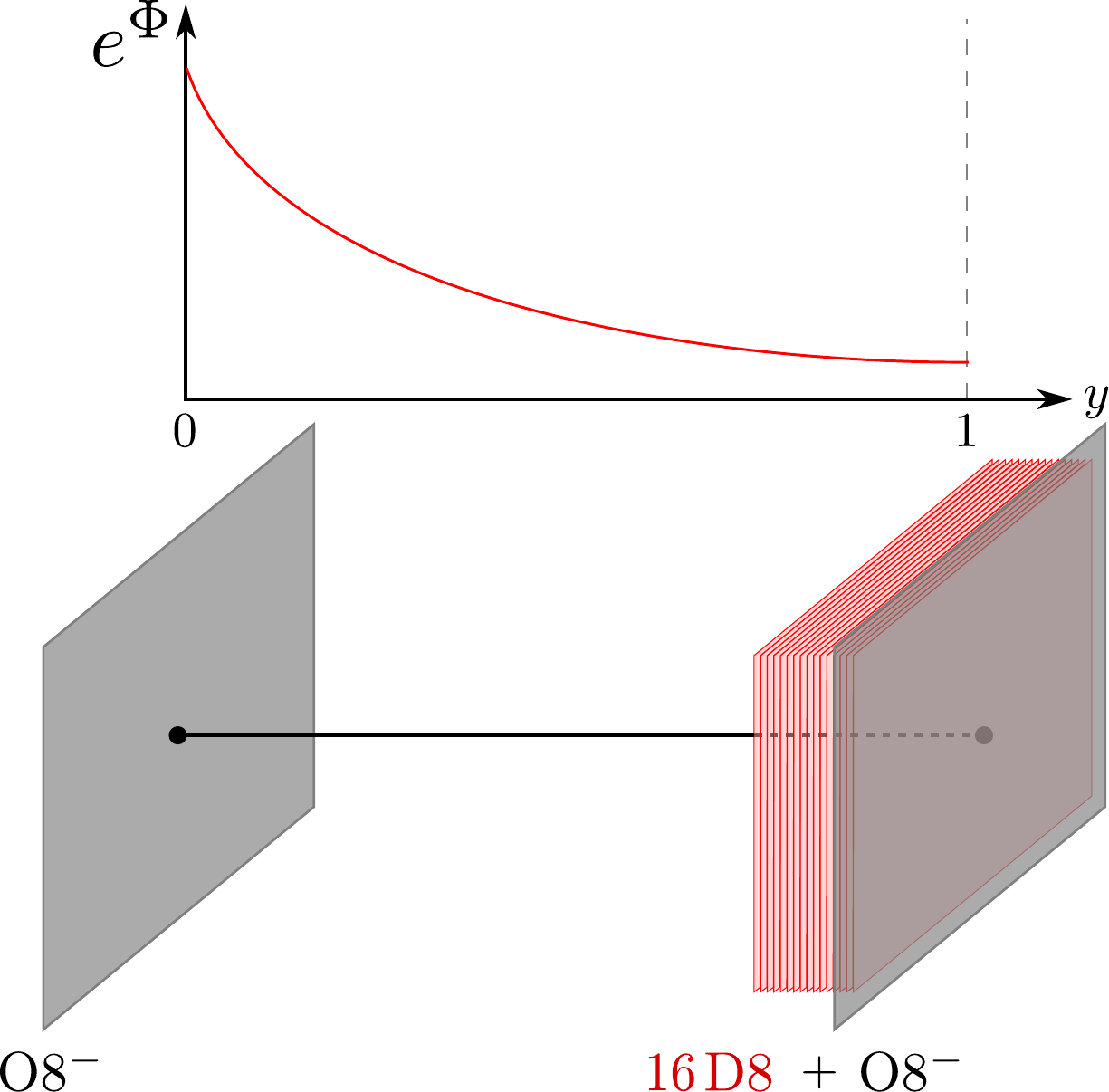}
    \caption{Brane arrangement in the massive type IIA orientifold with gauge groups $SO(32)$. The Romans mass is non-vanishing, and backreacts on the metric and 10d dilaton $\Phi$. In the limiting case of maximal warping, the dilaton diverges at the location of the opposite O8$^{-}$ plane, giving the enhancement $SO(32)\to SO(32)\times SU(2)$, see \cite{Polchinski:1995df,Aharony:2007du,Etheredge:2023odp}.
    \label{fig.typeIprime}}
\end{figure}

Equations of this type appear for example in massive type IIA supergravity, in which the scalar potential is the Romans mass given by the $F_0$ flux. 
In this case, the ten-dimensional Einstein-frame action is 
\begin{equation}\label{eq. action mIIA}
    S_{10}^{\rm (\rm Ein)}\supset\frac{1}{2\kappa_{10}^2}\int \dd^{10} x\sqrt{-G_{10}}\left\{\mathcal{R}_{10}-\frac{1}{2}(\partial\Phi)^2-e^{\frac{5}{2}\Phi}F_0^2\right\}\,+S_{\rm loc.},
\end{equation}
and, after canonically normalizing the 10d dilaton, $\hat{\varphi}=\frac{1}{\sqrt{2}}\Phi$,\footnote{\label{fn. string frame} Note that given the $D$-dimensional dilaton $\Phi_D$ (in \eqref{eq. action mIIA} we write $\Phi\equiv \Phi_{10}$ for clarity), transforming from the string frame to the Einstein frame,
\begin{equation}
    \frac{1}{2\kappa_D^2}\int\dd^{D}x\sqrt{-G_D^{(\rm s)}}e^{-2\Phi_{D}}\Big\{\mathcal{R}_D^{\rm (s)}+4(\partial\Phi_{D})\Big\}\to \frac{1}{2\kappa_D^2}\int\dd^{D}x\sqrt{-G_D}\Big\{\mathcal{R}_D-\frac{4}{D-2}(\partial\Phi_{D})^2\Big\}\,,
\end{equation}
one finds that $\mathsf{G}_{\Phi_{D}\Phi_{D}}=\frac{4}{D-2}$, thus arriving at the relation $\hat{\varphi}=\frac{2}{\sqrt{D-2}}\Phi_D$.}  the scalar potential reads
\begin{equation}\label{eq. romans pot}
    \hat{V}(\hat{\Phi})=\frac{1}{2}\kappa^{-2}_{10}F_0^2e^{\frac{5}{\sqrt{2}}\hat{\varphi}} \,,
\end{equation}
so $\gamma=\frac{5}{\sqrt{2}}$.
The codimension-one warped solutions \eqref{eq.met1d} of massive type IIA supergravity were first described by Polchinski and Witten in~\cite{Polchinski:1995df}, and have been the main focus of the related works~\cite{Etheredge:2023odp,Bedroya:2023tch}. These 9d Minkowski vacua are described by type I$'$ string theory, which is type IIA string theory on an $\mathbb{S}^1/\mathbb{Z}_2$ orientifold, where the charge of the O8$^-$ planes is canceled globally but not locally by 16 D8-branes, thus sourcing $F_0$ flux: the Romans mass. At strong coupling they are dual to heterotic string theory on a circle~\cite{Polchinski:1995df,Aharony:2007du}. Even if there is a non-vanishing potential in 10d, the 9d space is flat due to competing effects from the higher-dimensional potential and the localized sources, as explained in Appendix \ref{app:lower_potential}. Of course, this is at the cost of having a non-trivial warping and a non-trivial spatial profile for the dilaton $\Phi(y)$. The locations of the branes and orientifolds set the gauge group and the metric and dilaton profiles; see Figure~\ref{fig.typeIprime} for the arrangement of branes in the case of an $SO(32)$ gauge algebra and \cite{Aharony:2007du,Etheredge:2023odp} for more details on the $E_8\times E_8$ case.  In a nutshell, there are two non-compact moduli, which in the $SO(32)$ case are associated with the size of the interval and the value of the 10d dilaton $\Phi$ at the location of the O8$^-$ plane without D8-branes, see Figure~\ref{fig.typeIprime}. The pure decompactification limit is obstructed and must be accompanied by a weak coupling limit; i.e, one needs to send the dilaton to zero as the interval grows in order to avoid the dilaton blowing up at a regular point between the orientifolds.
In a generic infinite distance limit of the moduli space, the gradient $\partial_y\Phi$ goes to zero as the interval decompactifies, so one reaches the 10d Minkowski solution of massless type IIA. However, there is a particular direction along which the gradients do not vanish in the infinite distance limit, and one decompactifies instead to the 10d  time-dependent  solution of massive Type IIA in \eqref{eq. action mIIA} corresponding to a running dilaton (see \cite{Etheredge:2023odp}). This corresponds to a highly warped limit according to the notation introduced in Section \ref{ssec:highly_warped_limit}.

Other examples of warped solutions of the form \eqref{eq.met1d} in string theory come from tachyon-free non-supersymmetric models and their Dudas--Mourad vacua~\cite{Dudas:2000ff}. In those cases, the exponential rate of the potential is $\gamma=\frac{3}{\sqrt{2}}$ or $\frac{5}{\sqrt{2}}$ with $d=9$.\\

In the following, we will investigate the KK scaling in general codimension-one backgrounds with an exponential potential as in in \eqref{V1d} with arbitrary $\gamma$, thus including the above well-known cases for special values of $\gamma$.

    \subsection{Scaling solutions}\label{ssec:scaling}

We begin by considering \emph{scaling solutions} to \eqref{e.eom1d}. These are types of solutions with logarithmic scaling that include, among others, the Polchinski--Witten vacua.

    We can freely reparameterize the coordinate $y$, and one might naively use this freedom to fix $\sigma(y)$ completely, but the analysis of Section~\ref{ssec:warped} requires a fixed range for $y$; therefore, gauge fixing can at most determine $\sigma(y)$ up to a constant. We choose $\sigma(y)=\rho(y)+\sigma_0$. However, since we are considering compactifications with a flat $d$-dimensional spacetime, we can always rescale the macroscopic spacetime directions and add a constant term to $\rho(y)$, $\rho(y)\to\rho(y)+\rho_0$, so as to effectively implement the gauge $\sigma(y)=\rho(y)$.\footnote{\label{fn. exact lap} Note that with this gauge, the Laplace operator in \eqref{eq:laplacian_weighted} is exact, see \eqref{eq. op exp}, and thus we are not disregarding subleading KK momentum corrections. Notice also that in this case, we do not need (weighted) quantum ergodicity \eqref{eq. q erg} and thus the results for this type of solutions also apply to the low KK momentum states. While in this section we focus on the parametric dependence on the moduli, Appendix \ref{app:Full sol} contains a discussion on how warping affects the numerical value of the eigenvalues of the warped Laplacian.} The equations of motion in~\eqref{e.eom1d} become
    \begin{subequations}\label{eq.rho equals sigma}
    \begin{align}
         \hat{\varphi}''+(d-1) \hat{\varphi}'\rho'-\gamma V_0e^{2\rho+\gamma \hat{\varphi}}&=0\,,\\
        \rho''+(d-1)(\rho')^2+\frac{2V_0}{d-1}e^{2\rho+\gamma \hat{\varphi}}&=0\,,\\
        \rho''-(\rho')^2+\frac{1}{d-1}(\hat\varphi')^2&=0\,.
    \end{align}
    \end{subequations} 
    As in \cite{Polchinski:1995df,Etheredge:2022opl}, we consider a logarithmic ansatz for $\rho$ and $\hat{\varphi}$,
    \begin{equation}
     \hat{\varphi}(y)=\varphi_1\log[B(A+y)]\,,\quad \rho(y)=\rho_0+\rho_1\log[B(A+y)]\,,\quad\text{with }A,\,B>0\,,
    \end{equation}
    where the parameters $A$ and $B$ are to be promoted to moduli with dependence on the external coordinates $x^\mu$. Here, $\varphi_1$ and $\rho_1$ are numerical constants, and $\rho_0$ may also depend on the moduli as it encodes the rescaling of the coordinate $y$ (see the comments about gauge fixing at the beginning of this Section). This yields the following solution:
    \begin{subequations}\label{eq.sols rho sigma}
    \begin{align}
        \hat{\varphi}(y)&=-\frac{2 (d-1)\gamma}{ (d-1)\gamma ^2-4}\log[B(A+y)] \,, \label{eq. varphi scaling}\\
        \rho(y)&=\frac{4}{ (d-1)\gamma ^2-4}\log[B(A+y)]+\log B+\frac{1}{2}\log\left[\frac{2 (d-1) \left(\gamma ^2 (d-1)-4 d\right)}{\left[\gamma ^2 (d-1)-4\right]^2V_0}\right]\,,\label{eq. rho scaling}
    \end{align}
    \end{subequations}
    provided that $\gamma\neq\frac{2}{\sqrt{d-1}}$. Furthermore, requiring $\rho(y)$ to be real in \eqref{eq. rho scaling} imposes the following relation on $V_0$:
    \begin{equation}\label{e.restr1warp}
        {\rm sign}(V_0)= {\rm sign}\left(\gamma-2\sqrt{\frac{d}{d-1}}\right)={\rm sign}\left(\gamma-2\sqrt{\frac{D-1}{D-2}}\right)\,.
    \end{equation}
    This condition is the same that appears for the critical exponents for \emph{end-of-the-world branes} in $D$ dimensions found in~\cite[Section 2.2]{Angius:2022aeq}. For positive potentials, $V_0>0$, \eqref{e.restr1warp} implies the Trans-Planckian Censorship Conjecture bound \cite{Bedroya:2019snp}, $\gamma=\frac{|\nabla \hat{V}|}{\hat V}> 2\sqrt{\frac{D-1}{D-2}}$ (see also \cite{Blumenhagen:2023abk}). For $\gamma=2\sqrt{\scriptstyle\frac{d}{d-1}}$, condition \eqref{e.restr1warp} imposes $V_0=0$.\footnote{For $V_0=0$, the equations of motion \eqref{eq.rho equals sigma} result in two possible solutions: 
    \begin{subequations}\label{eq. V0 sols}
        \begin{equation}
            \hat{\varphi}_{\rm w}(y)=-\sqrt{\frac{d}{d-1}}\log[B(A+y)]\,,\qquad \rho_{\rm w}(y)=\frac{1}{d-1}\log[B(A+y)]+\log B\,,
        \end{equation}
        \begin{equation}
            \hat{\varphi}_{\rm uw}(y)=\hat{\varphi}_0\,,\qquad \rho_{\rm uw}(y)=\rho_0\,,
        \end{equation}
    \end{subequations}
    where the labels ``w'' and ``uw'' stand for ``warped'' and ``unwarped''. The constant term in $\rho_{\rm w}(y)$ and the sign in $\hat\varphi_{\rm w}(y)$ are not set by the equations of motion, but the choice in \eqref{eq. V0 sols} is the only one that reproduces the {moduli dependence in the limit} $\gamma\to2\sqrt{\frac{d}{d-1}}$; see \eqref{eq. rho scaling}.} Moreover, there are no scaling solutions for $\gamma=\frac{2}{\sqrt{d-1}}$ (see though Section \ref{ssec:codim_1_general} for more general solutions for all values of $\gamma$).

    We refer to the profiles in~\eqref{eq.sols rho sigma} as \emph{scaling solutions} because of their analogy to cosmological attractors that bear the same name. In fact, from~\eqref{eq.rho equals sigma}, one can extract an energy-like condition---the \emph{Hamiltonian constraint} of the dynamical system\footnote{See, e.g.,~\cite{Lust:2025vyz} for another recent use of this in codimension-one warped compactifications.}---in which the total energy is equally distributed among the different terms. These are also the typical solutions that arise with end-of-the-world branes, see \cite{Dudas:2000ff,Dudas:2002dg,Dudas:2004nd,Basile:2018irz,Antonelli:2019nar,Mourad:2021qwf,Mourad:2021roa,Buratti:2021fiv,Raucci:2022jgw,Angius:2022aeq,Angius:2022mgh,Basile:2022ypo,Blumenhagen:2023abk} for a non-comprehensive list of these backgrounds and \cite{Angius:2023uqk,Angius:2024zjv} for generalizations with several codimension-one boundaries.

    We are interested in the highly warped regime, as discussed in Section \ref{ssec:highly_warped_limit}. The gradients of the metric and dilaton fields for the scaling solutions with $\gamma\neq \frac{2}{\sqrt{d-1}}$ are
    \begin{equation}\label{eq. grad codim1}
        \partial_y\hat{\varphi}=-\frac{2 (d-1)\gamma}{ (d-1)\gamma ^2-4}\left(y+A\right)^{-1}\,,\quad \partial_y\rho=\frac{4}{ (d-1)\gamma ^2-4}\left(y+A\right)^{-1}\,.
    \end{equation}    
  It follows that $\partial_y\hat{\varphi}$ and $ \partial_y\rho$ do not vanish asymptotically in trajectories where $A\not\to\infty$. To identify the decompatification limits, we compute the volume of the warped interval: 
    \begin{equation}\label{eq. vol int}
        \int_0^1\dd y \,e^\rho=\underbrace{\tfrac{\gamma ^2 (d-1)-4}{\gamma ^2 (d-1)}\left[(A+1)^{\frac{\gamma ^2 (d-1)}{\gamma ^2 (d-1)-4}}-A^{\frac{\gamma ^2 (d-1)}{\gamma ^2 (d-1)-4}}\right]}_{>0}B^{\frac{\gamma ^2 (d-1)}{\gamma^2 (d-1)-4}}\,.
    \end{equation}
   
   We will focus on the \emph{warped decompactification limits} corresponding to 
   
    \begin{equation}\label{eq. highly warped limits scaling}
         B\to 0\quad \text{for }\, \gamma<\frac{2}{\sqrt{d-1}}\quad\text{and}\quad B\to \infty\quad \text{for }\, \gamma>\frac{2}{\sqrt{d-1}}\,,
    \end{equation}
    with finite $A$, since the volume in \eqref{eq. vol int} diverges in these limits. Note that when $A\to\infty$, $\partial_y\hat{\varphi},\,\partial_y\rho\to 0$ at infinite distance, which means decompactification to a Minkowski vacuum where the warping gets diluted. Here, the $A\geq 0$ modulus can be understood as an \emph{impact parameter}, associated with the value of $\hat{\varphi}$ at $y=0$. It can be explicitly seen that for finite $A$, all these highly warped trajectories describe parallel directions with the same asymptotic tangent vector. 
    In order to see this, we extract the kinetic term of the modulus $B$ from \eqref{eq:moduli_space_metric} and obtain\footnote{n Appendix \ref{app:compensators}, we show that the compensator fields do not change the moduli space metric for this type of solutions.}
    \begin{equation}\label{eq:metric anst1}
        \mathsf{G}_{BB}=\frac{(d-1)^2\gamma ^2  \left[(d-1)\gamma^2+4(d-2)\right]}{(d-2) \left[ (d-1)\gamma ^2-4\right]^2}B^{-2}\,,
    \end{equation}
    which is independent of $A$ and exact for any value of $B$ and $A$, so that all the trajectories with $B\to 0$ or $\infty$ and $A$ fixed have parallel tangent vectors (see also \cite[Appendix C]{Etheredge:2023odp}). By plugging \eqref{eq.sols rho sigma} into \eqref{eq:KK_mass_estimate}, we obtain the behavior of the KK masses for large momenta:
    \begin{equation}
        -\partial_B\log\frac{m_{\rm KK}}{M_{{\rm Pl},d}}\sim\frac{ (d-1)^2\gamma ^2}{(d-2) \left[(d-1)\gamma^2-4\right]}B^{-1}\,.
    \end{equation}
    We can then compute the exponential rate at which the KK modes become light with the geodesic field distance in the decompactification limits $B\to\infty$ or $B\to 0$ (depending on the value of $\gamma$, see \eqref{eq. highly warped limits scaling}) using \eqref{eq. 1 mod lambda}:\footnote{For $\gamma<\frac{2}{\sqrt{d-1}}$, the decompactification limit is given by $B\to 0$ (with fixed $A$) and therefore there is an additional sign due to the directionality of the derivative, which precisely cancels the factor ${\rm sign}((d-1)\gamma^2-4)$ that appears when evaluating \eqref{eq. 1 mod lambda}.}
    \begin{equation}
    \label{eq:kk_rate_case_1}
        \lambda_{\rm KK}=\sqrt{\frac{d-1}{d-2}}\left(1+\frac{4(d-2)}{(d-1)\gamma^2}\right)^{-1/2}\,.
    \end{equation}
    
Note that we are not computing the vector $\vec\zeta_{\rm KK}$ defined in \eqref{scalingvector}, nor its norm. These will generally have additional dependence on $A$ and other shape moduli, if present. The above exponential rate is instead the projection $\lambda_{\rm KK}=\hat{T}\cdot \vec\zeta_{\rm KK}$ (see \eqref{exprate}) along the trajectory where $B\to \infty$ or 0 (depending on the value of $\gamma$, see \eqref{eq. highly warped limits scaling}) while the other moduli are kept fixed. When applied to the massive type IIA case of Figure~\ref{fig.typeIprime}, with $d=9$ and $\gamma=\frac{5}{\sqrt{2}}$ (see \eqref{eq. action mIIA}), this reproduces the highly warped result found in~\cite{Etheredge:2023odp}: $\lambda_{\rm KK}=\frac{5}{\sqrt{28}}$.

  Similarly, using \eqref{eq. warped MplD}, we can compute the exponential rate at which $M_{{\rm Pl},d+1}$ becomes light in lower-dimensional Planck units:
\begin{equation}
\label{eq.ex-Mpl}
  \lambda_{{{\rm Pl},d+1}}=\frac{1}{\sqrt{(d-1)(d-2)}}\left(1+\frac{4(d-2)}{(d-1)\gamma^2}\right)^{-\frac{1}{2}}\,.
  \end{equation}

    For completeness, let us also consider a tower of KK modes associated with the decompactification from $D=d+1$ to $D'=d+1+n$, assuming that the extra $n$ dimensions are unwarped. In the $D=(d+1)$-dimensional theory, this KK tower behaves as
    \begin{equation}
    \label{mKKextra}
        \frac{m_{{\rm KK'},n}}{M_{{\rm Pl},d+1}}=\exp\left(-\sqrt{\frac{D+n-2}{n(D-2)}}\hat\varphi\right)=\exp\left(-\sqrt{\frac{d+n-1}{n(d-1)}}\hat\varphi\right)\,.
    \end{equation}
    Using \eqref{eqref. red tow warp 1}, this results in the following exponential rate for the additional KK tower along the highly warped limit $B\to\infty$ or $0$ (depending on the value of $\gamma$, see \eqref{eq. highly warped limits scaling}) with $A$ fixed, from the perspective of the lower $d$-dimensional theory:
    \begin{equation}
    \label{eq.ex-KK}
      \lambda_{\rm KK'}=\frac{1}{\sqrt{(d-1)(d-2)}}\left(1+\frac{4(d-2)}{(d-1)\gamma^2}\right)^{-1/2}\left(1-\frac{2(d-2)}{\gamma}\sqrt{\frac{d+n-1}{n(d-1)}}\right)\,.
    \end{equation}
   
     Replacing $d=9$, $n=1$, and $\gamma=\frac{5}{\sqrt{2}}$ in the above equation, one obtains $\lambda_{\rm KK'}=-\frac{1}{\sqrt{7}}$, precisely the exponential rate of M-theory KK modes in the type I$'$ Polchinski--Witten solution with gauge group $E_8\times E_8$, see~\cite[Figure 11]{Etheredge:2023odp}, dual to M-theory on a cylinder $\mathbb{S}^1/\mathbb{Z}_2\times\mathbb{S}^1$ \cite{Aharony:2007du}. However, as we will discuss in Section \ref{ss. results and bounds}, one should take this last result with a grain of salt, as the volume of the additional compact dimensions actually becomes small in $M_{{\rm Pl},d+1+n}$ units along the highly warped limits, so the supergravity approximation cannot really be trusted.\\

 Before closing this section, let us comment on additional highly warped decompactification limits beyond those in \eqref{eq. highly warped limits scaling}. When $\gamma<\frac{2}{\sqrt{d-1}}$, the trajectory $A\to 0$ with fixed $B$ appears to be an additional highly warped decompactification limit since the volume of the internal space in \eqref{eq. vol int} also diverges in that case. Moreover, one can check using \eqref{eq:moduli_space_metric} (see \eqref{eq. metric A0} in Appendix \ref{app. spooky solution} for the explicit expression) that it is an infinite distance limit when $\gamma<\frac{2}{\sqrt{d-1}}$ with $\gamma\neq 0$; while it remains at finite distance for $\gamma>\frac{2}{\sqrt{d-1}}$. However, this limit exhibits some unusual properties that we outline in Appendix \ref{app. spooky solution}. 
    Among other things, the profiles of $\rho$ and $\varphi$ blow up at $y=0$, and without a clear UV completion these solutions can be affected by higher-order corrections. The exponential rate of the KK towers also differs from \eqref{eq:kk_rate_case_1} and actually increases as $\gamma$ decreases, as discussed in Appendix \ref{app. spooky solution}. We will not discuss this case further here, but it would be interesting to study these features in the future. In fact, it shares certain similarities with Randall--Sundrum-like scenarios \cite{Randall:1999ee}, as we explain next.

For the limiting case $\gamma=0$, which corresponds to a negative cosmological constant in $D=d+1$ dimensions\footnote{This applies both to the case in which $\varphi$ is stabilized in higher dimensions and when it remains as a flat direction in the AdS vacuum.}, the modulus $B$ drops out of the expression of the scaling solutions \eqref{eq.sols rho sigma}, and our profiles are given by
    \begin{equation}\label{eq. pol scal bulk}
    \hat\varphi(y)=\hat{\varphi}_0\;,\quad \rho(y)=\sigma(y)=-\log(y+A)+\frac{1}{2}\log\left(\frac{d(d-1)}{2|V_0|}\right)\,.       
    \end{equation}
    Note that the higher-dimensional scalar $\hat\varphi$ does not depend on the internal coordinate $y$. The only decompactification limit is then given by $A\to 0$. As discussed in Appendix \ref{app. spooky solution}, this limit, $A\to 0$ with $\gamma=0$, happens to be at a finite distance, with $m_{\rm KK}$ becoming light polynomially and $\frac{M_{{\rm Pl},d+1}}{M_{{\rm Pl},d}}$ going to a constant. Note that unlike other finite-distance singularities such as the conifold \cite{Strominger:1995cz} (where a single state becomes massless), this would come accompanied by an \emph{infinite} tower of light modes, which goes against expectations coming from the \emph{Emergence Proposal} \cite{Heidenreich:2018kpg,Grimm:2018ohb,Palti:2019pca}. 
    Of course, this analysis assumes the existence of suitable localized sources that can cancel the higher-dimensional negative vacuum energy and yield a warped Minkowski vacuum. Hence, the above pathologies suggest that either such sources most likely do not exist in UV-complete models, or the limit $A\to 0$ is heavily modified by UV corrections. Analogous results would be obtained from Randall--Sundrum-like compactifications \cite{Randall:1999ee} (considering them as warped compactifications and not as braneworld scenarios), assuming the existence of five-dimensional scalar fields that would lead to scalar KK towers. See Appendix \ref{app. spooky solution} for further details on this special case, and \eqref{eq:RS_as_codim_1} for the details on the Randall--Sundrum metric as a solution to the codimension-one equations of motion \eqref{e.eom1d}.\\

    \subsection{General solutions}\label{ssec:codim_1_general}

    In this subsection, we discuss the most general set of solutions to~\eqref{e.eom1d} following the approach of~\cite{Basile:2022ypo}. 
    Codimension-one vacua from exponential potentials play an important role as vacuum solutions of tachyon-free non-supersymmetric string theories~\cite{Dudas:2000ff} because they are perturbatively stable~\cite{Basile:2018irz}, and the work of~\cite{Basile:2022ypo} generalizes them and points to a realization of the Distance Conjecture using a scaling argument. We will confirm and extend the results of~\cite{Basile:2022ypo} with the tools that we have developed in Section~\ref{ssec:warped}.

    We work in a convenient gauge for the coordinate $y$ that is different from the previous choice in Section \ref{ssec:scaling}. We can set $\sigma(y)=\sigma_0-\frac{\gamma}{2}\hat{\varphi}(y)$,\footnote{\label{ft:gauge_fixing_general}As in Section \ref{ssec:scaling}, $\sigma_0$ is required because the range of $y$ must be kept fixed. This is analogous to the appearance of $\sigma_0$ in the gauge fixing before~\eqref{eq.rho equals sigma}.} shift $\hat\varphi$ by a constant $\hat\varphi_0$ such that $\sigma(y)=-\frac{\gamma}{2}\hat{\varphi}(y)$, and then reabsorb $e^{2\sigma_0 +\gamma\hat\varphi_0}$ in $V_0$. This simplifies \eqref{e.eom1d}, leading to \begin{subequations}\label{eq_codim_1_general}
    \begin{align}
        \hat{\varphi}''+\frac{\gamma}{2}(\hat\varphi')^2+d\hat\varphi'\rho'-\gamma V_0&=0\,,\\
        \rho''+\frac{\gamma}{2}\hat{\varphi}'\rho'+d(\rho')^2+\frac{2V_0}{d-1}&=0\,,\\
        (\hat{\varphi}')^2-d(d-1)(\rho')^2-2V_0&=0\,.\label{eq:codim_1_general_third}
    \end{align}
    \end{subequations}
    Reabsorbing $\sigma_0$ and $\hat\varphi_0$ in $V_0$ ($V_0 e^{2\sigma_0+\gamma\hat\varphi_0}\to V_0$) turns $V_0$ into a modulus, whose spacetime dependence comes from the hidden $\sigma_0$ and $\hat\varphi_0$.

    The explicit form of the solution depends on the sign of $V_0$. If $V_0>0$, \eqref{eq:codim_1_general_third} implies that $\hat{\varphi}'$ and $\rho'$ can be parameterized as 
    \begin{equation}\label{eq:param_1}
        \hat{\varphi}'= \pm\sqrt{2V_0}\cosh\chi\,,\quad\rho'=\sqrt{\frac{2V_0}{d(d-1)}}\sinh\chi\,,
    \end{equation}
    in terms of a single function $\chi$.
    Equations~\eqref{eq_codim_1_general} reduce to the condition
    \begin{equation} \label{eq:chi_eq_1}
        \chi'+\sqrt{\frac{2dV_0}{d-1}}\cosh\chi\pm\gamma\sqrt{\frac{V_0}{2}}\sinh\chi=0\,.
    \end{equation}
    This equation leads to different solutions depending on the value of $\gamma$, and a complete analysis can be found in Appendix~\ref{app:codim_1_general}.

    If $V_0<0$, $\hat{\varphi}'$ and $\rho'$ admit an analogous parameterization:
    \begin{equation}\label{eq:param_2}
        \hat{\varphi}'=\sqrt{2|V_0|}\sinh\chi\,,\quad\rho'=\pm\sqrt{\frac{2|V_0|}{d(d-1)}}\cosh\chi \,,
    \end{equation}
    and the equations in~\eqref{eq_codim_1_general} are equivalent to
    \begin{equation}\label{eq:chi_eq_2}
        \chi'+\gamma\sqrt{\frac{|V_0|}{2}}\cosh\chi\pm\sqrt{\frac{2d|V_0|}{d-1}}\sinh\chi=0\,.
    \end{equation}
    Again, we discuss the full solutions in Appendix~\ref{app:codim_1_general}. 

    Both solutions, eqs.~\eqref{eq:codim_1_V_pos} and~\eqref{eq:codim_1_V_neg}, have highly-warped decompactification limits that correspond to sending $V_0$ to zero, with the impact parameter (which in Appendix~\ref{app:codim_1_general} is the constant $C$) fixed. In fact, the proper length of the interval for small $V_0$ scales as 
    \begin{equation}
    \int_0^1 e^{\sigma(y)}\dd y=\int_0^1 e^{-\frac{\gamma}{2}\hat\varphi(y)}\dd y\sim (\sqrt{|V_0|})^{-\frac{\gamma\sqrt{d-1}}{\gamma\sqrt{d-1}+2\sqrt{d}}}\,,
    \end{equation}
    so that $|V_0|\to0$ is a decompactification limit.
   
    In this limit, all fields behave logarithmically in $y$, regardless of the value of $\gamma$ and the sign of $V_0$. 
    If one is not interested in the full sliding of the KK towers as in~\cite{Etheredge:2023odp} but only in the value of $\lambda_{\rm KK}$ in the highly-warped regime, all the other moduli are irrelevant and one can focus on the $V_0\to 0$ limit of the dilaton and metric fields. As with the scaling solutions of the previous subsection, one can show that highly warped trajectories with different (finite) impact parameters are asymptotically parallel and have the same unit tangent vector. We illustrate this with a specific example at the end of Appendix \ref{app:codim_1_general}.

    From the solutions \eqref{eq:codim_1_V_pos} and~\eqref{eq:codim_1_V_neg} in Appendix~\ref{app:codim_1_general}, we then have the following universal limiting behavior as $V_0\to 0$: 
    \begin{equation}\label{eq:log_asymptotic}
        \rho \sim \rho_1 \log\left(\sqrt{|V_0|} \, y\right) \qquad \text{and} \qquad \hat{\varphi} \sim \varphi_1 \log\left(\sqrt{|V_0|}\,y\right) \,,
    \end{equation}
    where $\rho_1$ and $\varphi_1$ are numerical constants that do not contain the modulus $V_0$.
    The metric for $V_0$ is then\footnote{Since the solutions studied in this section have the same limiting behavior as those from Section \ref{ssec:scaling} in the decompactification limit $V_0\to0$, the compensator fields do not contribute to the moduli space metric; see Appendix \ref{app:compensators}.}
    \begin{equation}
        \mathsf{G}_{V_0 V_0} = \frac{\varphi_1^2}{4}\left(1+\frac{d-1}{d-2}\frac{\gamma^2}{4}\right)\frac{1}{V_0^2}+\dots \,,
    \end{equation}
    up to subleading corrections, and the large-momentum behavior of the KK masses from~\eqref{eq:KK_mass_estimate} leads to
    \begin{equation}\label{eq:codim_1_exp_kk_scaling}
        \d_{|V_0|}\log m_{\rm KK}=\frac{\gamma}{4}\frac{d-1}{d-2}\frac{\varphi_1}{|V_0|}+\dots \, .
    \end{equation}
    For all cases, the exponential rate of the KK modes becoming light in the highly warped limit is then\footnote{Note that to obtain \eqref{eq:codim_1_exp_rate}, one must take the $V_0$ derivative with the positive sign, as in \eqref{eq:codim_1_exp_kk_scaling}, and not with the negative sign as in \eqref{eq. 1 mod lambda}. This is because the decompactification limit is $V_0\to0$, so that the modulus $\varphi$ of \eqref{eq. 1 mod lambda} can be taken to be $\varphi=V_0^{-1}$ and the change of variables carries an additional sign.}
    \begin{equation}\label{eq:codim_1_exp_rate}
        \lambda_{\rm KK}=\sqrt{\frac{d-1}{d-2}}\left(1+\frac{4(d-2)}{(d-1)\gamma^2}\right)^{-\frac{1}{2} } \,.
    \end{equation}
    This is the same value obtained for scaling solutions; see \eqref{eq:kk_rate_case_1}. Similarly, the same expressions as in \eqref{eq.ex-Mpl} and \eqref{eq.ex-KK} can be recovered for the scaling of the higher-dimensional Planck mass and higher-dimensional KK tower.

    \subsection{Some comments on the string scaling\label{ss.string scaling}}

 The other scale that is physically relevant in asymptotic regimes of string theory compactifications is the string scale $M_{\rm str}=\ell_{\rm str}^{-1}=(2\pi\sqrt{\alpha'})^{-1}$ (equivalently, the mass of the first string oscillator mode $m_{\rm osc}\sim M_{\rm str}$). 
 
 In unwarped compactifications, with ${\rm vol}(X_n)\gtrsim\ell_{\rm str}^n$ and small string coupling in the perturbative regime, strings feel the same local geometry along compact and non-compact dimensions, and we can still trust the usual quantization in flat space. Therefore, given the string scale in higher-dimensional Planck units, $m_{\rm osc}=M_{{\rm Pl,D}}e^{\frac{2}{D-2}\Phi_D}$, where $\Phi_D$ is the higher-dimensional dilaton, one has 
    \begin{equation}
    \label{eq. stringunwarped}
        \frac{m_{\rm osc}}{M_{{\rm Pl,}d}}\sim \frac{m_{\rm osc}}{M_{{\rm Pl},D}}\frac{M_{{\rm Pl},D}}{M_{{\rm Pl,d}}}\sim e^{\frac{2}{d+n-2}\Phi_D-\frac{n}{d-2}\rho}\qquad\text{for $X_n$ unwarped}\,,
    \end{equation}
     by rescaling the Planck mass. This is equivalent to $m_{\rm osc}=M_{{\rm Pl,d}}e^{\frac{2}{d-2}\Phi_d}$ upon defining the lower-dimensional dilaton $\Phi_d=\Phi_D-\frac{1}{2}\log({\rm vol}(X_n)/\ell_{\rm str}^n)$. \\

    However, the situation becomes more complicated in warped compactifications, where the string dilaton gets a spatial dependence and may become strongly coupled in certain regions. In this case, there is no suitable worldsheet description, and the quantization of the string might differ from the case of flat space. If the warping gets diluted in the decompactification limit, we expect the flat space result to provide a good approximation for the perturbative spectrum of the string; however, the warping does not get diluted in the cases of interest in this paper. Therefore, we do not expect $\frac{m_{\rm osc}}{M_{{\rm Pl,}d}}$ to have necessarily an expression in terms of integrals of $\rho$, $\sigma$, and $\Phi_D$ over the compact space, and we cannot apply~\eqref{eqref. red tow warp} to the oscillator modes.

  For example, in the massive type IIA case of Figure~\ref{fig.typeIprime}, the scaling of the string scale was computed in \cite[Eq. A.33a]{Etheredge:2023odp} using a chain of string dualities:\footnote{In \cite{Etheredge:2023odp}, the expressions were given in terms of the warped metric in the ten-dimensional string frame; for clarity, here we present them in the Einstein frame.}
    \begin{equation}\label{eq. typeIprime string}
        \frac{m_{\rm osc}}{M_{{\rm Pl},9}}\sim\left(\int\dd y\, e^{\frac{1}{2}\Phi+2\rho} \right)^{1/4}\left(\int\dd y\, e^{8\rho}\right)^{-11/28}\left(\sum_i \left.e^{\frac{1}{4}\Phi+5\rho}\right|_{y_i}\right)^{1/2}\,,
    \end{equation}
    where $y_i$ denote the positions of the D8 branes and $\Phi\equiv\Phi_{10}$ the (non-canonically normalized) 10d dilaton as in \eqref{eq. action mIIA}, see Footnote \ref{fn. string frame}.\footnote{This also holds in more general cases, such as for the $E_8\times E_8$ gauge group, where different stacks of D8 branes are located at various points in the interval.} This contains a localized contribution from the branes, and the final expression does not take a closed form purely in terms of an integral over the compact space.
    
    Obtaining a general expression for the string scale for general warped backgrounds is well beyond the scope of this paper, especially in non-supersymmetric setups. However, despite these caveats, we now argue that for codimension-one backgrounds, it is possible to \emph{guess} the scaling of the string modes in the limit of maximum warping.\\

    For simplicity, take a scaling solution with $\rho=\sigma$ (in highly warped limits, all codimension-one solutions have similar behaviors, and therefore this argument applies more generally). The total powers of $\Phi_{D}$ and $\rho$ in \eqref{eq. typeIprime string} are the same as in the unwarped case \eqref{eq. stringunwarped} after setting $d=9$ and $n=1$.
    Requiring this for any warped codimension-one background suggests the following ansatz:
    \begin{equation}\label{eq. gen mosc}
        \frac{m_{\rm osc}}{M_{{\rm Pl},d}}\sim \prod_I\left(\int\dd y\,\exp\left\{f_I\Phi_{D}+r_I\rho\right\}\right)^{\alpha_I}\times\prod_J\bigg(\sum_{i_J}\left.\exp\left\{\hat{f}_I\Phi_{D}+\hat{r}_I\rho\right\}\right|_{y_{i_J}}\bigg)^{\beta_J}\,,
    \end{equation}
    with the following conditions on the exponents:
    \begin{equation}
        \sum_I\alpha_If_I+\sum_J\beta_J\hat{f}_J=\frac{2}{d+n-2}\qquad \text{and}\qquad \sum_I\alpha_Ir_I+\sum_J\beta_J\hat{r}_J=-\frac{n}{d-2}\,.
    \end{equation}
For $n=1$, i.e., $D=d+1$, the canonically normalized scalar is given by $\hat{\varphi}=\frac{2}{\sqrt{D-2}}\Phi_{D}=\frac{2}{\sqrt{d-1}}\Phi_{d+1}$ (see Footnote \ref{fn. string frame}), and the codimension-one scaling solutions in \eqref{eq.sols rho sigma} behave as
    \begin{subequations}
        \begin{align}
            \Phi_{d+1}(y)&=\frac{\sqrt{d-1}}{2}\hat{\varphi}=-\frac{(d-1)^{3/2}\gamma}{(d-1)\gamma^2-4}\left[\log B+\log (A+y) \right]\,,\\
            \rho(y)&=\frac{(d-1)\gamma^2}{(d-1)\gamma^2-4}\log B+\frac{4}{(d-1)\gamma^2-4}\log (A+y)\,,
        \end{align}
    \end{subequations}
      in the highly warped limits corresponding to $B\to\infty$ or 0 (depending on the value of $\gamma$) with fixed $A$.

    Note that the dependence on $B$ is factorized in both $\Phi_{d+1}$ and $\rho$, and can be expressed as a common prefactor in $\frac{m_{\rm osc}}{M_{{\rm Pl},d}}$. The remaining $y$-integrals in \eqref{eq. gen mosc} only enter as a numerical factor, which is irrelevant for the scaling of the string mass. Therefore, putting everything together, we obtain
    \begin{equation}
        -\partial_B\log\frac{m_{\rm osc}}{M_{{\rm Pl},d}}\sim \frac{\gamma[2(d-2)\sqrt{d-1}+(d-1)\gamma]}{(d-2)[(d-1)\gamma^2-4]}B^{-1}\,
    \end{equation}
    in the limit $B\to \infty$ or 0 with $A$ fixed, regardless of the specific values for $\{f_I,r_i,\alpha_i,\hat f_I,\hat r_I,\beta_J\}$ in \eqref{eq. gen mosc}. Using \eqref{eq:metric anst1} and \eqref{eq. 1 mod lambda}, this leads to the  following exponential rate for the string scale in Planck units in the highly warped limit,
    \begin{equation}\label{eq. warped string}
        \lambda_{\rm osc}=\frac{1}{\sqrt{(d-1)(d-2)}}\frac{2(d-2)+\sqrt{d-1}\gamma}{\sqrt{4(d-2)+(d-1)\gamma^2}}\,.
    \end{equation}
    Note again that this is not the norm of the $\vec{\zeta}_{\rm osc}$ scaling vector, but only the projection along the appropriate highly warped trajectory with $B\to \infty$ or 0 and $A$ fixed. There could be an additional component along the $A$ direction, but since we keep $A$ fixed, the exponential rate is not affected by it along the specific trajectory that we are considering. We discuss the implications of $\lambda_{\rm osc}$ (and its relation with $\lambda_{\rm KK}$ in \eqref{eq:kk_rate_case_1}) in the following subsection.

    As a double-check, this expression reproduces $\lambda_{\rm osc}=\frac{3}{4\sqrt{7}}$ after replacing $\gamma=\frac{5}{\sqrt{2}}$ and $d=9$ in \eqref{eq. warped string}, which is the scaling found for the string scale of Type I$'$ in \cite{Etheredge:2023odp} using a different argument.

    \subsection*{Winding the string along the warped dimension}
    
    The tension of the string living in the higher $(d+1)$-dimensional spacetime is
    \begin{equation}
        \mathcal{T}=M_{{\rm Pl,}d+1}^2 e^{\frac{2}{\sqrt{d-1}}\hat\varphi}\,,
    \end{equation}
    where $\hat\varphi=\frac{2}{\sqrt{d-1}}\Phi_{d+1}$ is the canonically normalized dilaton in $(d+1)$ dimensions, see Footnote \ref{fn. string frame}. We can therefore wrap this string along the warped dimension, resulting in a particle in $d$ dimensions with mass
    \begin{equation}
        m_{\rm w}=M_{{\rm Pl,}d+1}\int\dd y \, e^{\sigma+\frac{2}{\sqrt{d-1}}\hat\varphi}=M_{{\rm Pl,}d}\left[\int \dd y\ e^{d(\rho-\theta)+3\sigma+\frac{4}{\sqrt{d-1}}\hat\varphi}\right]^{\frac{1}{2}}\,,
    \end{equation}
    where we have used \eqref{eqref. red tow warp 1} to express the mass in terms of the lower-dimensional Planck units. Note that nothing ensures the presence of a full tower of winding modes, as they will be generically non-BPS, so their existence as light states in the small volume limit will depend on the specific case under consideration. However, regardless of this, in the highly warped decompactification limit, their mass would scale as
    \begin{equation}
        -\partial_B\log\frac{m_{\rm w}}{M_{{\rm Pl},{d}}}=\frac{\gamma  \left[\gamma  (d-1) (d-3)-4 (d-2) \sqrt{d-1}\right]}{(d-2) \left(\gamma ^2 (d-1)-4\right)}B^{-1}\,,
    \end{equation}
    which, using the metric in \eqref{eq:metric anst1}, gives the exponential rate
    \begin{equation}\label{eq. first winding}
        \lambda_{\rm w}=\frac{4(d-2)-\gamma  \sqrt{d-1} (d-3)}{\sqrt{(d-2) (d-1) \left[\gamma ^2 (d-1)+4 (d-2)\right]}}\,.
    \end{equation}
    For the type I$'$ solution of Figure~\ref{fig.typeIprime}, with $\gamma=\frac{5}{\sqrt{2}}$ and $d=9$, this gives $\lambda_{\rm w}=-\frac{1}{\sqrt{7}}$, which reproduces the exponential rate of the KK modes of the nine-dimensional T-dual Type I theory, see \cite[Figure 7]{Etheredge:2023odp}.

    \subsection{Results and Swampland Bounds\label{ss. results and bounds}}

    In this subsection, we summarize our results for the scaling of Kaluza--Klein and string modes in highly warped limits, discuss their physical implications and their interplay with Swampland bounds. As introduced in Section \ref{ssec:highly_warped_limit}, these correspond to limits for which the gradients of the warped profiles $\rho(y)$, $\sigma(y)$, and $\hat{\varphi}(y)$ do not vanish when taking the infinite-distance limit. 
    
    In codimension-one backgrounds, this set of geodesic trajectories are characterized by sharing the same asymptotic direction and differ only by the value of an impact parameter. Therefore, each tower will have the same exponential mass decay rate along any of these highly warped  limits, which can be computed using \eqref{eq. 1 mod lambda}. 
    Note that here we are not computing the full $\zeta$-vector $\vec{\zeta}_{\rm KK}$ as in \eqref{scalingvector}, but rather its projection along the above highly warped direction, $\lambda_{\rm KK}=\hat{T}\cdot\vec\zeta_{\rm KK}$; see \eqref{exprate}. The derivation of the vectors will be the topic of Section~\ref{ss. taxonomy}.

\subsubsection*{Exponential rate of warped KK tower}
    Our main result is the asymptotic exponential rate for the KK tower  $\lambda_{\rm KK}$ in~\eqref{eq:kk_rate_case_1} and~\eqref{eq:codim_1_exp_rate}. The result is universal for codimension-one warped backgrounds of the form \eqref{eq.met1d}: both the scaling solutions in Section \ref{ssec:scaling} and the more general solutions in Section~\ref{ssec:codim_1_general} lead to the same expression,
    \begin{equation}\label{eq:kk_rate disc}
        \boxed{\lambda_{\rm KK}=\sqrt{\frac{d-1}{d-2}}\left(1+\frac{4(d-2)}{(d-1)\gamma^2}\right)^{-\frac{1}{2} } \,,}
    \end{equation}
in terms of the spacetime dimension $d$ and the exponential rate $\gamma$ of the higher-dimensional potential $V(\hat\varphi)=V_0e^{\gamma\hat{\varphi}}$ in $D=d+1$ dimensions.

    The first consequence of~\eqref{eq:kk_rate disc} is that the exponential rate is always bounded from above by the rate associated with a homogeneous decompactification of a single dimension,
    \begin{equation}\label{eq:codim_1_decpt_1_bound}
        \lambda\leq\sqrt{\frac{d-1}{d-2}}\,,
    \end{equation}
    which is saturated in the limit $\gamma\to \infty$. This implies that the warped KK towers always become light at a slower rate in terms of the field distance than in the unwarped case. This fits with bottom-up cosmological considerations supporting this bound~\cite{Casas:2024oak} 
    (see also \cite{Etheredge:2022opl,Rudelius:2022gbz,Calderon-Infante:2023ler} for earlier proposals for this bound based on top-down unwarped compactifications).\\

    The smaller the value of $\gamma$, the smaller the KK exponential rate $\lambda_{\rm KK}$. A natural question then is: how small can it become? Could it violate the Sharpened Distance Conjecture \cite{Etheredge:2022opl} given by $\lambda\geq\frac{1}{\sqrt{d-2}}$?

      For the general ansatz of  Section~\ref{ssec:codim_1_general} and Appendix~\ref{app:codim_1_general}, we find solutions for any value of $\gamma$ and sign of $V_0$, leading to \eqref{eq:kk_rate disc}. Therefore, we could in principle get arbitrarily small values for $\lambda_{\rm KK}$ starting from a higher-dimensional potential with an arbitrarily small $\gamma$: $\lambda_{\rm KK}\to 0$ as $\gamma\to 0$. 

    Using \eqref{eq:kk_rate disc}, one can check that the sharpened bound for the Distance conjecture is then only satisfied if $\gamma$ itself is lower bounded as follows:
    \begin{equation}
          \boxed{\lambda_{\rm KK}\geq\frac{1}{\sqrt{d-2}} \quad \text{only if}\quad   \gamma \geq \frac{2}{\sqrt{d-1}}=\frac{2}{\sqrt{D-2}}\,.}
    \end{equation}
    This lower bound for $\gamma$ corresponds precisely to the (strong) de Sitter conjecture~\cite{Rudelius:2021azq}, forbidding accelerated expansion!\footnote{This interpretation works for a positive potential in higher dimensions. Otherwise, if the potential is negative, the bound is the same, but it does not have the interpretation of an accelerating cosmology.} In other words, if we start with a D-dimensional slowly varying scalar potential exhibiting asymptotic accelerated expansion, and compactify one dimension down to a warped Minkowski vacuum, then the KK tower associated with decompactifying this warped direction violates the Sharpened Distance Conjecture. Notice that this lower bound for $\lambda$ is saturated for a tower of critical string oscillator modes, so violating it  implies having a KK tower that decays at an even slower rate. It would also necessarily violate the species-tower pattern of \cite{Castellano:2023stg,Castellano:2023jjt}. 
    
    We will later further comment on potential pathologies of such violation, but for the moment we just want to emphasize the correlation we have found between the exponential rates of the warped KK tower and the higher-dimensional potential. This is a surprising result, linking Swampland constraints in different dimensions. Similar constraints have been found when comparing a KK tower and the potential in the same number of dimensions \cite{Rudelius:2022gbz,Montero:2022prj,Andriot:2020lea,Casas:2024oak}, since that case is motivated by the Higuchi bound. However, here the lower-dimensional solution is Minkowski, so there is no potential in lower dimensions and the above correlation has nothing to do with the Higuchi bound as far as we can see.  Moroever, consistency under dimensional reduction can also motivate bounds for the Distance or de Sitter conjectures independently (see e.g.~\cite{Etheredge:2022opl,Hebecker:2023qke}), while here we are connecting two distinct conjectures in different dimensions.\\

    The above result applies to the general solutions discussed in Section~\ref{ssec:codim_1_general}. In contrast, for the scaling solutions of Section~\ref{ssec:scaling}, the situation is slightly different: even though \eqref{eq:kk_rate disc} still holds, scaling solutions only exist for particular values of $\gamma$ depending on the sign of the $D=d+1$-dimensional potential (see \eqref{e.restr1warp}).
    
    The resulting bounds are:
    \begin{subequations}\label{eq:codim_1_bounds}
    \begin{align}
        \lambda_{\rm KK,1}^{\rm homo}=\sqrt{\tfrac{d-1}{d-2}}&>\lambda_{\rm KK}>\sqrt{\tfrac{d}{2(d-2)}}=\lambda_{\rm KK,2}^{\rm homo} &&\text{for}\quad V_0>0\,,\gamma\in\left(2\sqrt{\tfrac{d}{d-1}},\infty\right)\\
        \lambda_{\rm KK,2}^{\rm homo}=\sqrt{\tfrac{d}{2(d-2)}}&>\lambda_{\rm KK}>\tfrac{1}{\sqrt{d-2}}=\lambda_{\rm string} &&\text{for}\quad V_0<0\,, \gamma\in\left(\tfrac{2}{\sqrt{d-1}},2\sqrt{\tfrac{d}{d-1}}\right)\,,\\
       \lambda_{\rm string}=\tfrac{1}{\sqrt{d-2}}&> \lambda_{\rm KK}>0 &&\text{for}\quad V_0<0\,, \gamma\in\left(0,\tfrac{2}{\sqrt{d-1}}\right)\,,\label{eq. 344c}
    \end{align}
    \end{subequations}
    where we respectively denote by $\lambda_{{\rm KK},n}^{\rm homo}=\sqrt{\frac{d+n-2}{n(d-2)}}$ and $\lambda_{\rm string}=\frac{1}{\sqrt{d-2}}$ the exponential rates associated with unwarped decompactifications of $n$-dimensions and with unwarped emergent string limits. 
    
    If the potential that generates the warping has $\gamma\in\left(2\sqrt{\tfrac{d}{d-1}},\infty\right)$ (which for our scaling solutions implies that the potential is positive), the exponential rate of the KK tower always lies between the rates of homogeneous decompactifications of one and two dimensions. If the potential has $\gamma\leq 2\sqrt{\frac{d}{d-1}}$ (which for scaling solutions requires a negative potential), the exponential rate $\lambda_{\rm KK}$ becomes smaller than that of the homogeneous decompactification of two dimensions, becoming effectively zero for $\gamma\to 0$. As it is clear in \eqref{eq. 344c}, for negative potential we find again that $\lambda_{\rm KK}$ violates the sharpened bound only if $\gamma<\frac2{\sqrt{d-1}}$. As discussed in \eqref{eq. pol scal app} from Appendix \ref{app. spooky solution}, for $\gamma=0$ the KK tower also becomes light, but does so polynomially with respect to the moduli space distance (see discussion around  \eqref{eq. pol scal bulk}).
    
  \subsubsection*{Exponential rate of Planck scale and other KK towers}

  The higher-dimensional Planck mass also decreases exponentially with the field distance (in lower-dimensional Planck units). According to \eqref{eq.ex-Mpl}, the exponential rate is given by
  \begin{equation}
  \label{eq. MPl NEW}
   \boxed{\lambda_{{\rm Pl,d+1}}=\frac{1}{\sqrt{(d-1)(d-2)}}\left(1+\frac{4(d-2)}{(d-1)\gamma^2}\right)^{-\frac12}\,,}
   \end{equation}
   which is valid for both the scaling and the general solutions of Section \ref{ssec:codim_1_general}. The ratio between the exponential rates of the KK modes and the $d+1$-dimensional Planck mass is given by
   \begin{equation}\label{eq. d-1 ratio}
       \frac{\lambda_{{{\rm Pl},d+1}}}{\lambda_{\rm KK}}=\frac{1}{d-1}\,,
   \end{equation}   
   which is independent of $\gamma$ and thus the same as in the unwarped case. The expectation for this scaling is motivated by the definition of the species scale (identifying $\Lambda_{\rm QG}\equiv M_{{\rm Pl},d+1}$) in terms of the number of light states and Weyl's law\footnote{\label{fn. d-1 ratio}In the presence of a large number $N$ of light species below the QG cut-off $\Lambda_{\rm QG}$ (also known as \emph{species scale}), it is expected that $\Lambda_{\rm QG}\sim M_{{\rm Pl},d}N^{-\frac{1}{d-2}}$, see e.g.~\cite{Han:2004wt,Dvali:2007wp, Dvali:2007hz}. From Weyl's Law \eqref{eq. WEYLS LAW}, \cite{Weyl1911}, the degeneracy of the $l$-th eigenvalue of the Laplacian on a compact $n$-manifold scales as $\mathcal{O}(l^{n-1})$, so $m_{{\rm KK}}^{(l)}\approx l\,m_{\rm KK}\sim k(l)^{1/n}m_{\rm KK}$, where $k(l)$ is the number of eigenvalues (multiplicity included) below or equal to $m_{{\rm KK}}^{(l)}$, as in \eqref{eq. WEYLS LAW}. Assuming that this is also a good approximation for low KK momentum, one gets that the number of species below $\Lambda_{\rm QG}$ scales as $N\sim (\Lambda_{\rm QG}/m_{\rm KK})^n$. Given $\Lambda_{\rm QG}\sim M_{{\rm Pl},d}e^{-\lambda_{\rm QG}\Delta\varphi}$ and $m_{\rm KK}\sim M_{{\rm Pl},d}e^{-\lambda_{\rm KK}\Delta\varphi}$ in terms of the geodesic distance along an asymptotic limit $\Delta\varphi\to \infty$, this implies $N\sim e^{n(\lambda_{\rm KK}-\lambda_{\rm QG})\Delta\varphi}\sim e^{(d-2)\lambda_{\rm QG}}\to \infty$ as we decompactify. This yields the expectation that $\frac{d-2+n}{n}\lambda_{\rm QG}=\lambda_{\rm KK}$ which reduces to $(d-1)\lambda_{\rm QG}=\lambda_{\rm KK}$ for one compact dimension. Upon identifying $\Lambda_{\rm QG}=M_{{\rm Pl},d+1}$, this yields \eqref{eq. d-1 ratio}.
    } and, as we have seen, it holds even along the highly warped limits. \\
   
   Consider now the case where $\hat{\varphi}$ corresponds to a higher-dimensional radion, signaling $n$ further compact (unwarped) dimensions. In that case, we should already have in the $D$-dimensional theory additional KK towers scaling with the field distance in $\hat{\varphi}$ (see \eqref{mKKextra}). When dimensionally reducing one dimension to get our warped Minkowski solution, these additional KK towers acquire dependence on the warp factors due to the rescaling of the Planck mass. Using \eqref{eq. MPl NEW}, we get that the exponential mass decay rate  (in $M_{{\rm Pl,}d}$ units) of these KK towers associated with the unwarped extra dimensions is given by \eqref{eq.ex-KK}:
 \begin{equation}\label{eq. KK NEW}
       \boxed{\lambda_{{\rm KK'},n}=\frac{1}{\sqrt{(d-1)(d-2)}}\left(1+\frac{4(d-2)}{(d-1)\gamma^2}\right)^{-\frac12}\left(1-\frac{2(d-2)}{\gamma}\sqrt{\frac{d+n-1}{n(d-1)}}\right)\,.}
    \end{equation}
    This reproduces the expected  $\lambda_{\rm KK'}=\lambda_{{\rm Pl},d+1}=\frac{1}{\sqrt{(d-1)(d-2)}}$ \cite{Etheredge:2024tok} in the unwarped $\gamma\to \infty$ limit,\footnote{\label{fn. Unwarped rescaling} In the unwarped limit, the moduli space directions from decompactifying $d\to D=d+1$ and $D\to D'=D+n$ factorize. Since $\frac{m_{\rm KK'}}{M_{{\rm Pl},d}}=\frac{m_{\rm KK'}}{M_{{\rm Pl},D}}\frac{M_{{\rm Pl},D}}{M_{{\rm Pl},d}}$, along the $d\to D=d+1$ limit, the only scaling that $\frac{m_{\rm KK'}}{M_{{\rm Pl},d}}$ experiences is that of $M_{{\rm Pl},D}$ in terms of $M_{{\rm Pl},d}$, which for unwarped compactifications is known to follow \cite{Calderon-Infante:2023ler,vandeHeisteeg:2023ubh,Castellano:2023jjt,Etheredge:2024tok}
    \begin{equation}
        \lambda_{{\rm Pl},d+1}=\frac{1}{\sqrt{(d-1)(d-2)}}\,.
    \end{equation}
    This is not necessarily the case in warped compactifications, since the higher-dimensional radion can also have dependence along the warped direction, and thus be affected upon decompactification of the latter.
    } while it becomes smaller for $\gamma$ finite.
    
\begin{figure}[hbt!]
    \centering
    \includegraphics[width=\linewidth]{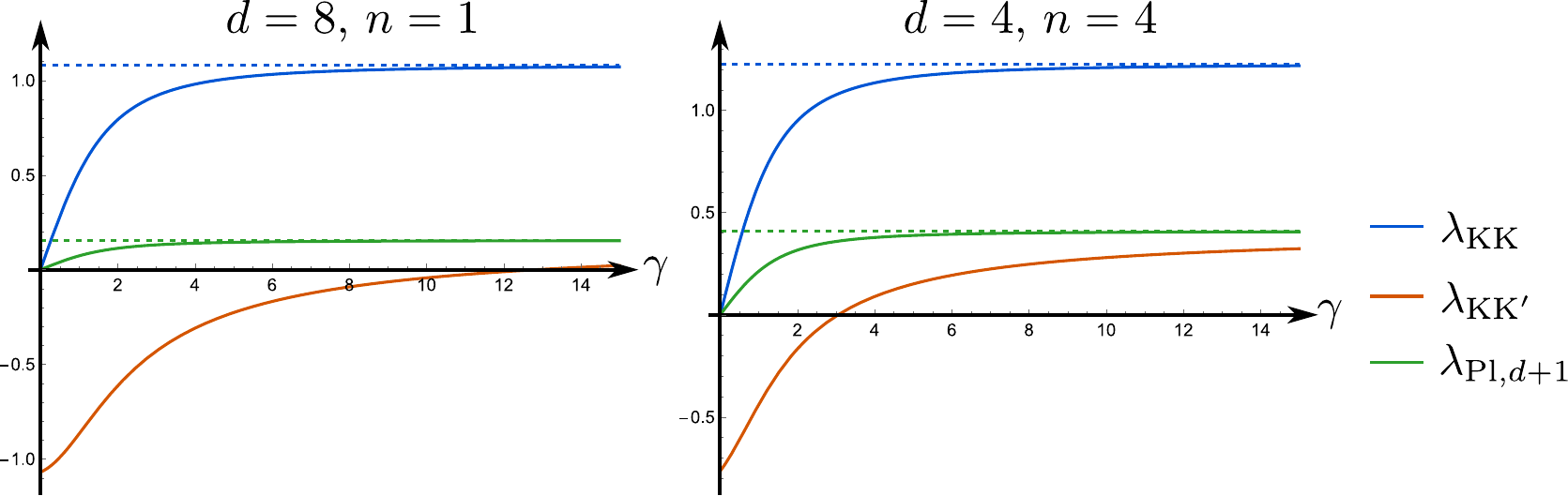}
    \caption{Comparison of exponential rates for the KK modes of the warped one-dimensional compact dimension,  the tower of KK$'$ modes in $d+1$ dimensions (assuming that the higher-dimensional scalar field is a radion measuring the volume of $n$ additional compact unwarped dimensions), and the Planck mass $M_{{\rm Pl},d+1}$. We illustrate the results in two cases, with $(d,n)=(8,1)$ and $(d,n)=(4,4)$. The exponential rates correspond to the highly warped limits with fixed impact parameter. The unwarped values, $\sqrt{\frac{d-1}{d-2}}$ for $\lambda_{\rm KK} $ and $\frac{1}{\sqrt{(d-1)(d-2)}}$ for $\lambda_{{\rm Pl},d+1}$, are in dashed lines and are recovered as $\gamma\to\infty$. 
    \label{fig.ratesCoDim1KK}}
    \end{figure}

    In Figure~\ref{fig.ratesCoDim1KK}, we plot the results for the exponential rates of the warped KK tower $\lambda_{\rm KK}$, the unwarped KK towers $\lambda_{\rm KK'}$, and the $(d+1)$-dimensional Planck scale $\lambda_{{\rm Pl},d+1}$ as functions of $\gamma$. The exponential rate at which the $(d+1)$-dimensional Planck mass becomes light decreases until it reaches an effective $\lambda_{{\rm Pl},d+1}=0$ for $\gamma=0$, as $\frac{M_{{\rm Pl},d+1}}{M_{{\rm Pl},d}}$ asymptotes to a constant rather than going to zero along this limit, see \eqref{eq. pol scal app} from Appendix \ref{app. spooky solution}.\\
    
    Moreover, as is evident both in Figure~\ref{fig.ratesCoDim1KK} and when comparing \eqref{eq. KK NEW} with \eqref{eq. MPl NEW}, the following hierarchy is preserved for warped solutions:
    \begin{equation}
        \frac{1}{\sqrt{(d-1)(d-2)}}>\lambda_{{\rm Pl},d+1}>\lambda_{\rm KK'}\,.
    \end{equation}
    This means that, along this type of limit, the higher-dimensional KK tower becomes parametrically heavy in $M_{{\rm Pl},d+1}$ units,\footnote{It also becomes parametrically heavier than $M_{{\rm Pl},d}$ for $\gamma <2(d-2)\sqrt{\frac{d+n-1}{n(d-1)}}$, which makes $\lambda_{\rm KK'}$ negative. As far as we can see, there is no straightforward higher-dimensional interpretation of this particular numerical value for $\gamma$.} so that this tower is never present in the resulting $(d+1)$-dimensional theory to which we decompactify, and we cannot perform a subsequent decompactification to $d+1+n$ dimensions. More drastically, from \eqref{eq. varphi scaling}, it follows that in the highly warped limits in \eqref{eq. highly warped limits scaling}, one gets $\hat{\varphi}\to -\infty$. This means that the higher-dimensional volume $V_n\sim \exp\big({\scriptstyle\sqrt{\frac{n(d-1)}{d+n-1}}}\hat{\varphi}\big)$ (see \eqref{eq. KK unwarped}) becomes parametrically small in $M_{{\rm Pl},d+n+1}$ units, so that the supergravity approximation should not be trusted, and the $\lambda_{{\rm KK'},n}$ value from \eqref{eq. KK NEW} might not be valid.

    Nevertheless, \eqref{eq. KK NEW} reproduces the correct exponential rate of (BPS) M-theory KK modes in the 9d type I$'$ Polchinski--Witten solution with gauge group $E_8\times E_8$, dual to M-theory on a cylinder $\mathbb{S}^1/\mathbb{Z}_2\times\mathbb{S}^1$ \cite{Aharony:2007du}. When moving along the highly warped limit of Type I$'$ (which coincides with the self-dual line), the KK modes associated with the M-theory interval have an exponential rate (in $M_{\rm Pl,9}$ units) equal to $\lambda_{\rm KK'}=-\frac{1}{\sqrt{7}}$ (see~\cite[Figure 11]{Etheredge:2023odp}).\footnote{Note that the expression of $m_{\rm KK',1}$ found in~\cite[Eq. A.34b]{Etheredge:2023odp} does not follow \eqref{eqref. red tow warp 1} (with $\lambda_{d+1}=\frac{3}{\sqrt{2}}$), but is obtained through a series of dualities:
    \begin{equation}\label{eq. mkkh}
        \frac{m_{\rm KK',1}}{M_{\rm Pl,9}}\sim\left(\int_0^{y_i^*}\dd y\,e^{2\rho+\frac{1}{\sqrt{2}}\hat{\varphi}}\right)^{-1}\left(\int\dd y\, e^{8\rho}\right)^{6/7}\left(\sum_{i}\left.e^{5\rho+\sqrt{2}\hat{\varphi}}\right|_{y_i}\right)^{-1}\,,
    \end{equation}
    where $\{y_i\}_{i=1}^{16}$ are the position of the D8-branes along the interval and $y_i^*$ is the location of the first D8-brane not in the boundary. However, \eqref{eq. mkkh} and \eqref{eqref. red tow warp 1} feature the same powers of $\rho$ and $\hat{\varphi}$ (respectively $-\frac{1}{7}$ and $\frac{3}{\sqrt{2}}$), and thus have the same exponential rate long the highly warped limit, following an argument similar to that for the string scaling $\lambda_{\rm osc}$ in Section \ref{ss.string scaling}.} This is precisely the value obtained from \eqref{eq. KK NEW} by replacing $d=9$, $n=1$, and $\gamma=\frac{5}{\sqrt{2}}$. The exponential rate is negative, implying that the tower is actually heavy in lower-dimensional Planck units. This matches the fact that, in the highly warped decompactification limit of type I$'$ string theory, one recovers 10d massive type IIA, which has no straightforward interpretation as an M-theory compactification (so there is no additional light KK tower). Our results suggest that this is the case for all warped compactification where the higher-dimensional modulus is a radion and the highly warped limit can be taken.

\subsubsection*{Exponential rate of string tower}

\begin{figure}[hbt!]
    \centering
    \includegraphics[width=0.95\linewidth]{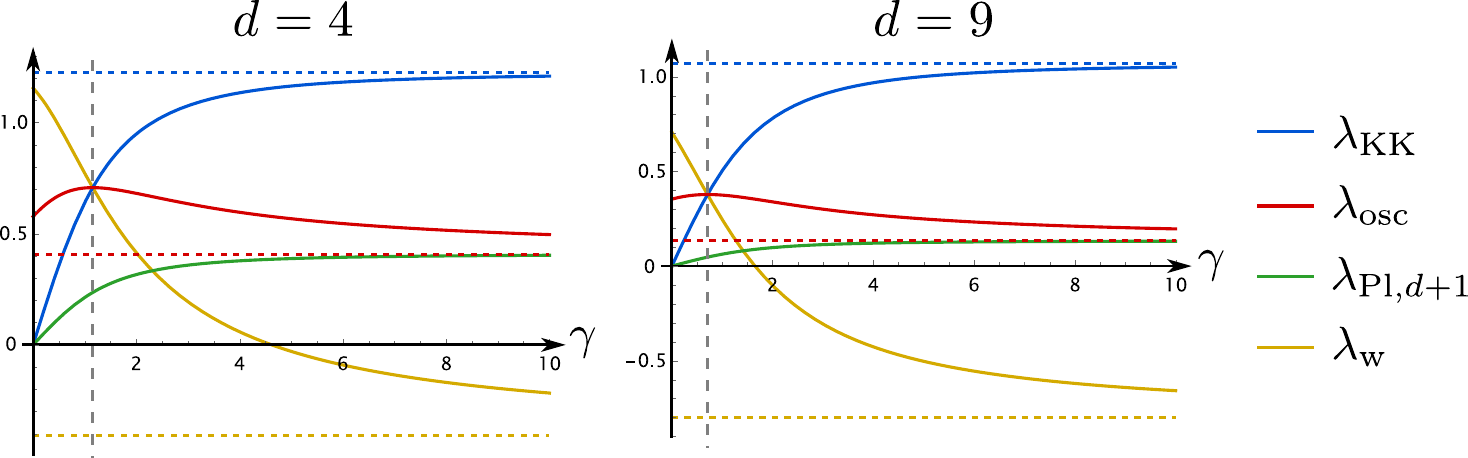}
    \caption{Comparison of exponential rates for the KK modes, the string oscillation modes, and the winding modes from the fundamental string (assuming that the higher-dimensional scalar field is the higher-dimensional dilaton $\Phi_{d+1}$) and the Planck mass $M_{{\rm Pl},d+1}$ for warped codimension-one compactifications with $d=4$ and $d=9$. The rates correspond to the highly warped limits with fixed impact parameter. The unwarped values, $\sqrt{\frac{d-1}{d-2}}$, $\frac{1}{\sqrt{(d-1)(d-2)}}$, and $\frac{3-d}{\sqrt{(d-1)(d-2)}}$ are in dashed lines, and are recovered as $\gamma\to\infty$. The dashed vertical gray line corresponds to $\gamma=\frac{2}{\sqrt{d-1}}$.
    \label{fig.ratesCoDim1}}
    \end{figure}

    When the higher-dimensional scalar is the $D$-dimensional dilaton,\footnote{The canonically normalized $D=d+1$-dimensional dilaton would be $\hat{\Phi}_{d+1}=\frac{2}{\sqrt{d-1}}\Phi_{d+1}$ where $\Phi_{10}=\log g_s$, see Footnote \ref{fn. string frame}.} our results have further implications. In the highly warped decompactification limit, we argued in Section \ref{ssec:highly_warped_limit} (see \eqref{eq. warped string})  that the string oscillator modes become light with the rate
    \begin{equation}\label{eq. warped string disc}
       \boxed{ \lambda_{\rm osc}=\frac{1}{\sqrt{(d-1)(d-2)}}\frac{2(d-2)+\sqrt{d-1}\gamma}{\sqrt{4(d-2)+(d-1)\gamma^2}}\,.}
    \end{equation}
    We want to remark that this result for the string modes comes from an educated guess, unlike the result for KK modes in \eqref{eq:kk_rate disc} which is obtained via explicit computation. In any case, it is interesting to study its implications, as it recovers the correct result in Type I$'$ Polchinski--Witten solution \cite{Etheredge:2023odp}.
    This value differs from the unwarped expectation $\lambda_{\rm osc}=\frac{1}{\sqrt{(d-1)(d-2)}}$ for the decompactification of a single direction \cite{Calderon-Infante:2023ler,Castellano:2023jjt,Etheredge:2024tok}, where the $(d+1)$-dimensional dilaton $\Phi_{d+1}$ is perpendicular to the decompactification direction, and thus only scales due to the Planck mass rescaling, see Footnote \ref{fn. Unwarped rescaling}. Note that this unwarped result is indeed recovered as $\gamma\to\infty$ and sets a lower bound for $\lambda_{\rm osc}$ in~\eqref{eq. warped string disc}, so that $\lambda_{\rm osc}\geq \frac{1}{\sqrt{(d-1)(d-2)}}$ . The upper bound is instead set at $\lambda_{\rm osc}\leq \frac{1}{\sqrt{d-2}}$ which is saturated for $\gamma=\frac{2}{\sqrt{d-1}}$, i.e. the value of $\gamma$ that saturates the strong de Sitter bound in $D$ dimensions \cite{Rudelius:2021azq}. This is interesting as it implies that both strings and KK modes decay at the same rate
    \begin{equation}
        \lambda_{\rm osc}=\lambda_{\rm KK}=\frac{1}{\sqrt{d-2}}\qquad\text{for }\,\gamma=\frac{2}{\sqrt{d-1}}\ .
    \end{equation}
    Hence, the highly warped limit (see \eqref{eq. highly warped limits scaling}) corresponds to an \emph{emergent string limit} for this special value of $\gamma$. When $\gamma$ is smaller than this critical value, $\lambda_{\rm osc}>\lambda_{\rm KK}$, as can be seen in Figure~\ref{fig.ratesCoDim1}. This would correspond to an emergent string limit where the string oscillator modes are not accompanied by KK modes becoming light at the same rate! This seems very strange and definitely not possible within string perturbation theory in flat space compactifications. Recall that the perturbative spectrum of the fundamental string includes the massless fields and all their massive KK excitations, so that their masses in Planck units share the same overall factor controlled by the dilaton. This implies that KK modes and string oscillator modes become light at the same rate in emergent string limits. In contrast, when considering directions along which the KK modes become heavier than the string modes, one expects to find additional towers (like winding modes) that become lighter than the string and hint at some duality.
    
   In \eqref{eq. first winding}, we computed the naive expectation for the mass of winding modes coming from the fundamental string wrapping the warped dimension (if they exist), obtaining
    \begin{equation}\label{eq. winding}
       \boxed{ \lambda_{\rm w}=\frac{4(d-2)-\gamma  \sqrt{d-1} (d-3)}{\sqrt{(d-2) (d-1) \left[\gamma ^2 (d-1)+4 (d-2)\right]}}\,.}
    \end{equation}
    As in previous cases, this recovers the unwarped exponential rate $\lambda_{\rm w}=\frac{3-d}{\sqrt{(d-1)(d-2)}}$ when $\gamma\to\infty$,\footnote{To see this, consider a wrapped fundamental string with tension $\mathcal{T}_{\rm F1}\sim M_{{\rm Pl,}d+1}^2 e^{\frac{4}{d-1}\Phi_{d+1}}$ on a circle of radius $R$ in $M_{{\rm Pl,}d+1}$ units. Then
\begin{equation}
    \frac{m_{\rm w}}{M_{{\rm Pl},d}}\sim\frac{\mathcal{T}_{\rm F1} R}{M_{{\rm Pl,}d+1}M_{{\rm Pl,}d}}\sim e^{\frac{4}{d-1}\Phi_{d+1}} R^{\frac{d-3}{d-2}}\sim\exp\left\{\frac{2}{\sqrt{d-1}}\hat{\varphi}+\frac{d-3}{\sqrt{(d-1)(d-2)}}\hat{\sigma}\right\}\,,
\end{equation}
where we have used that $M_{{\rm Pl},d}=M_{{\rm Pl},d+1}R^{\frac{1}{d-2}}$ and that the canonically normalized moduli are given by $\Phi_{d+1}=\frac{\sqrt{d-1}}{2}\hat{\varphi}$ (see Footnote \ref{fn. string frame}) and $R=\exp\left\{\sqrt{\frac{d-2}{d-1}}\hat{\sigma}\right\}$ (see \eqref{eq:sigma_hat}). While the $\zeta$-vector $\vec{\zeta}_{\rm w}=\left(-\frac{2}{\sqrt{d-1}},-\frac{d-3}{\sqrt{(d-1)(d-2)}}\right)$ has the expected $\sqrt{\frac{d-1}{d-2}}$ norm (since it is dual to the tower of KK modes), when projecting along the decompactification direction $\hat{\sigma}\to\infty$ (with fixed $\hat{\varphi}$) then $\lambda_{\rm w}=-\frac{d-3}{\sqrt{(d-1)(d-2)}}$.
    
    } which also sets the lower bound for $\lambda_{\rm w}$. In the Type I$'$ case from \cite{Etheredge:2023odp}, we see that for the $SO(32)$ ($E_8\times E_8$) case, there is a tower of winding modes corresponding to a KK tower in the dual Type I on $\mathbb{S}^1$ (M-theory on $\mathbb{S}^1\times [0,1]$). Along the highly warped trajectory, we recover $\lambda_{\rm w}=-\frac{1}{\sqrt{7}}$ (see \cite[Figures 7 and 11]{Etheredge:2023odp}), the same value obtained from \eqref{eq. winding} for $d=9$ and $\gamma=\frac{5}{\sqrt{2}}$.
    
    There is a critical value of $\gamma$, $\gamma_{\rm w,0}=\frac{4(d-2)}{\sqrt{d-1}(d-3)}$, such that when $\gamma<\gamma_{\rm w,0}$, the winding modes become light in $M_{{\rm Pl},d}$ units. For $\gamma\in(\frac{2}{\sqrt{d-1}},\gamma_{\rm w,0})$, this is not necessarily a problem because there is a parametric separation between the winding modes and the string oscillator scale. However, for $\gamma=\frac{2}{\sqrt{d-1}}$, we find that the three scales above become light at the same rate, $\lambda_{\rm KK}=\lambda_{\rm osc}=\lambda_{\rm w}=\frac{1}{\sqrt{d-2}}$; and for $\gamma<\frac{2}{\sqrt{d-1}}$, we find that the previous hierarchy is inverted, so that $\lambda_{\rm w}>\lambda_{\rm osc}>\lambda_{\rm KK}$. This seems very strange since we are taking a decompactification limit, so how come the winding modes become lighter than the KK modes? It suggests that we can no longer trust our results for the KK modes in this regime, as the internal volume becomes small in string units in the decompactification limit. Hence, starting with a higher-dimensional potential that exhibits accelerated expansion, so that $\gamma<\frac{2}{\sqrt{d-1}}=\frac{2}{\sqrt{D-2}}$, the highly warped decompactification limit is not geometric in the sense that the volume, even if large in Planck units, seems to be necessarily small in string units. 

\subsubsection*{Final comments}

We have found through explicit computation that the exponential mass decay rate of the KK modes becomes smaller as the exponential decay rate $\gamma$ of the higher-dimensional potential decreases. From the perspective of the Swampland program, we have found an interesting correlation between the Sharpened DC and the Strong de Sitter conjecture, implying that $\lambda_{\rm KK}\geq \frac1{\sqrt{d-2}}$ only if $\gamma\geq \frac{2}{\sqrt{d-1}}$, so that the higher-dimensional potential, if positive, does not exhibit accelerated expansion.

It would be very interesting to find an independent argument on why the KK modes must satisfy the Sharpened DC bound, since it could then be used---with the results of this paper---to provide an argument for no asymptotic accelerated expansion in string theory.

If the higher-dimensional scalar corresponds to the volume of further internal dimensions, nothing pathological seems to happen when violating these bounds. However, the situation is different if the higher-dimensional scalar is the string dilaton. In that case, we have found that the string oscillator modes become actually lighter than the KK modes when $\gamma<\frac{2}{\sqrt{d-1}}$.

The limits studied in the literature where the string seems to become parametrically lighter than the accompanying tower of KK modes have been shown to be pathological \cite{Lee:2019wij,Baume:2019sry,Klaewer:2020lfg} and heavily corrected by quantum effects. Note, however, that this is not the case if the limit does not correspond to a weakly coupled Einstein gravity theory; e.g.~the strongly curved limits with small 't Hooft coupling in AdS$_5$/CFT$_4$ constructions studied by \cite{Calderon-Infante:2024oed} exhibit parametrically lighter string modes of non-critical strings. For the Einstein-gravity flat-space warped compactifications under consideration, our results seem to indicate that this asymptotic regime becomes pathological in the sense that the warped compactification becomes non-geometric; namely, the internal volume is always small in string units, even if it is large in Planck units. Hence, we cannot really trust our geometric computation of the KK modes in that case. It is interesting that this change between a geometric and a non-geometric string compactification in our setup is controlled by whether the higher-dimensional theory describes a decelerating or accelerating cosmology.

\subsection{Update of taxonomy rules in the presence of warping\label{ss. taxonomy}}

Until now, we have focused on computing the exponential mass decay rates of various towers of states in highly warped limits using \eqref{eq. 1 mod lambda}. However, it is also interesting to determine the scaling vectors $\vec\zeta$ (see \eqref{scalingvector}) of these towers. In \cite{Etheredge:2024tok}, it was shown that these vectors obey a set of taxonomy rules, which strongly constrain how different towers can be arranged in the moduli space. In particular, in the unwarped case, the scaling vectors of two towers satisfy the following taxonomy rule:
\begin{equation}\label{eq. taxonomy}
\vec{\zeta}_{{\rm KK},n}\cdot\vec\zeta_{{\rm KK}',m}=\frac{1}{d-2}+\frac1n\delta_{nm}\,,
\end{equation}
where $n$ and $m$ are the numbers of decompactifying dimensions, and $n=\infty$ formally corresponds to the case of string oscillator modes.\footnote{Similarly, when considering the product between the $\zeta$-vectors associated with a KK tower and a higher-dimensional Planck mass, one obtains \cite{Etheredge:2024tok}
\begin{equation}\label{eq. taxonomy Planck}
    \vec{\zeta}_{{\rm KK},n}\cdot\vec\zeta_{{\rm Pl},d+m}=\frac{1}{d-2}-\frac{1}{d+m-2}(1-\delta_{n,m})\,,
\end{equation}
from which one can recover the species-tower pattern of \cite{Castellano:2023stg,Castellano:2023jjt}. See also \cite{Etheredge:2025ahf} for taxonomy rules relating the tensions of extended objects.} This reproduces the standard unwarped lengths of the KK and string vectors in \eqref{eq. KK rate}, while also fixing the angles between them. This was used in \cite{Etheredge:2024tok} to start a taxonomic classification of the global arrangement of perturbative regimes in  moduli space and the dualities relating them. The rule was derived from a bottom-up perspective, assuming the Emergent String Conjecture, and verified in a plethora of string theory examples. However, as already pointed out in \cite{Etheredge:2024tok}, this taxonomy rule no longer holds in highly warped limits. In this subsection, we investigate how this taxonomy rule is modified in highly warped one-dimensional compactifications.

As discussed in Section \ref{ssec:highly_warped_limit}, both for scaling and general solutions, highly warped trajectories with different impact parameters are always parallel asymptotically, and thus have the same asymptotic unit tangent vector $\hat{T}$, see Section \ref{ssec:scaling} after \eqref{eq:metric anst1} and Appendix \ref{app:codim_1_general}. This implies that the projection of the tower scaling vector $\vec\zeta$ on these trajectories (which gives us the exponential mas decay rate) will be the same. However, the scaling vectors can in general have an additional perpendicular component to $\hat{T}$, which will depend on the value of the impact parameter. In the scaling solutions from Section \ref{ssec:scaling}, the highly warped limits correspond to $B\to \infty$ or $0$ (depending on the value of $\gamma$, see \eqref{eq. highly warped limits scaling}) with $A$ fixed, where $A$ plays the role of the impact parameter. For simplicity, we use this notation from now on, but keeping in mind that it also applies to the general solutions in Section \ref{ssec:codim_1_general} (see also Appendix \ref{app:codim_1_general}), where $V_0^{-1}$ and $C$ play the role of $B$ and $A$ respectively.

For general values of $A$ and $B$, the moduli space metric $\mathsf{G}=\left({\begin{smallmatrix}    \mathsf{G}_{BB}&\mathsf{G}_{BA}\\\mathsf{G}_{BA}&\mathsf{G}_{AA}
\end{smallmatrix}}\right)$ is not diagonal, and to work with a properly orthonormal basis $(\hat{B},\hat{A})$, we introduce the \emph{zweibein} ${\mathsf{e}^{a}}_{\varphi}$, with ${\mathsf{e}^{ a}}_{\varphi^i}{\mathsf{e}^{ b}}_{\varphi^j}\delta_{ab}=\mathsf{G}_{\varphi^ i\varphi^j}$, and
\begin{equation}
    \zeta^a=-\mathsf{e}^{a \varphi^ i}\partial_{\varphi^i}\log\frac{m(\vec\varphi)}{M_{{\rm Pl},d}}\,.
\end{equation}
Using the $SO(2)$ (we take $\det\mathsf{e}>0$) ambiguity in defining ${\mathsf{e}^{a}}_{\varphi}$, we can choose the first component of the above vectors to be in the $B$ direction, i.e. $\hat{B}\propto B$, parallel to the highly warped limits. As it is clear from \eqref{eq:KK_mass_estimate} and \eqref{eq.sols rho sigma}, for the highly warped limits $B\to \infty$ or $0$ (see \eqref{eq. highly warped limits scaling}) with \emph{maximal warping} (effectively, zero impact parameter $A=0$), $\frac{m_{\rm KK}}{M_{{\rm Pl},d}}$ is only a function of $B$, and thus, in the orthonormal basis, $\vec\zeta_{\rm KK,1}$ is parallel to the unit tangent vector $\hat{T}$, with the norm given by \eqref{eq:kk_rate disc}. However, for general values of the impact parameter $A>0$ there is some additional dependence and the $\hat{A}$ component is non-zero. In the $A\to \infty$ limit, one should instead recover the unwarped result, i.e., $|\vec\zeta_{\rm KK,1}|=\sqrt{\frac{d-1}{d-2}}$. Therefore, the vector $\vec\zeta_{\rm KK,1}$ is going to \emph{slide} as a function of the impact parameter in order to interpolate between the result at small and large values of $A$:
\begin{equation}\label{eq. warp KK BA}
    \vec\zeta_{\rm KK,1}=(\zeta_{\rm KK,1}^{\hat{B}},\zeta_{\rm KK,1}^{\hat A})=\Bigg(\sqrt{\frac{d-1}{d-2}}\left(1+\frac{4(d-2)}{(d-1)\gamma^2}\right)^{-\frac{1}{2} }\,,\zeta_{\rm KK,1}^{\hat A}(\hat{A})\Bigg)\,,
\end{equation}
where $\zeta_{\rm KK,1}^{\hat A}({A})$ is a continuous function such that 
\begin{equation}\label{comp}
    \zeta_{\rm KK,1}^{\hat A}({A})\to\left\{\begin{array}{ll}
       0  &\text{for }A\to 0  \\
       \frac{4 (d-1)}{\gamma ^2 (d-1)+4 (d-2)}  & \text{for }A\to\infty
    \end{array}\right.\,.
\end{equation}
The general expression for $\zeta_{\rm KK,1}^{\hat A}({A})$ is model-dependent, since \eqref{eq.sols rho sigma} admits different boundary conditions and can be defined piece-wise over the warped direction, and can be quite involved. We refer the reader to \cite[Appendix C.2]{Etheredge:2023odp} for the explicit dependences in the case of the Polchinski--Witten Type I$'$ warped solutions, both for $SO(32)$ and $E_8\times E_8$ gauge groups.  As argued in \cite[Appendix B]{Etheredge:2024tok}, this sliding always seems to occur perpendicular to the tangent vector $\hat{T}$ (which is the same for all highly warped trajectories, independent of the impact parameter), and thus does not affect the exponential rate $\lambda=\hat{T}\cdot\vec\zeta$.

Similarly, we expect that the $\zeta$-vectors of other towers---whose masses have explicit dependence on the warped profiles $\{\rho(y),\sigma(y),\hat{\varphi}(y)\}$, such as $\vec\zeta_{{\rm KK'},n}$ or $\vec{\zeta}_{\rm osc}$---slide as a function of the impact parameter in highly warped limits. However, as we describe now in more detail, their masses could have an explicit dependence on $A$ that does not disappear even in the maximal warped direction with zero impact parameter, and thus the $\hat{A}$ component of their $\zeta$-vectors does not need to vanish for $A=0$.\\

In this work, we will not attempt to fully generalize the taxonomy rule \eqref{eq. taxonomy} for any trajectory with a finite value of the impact parameter $A$, as the sliding functions $\zeta^{\hat A}$ are model-dependent.  Instead, we will provide the full expression of the $\zeta$-vectors and the taxonomy rule for the particular highly warped limit corresponding to $A=0$ (maximal warping), where we know that $\vec\zeta_{\rm KK,1}\propto\hat{T}$.

\begin{figure}[hbt!]
    \centering
    \includegraphics[width=0.8\linewidth]{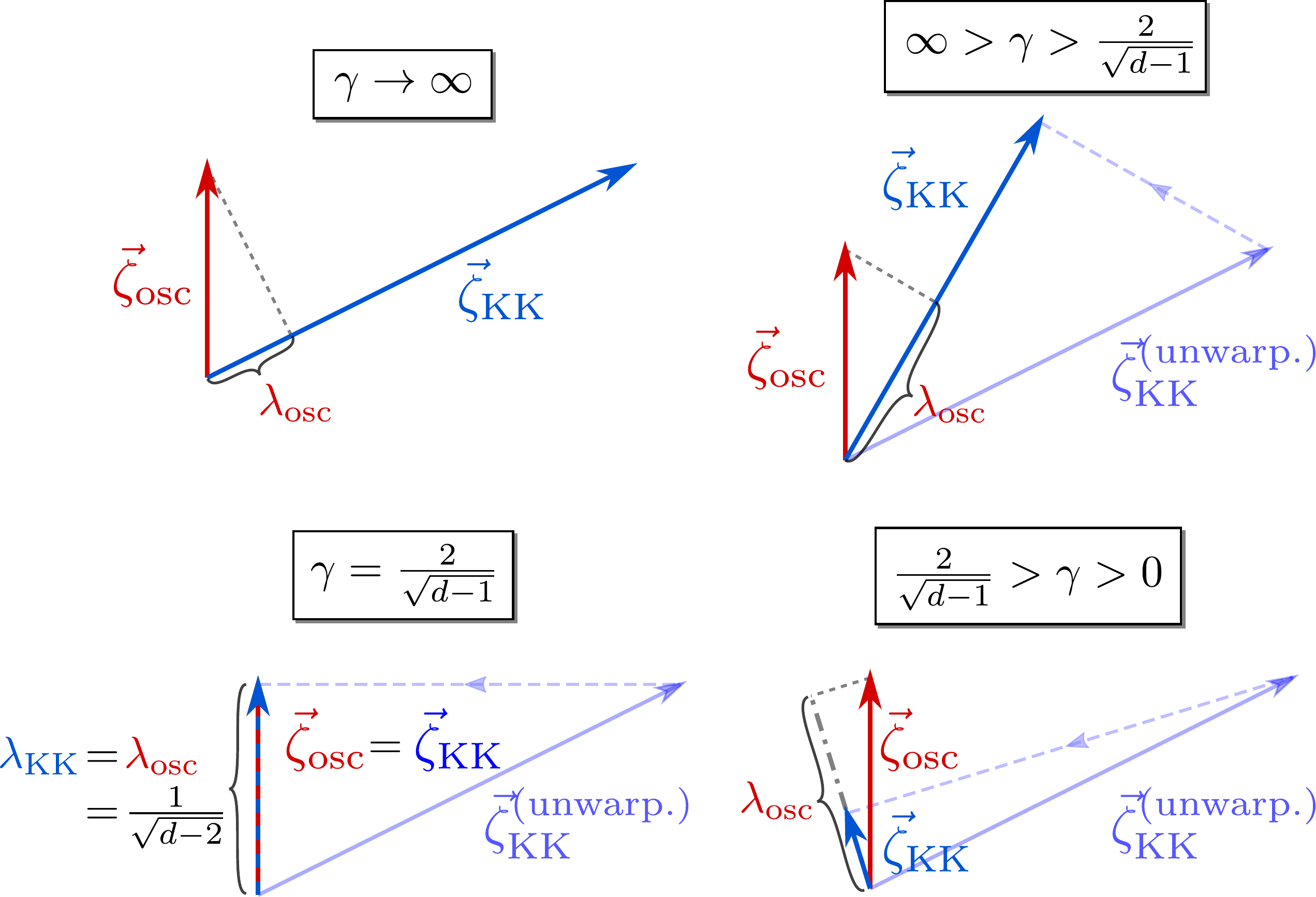}
    \caption{$\zeta$-vectors for the KK modes (in blue) and the string oscillator modes (in red) in the maximum warping regimes with vanishing impact parameter. Since in this regime the tangent vector $\hat{T}\propto \partial_{\hat{B}}$ is proportional to $\vec\zeta_{\rm KK}$, we have that $\lambda_{\rm KK}=|\vec\zeta_{\rm KK}|$ and that $\lambda_{\rm osc}=\hat{T}\cdot\vec{\zeta}_{\rm osc}$ is the projection of $\vec{\zeta}_{\rm osc}$ over the $\vec\zeta_{\rm KK}$ direction. In lighter blue we depict for reference the scaling vectors of the KK modes in the unwarped case, $\vec\zeta_{\rm KK}^{\rm (unwarp.)}$, as well as the sliding direction as we move in the highly warped direction. 
    }
    \label{fig.sliding}
    \end{figure}

We can distinguish between two cases, depending on whether the higher-dimensional modulus $\hat{\varphi}$ corresponds to a volume modulus (so the additional tower is one of KK modes) or the $(d+1)$-dimensional dilaton (with a tower of string oscillator modes):
\begin{itemize}
    \item \textbf{Two KK towers $\vec\zeta_{\rm KK,1}$ and $\vec\zeta_{{\rm KK'},n}$}: Consider first the case of a higher-dimensional radion $\hat{\varphi}$ associated with $n$ (unwarped) internal dimensions. In the $D=(d+1)$-dimensional theory, such a tower has an exponential rate $\sqrt{\frac{D+n-2}{n(D-2)}}=\sqrt{\frac{d+n-1}{n(d-1)}}$. After compactification on the warped interval, we can rewrite the profile \eqref{eq.sols rho sigma} for $\hat{\varphi}$ as
    \begin{equation}
        \hat{\varphi}(y)=\underbrace{-\frac{2 (d-1)\gamma}{ (d-1)\gamma ^2-4}\log B}_{\hat\varphi_0}-\frac{2 (d-1)\gamma}{ (d-1)\gamma ^2-4}\log(A+y)\,,
    \end{equation}
    and similarly for the general solutions from Appendix \ref{app:codim_1_general}. Here, $\hat{\varphi}_0$ acts as a ``background'' value for $\hat{\varphi}(y)$, controlling the size of the $n$ additional dimensions, on top of which the warping induces a $y$-dependence. However, in the highly warped limits we find $\hat{\varphi}_0\to-\infty$, with the $y$-dependent part only amounting for a finite correction, so the mass $\frac{m_{{\rm KK}',n}}{M_{{\rm Pl,}d}}$ effectively ceases to depend on the impact parameter, the same way it occurs for the warped $\frac{m_{\rm KK,1}}{M_{{\rm Pl,}d}}$ and $\frac{M_{{\rm Pl,}d+1}}{M_{{\rm Pl,}d}}$. This would seem to imply that, along the highly warped limits with vanishing impact parameter, the three vectors $\vec\zeta_{\rm KK,1}$, $\vec\zeta_{{\rm KK}',n}$ and $\vec{\zeta}_{{\rm Pl},d+1}$ are parallel. However, we know that this is not the case in the Type I$'$ example, where the additional KK tower of the M-theory interval does not slide and exhibits a fixed non-vanishing perpendicular component, see \cite{Etheredge:2023odp}. As explained in the previous subsection, the highly warped limit pushes us towards a regime in which the volume of the unwarped dimensions becomes small, so we cannot really trust the supergravity approximation. It is not surprising then that it does not give the correct result for the full vector $\vec\zeta_{{\rm KK}',n}$, although the value for the exponential rate in \eqref{eq. KK NEW} still reproduces the correct result in known cases.
    In any case, since $\vec\zeta_{\rm KK,1}$ has vanishing perpendicular component when $A=0$ (see \eqref{comp}), the taxonomy rules for vanishing impact parameter will necessarily read
    \beq
    \vec\zeta_{\rm KK,1}\cdot \vec\zeta_{\rm Pl,d+1}= \frac1{d-1}|\vec\zeta_{\rm KK,1}|^2  \,,
    \eeq
    \beq\vec\zeta_{\rm KK,1}\cdot \vec\zeta_{\rm KK'}= \frac1{d-1}|\vec\zeta_{\rm KK,1}|^2 \left(1-\frac{2(d-2)}{\gamma}\sqrt{\frac{d+n-1}{n(d-1)}}\right)\,.
    \eeq
    These products, however, can change for non-vanishing impact parameter, so they are only valid for $A=0$.

    \item \textbf{A KK tower $\vec\zeta_{\rm KK,1}$ and string oscillator modes $\vec\zeta_{\rm osc}$}: The other possibility is that $\hat\varphi$ is the (canonically normalized) $(d+1)$-dimensional dilaton controlling the tension of a critical string living in the $D=(d+1)$-dimensional theory. We already computed the value of the exponential rate $\lambda_{\rm osc}$ for highly warped limits in \eqref{eq. warped string disc}, corresponding to the projection of the vector $\vec\zeta_{\rm osc}$ onto the direction $\vec\zeta_{\rm KK}\propto \hat T$. It is interesting to note that precisely this $\lambda_{\rm osc}$ value \eqref{eq. warped string disc} is the same as one obtains by projecting a fixed $\vec\zeta_{\rm osc}$ (which would not change between warped and unwarped limits) along $\vec\zeta_{\rm KK}\propto \hat T$, which, as we have already explained, slides as a function of the impact parameter. This suggests the following result (which agrees with the example in \cite{Etheredge:2023odp}): along the highly warped direction, only the vector $\vec{\zeta}_{\rm KK,1}$ slides with the impact parameter, while $\vec\zeta_{\rm osc}$ has a fixed norm and angle with respect to $\vec\zeta_{\rm KK,1}^{\rm (unwarp.)}$.\footnote{Note that, from \eqref{eq. gen mosc} (which particularizes to \eqref{eq. typeIprime string} in the Type I$'$ string, see \cite{Etheredge:2023odp}), the expression for $\frac{m_{\rm osc}}{M_{{\rm Pl,}d}}$ is not simply given by integrals of the warped profiles along the compact dimensions, but also has localized contributions. While the former loses all dependence on $A$ in the highly warped limit, this is not the case for the latter, where we can have explicit factors of $A$ which remain even when taking the limit $B\to \infty$.} In the $(\hat{B}\propto B,\hat{A})$ orthonormal basis, this would read
    \begin{equation}\label{eq. zeta osc}
        \vec\zeta_{\rm osc}=\left(\frac{1}{\sqrt{(d-1)(d-2)}}\frac{2(d-2)+\sqrt{d-1}\gamma}{\sqrt{4(d-2)+(d-1)\gamma^2}},\frac{\frac{2}{\sqrt{d-1}}-\gamma}{\sqrt{4(d-2)+(d-1)\gamma^2}}\right)\,,
    \end{equation}
    in the $B\to\infty$ or $0$, $A=0$ trajectory \eqref{eq. highly warped limits scaling}, with $\vec\zeta_{\rm KK,1}$ as in \eqref{eq. warp KK BA}. Indeed, one has $|\vec\zeta_{\rm osc}|=\frac{1}{\sqrt{d-2}}$ and
    \begin{equation}
        \vec{\zeta}_{\rm KK,1}\cdot\vec\zeta_{\rm osc}=\frac{\sqrt{d-1}\gamma\big[2(d-2)+\sqrt{d-1}\gamma\big]}{(d-2)\big[4(d-2)+\sqrt{d-1}\gamma^2\big]}\,,
    \end{equation}
    from which one can infer the angle between the two vectors, recovering \eqref{eq. taxonomy} in the $\gamma\to\infty$ unwarped limit. For $\gamma=\frac{2}{\sqrt{d-1}}$, the vectors $\vec\zeta_{\rm KK,1}$ and $\vec\zeta_{\rm osc}$, \eqref{eq. warp KK BA} and \eqref{eq. zeta osc}, are aligned and actually coincide. This is illustrated in Figure~\ref{fig.sliding}, from which one can see that for $\gamma<\frac{2}{\sqrt{d-1}}$ the leading tower is that of string oscillator modes, while the KK tower is subleading.    
\end{itemize}

We conclude our discussion of the taxonomy rules with the following comment. From the $d$-dimensional perspective, we have a Minkowski vacuum with the moduli $\hat B$ and $\hat{A}$ that control the mass scale of some towers becoming light. We could then further compactify our theory on an (unwarped) $k$-manifold, whose volume is controlled by some canonically normalized radion $\hat\Upsilon$, down to $d'=d-k>2$ dimensions. Following \eqref{eq. Mpl unwarped}, we find that the relation between the $d$- and $d'$-dimensional Planck masses is given by
\begin{equation}
    \frac{M_{{\rm Pl,}d}}{M_{{\rm Pl},d-k}}=\exp\left\{-\sqrt{\frac{k}{(d-k-2)(d-2)}}\hat\Upsilon\right\}\,.
\end{equation}
After compactification, the $\vec\zeta$ vector of the higher-dimensional towers pick up a dependence on $\hat\Upsilon$ from the rescaling of the Planck mass, and we find a new tower $\vec\zeta_{{\rm KK'',}k}$, so that in this $d'=(d-k)$-dimensional theory,
\begin{equation}
    \vec\zeta_{i}=\left(\sqrt{\frac{k}{(d-k-2)(d-2)}},\vec\zeta_i^{(d)}\right)\,,\quad\vec\zeta_{{\rm KK'',}k}=\left(\sqrt{\frac{d-2}{k(d-k-2)}},\vec 0\right)\,,
\end{equation}
in the $(\hat{\Upsilon},\hat{B},\hat{A})$ basis. One could try to generalize these expressions to the case when the $k$-manifold is a warped codimension-one background; we will refrain from exploring this case because it would involve a plane of highly warped limits (spanned by the highly warped directions of both warped intervals), and the resulting expressions would become quite involved.

\section{Higher codimension backgrounds}\label{sec:higher_cod}

In our results thus far, scalar potentials have been the source of warping. In codimension-one setups we have addressed the problem in full generality for exponential potentials, but in higher codimensions it becomes considerably more complicated to even find solutions to the system of equations. Hence, we will have to focus on concrete classes of examples.

For instance, we can try to apply the tools of Section~\ref{ssec:codim_1_general} to warped compactifications with fluxes, due to their relevance in the literature of string compactifications. In this case, we expect that fluxes do not affect the universal gravitational sector (namely, Einstein+dilaton), so we can directly apply the tools developed in this paper. In particular, we are going to focus on examples involving spacetime profiles of BPS branes in type II string theories. 
Our analysis will be less general than in the codimension-one setups, covering systems that are of direct interest in string theory constructions and not exploring the complete space of solutions to the equations of motion. We will argue that, for this class of solutions, the asymptotic exponential decay rate of the KK tower is not affected by the warping for solutions with codimension $n > 2$.\\

Consider a canonically normalized scalar $\hat{\varphi}$ coupled to some $p$-brane with tension $\mathcal{T}_{p}$, electrically charged under an $A_{p+1}$ form with charge $q$:
\begin{align}
    S_{d+n}\supset&\frac{1}{2\kappa_{d+n}^2}\int\dd^{d+n}x\sqrt{-G_{d+n}}\left\{\mathcal{R}_{d+n}-(\partial\hat{\varphi})^2-\frac{e^{\alpha\hat{\varphi}}}{2(p+2)!}F_{p+2}^2\right\}\notag\\
    &-\mathcal{T}_p\int_{\Sigma_{p+1}}\dd^{p+1}x\sqrt{-{G}_{d}}-q \int_{\Sigma_{p+1}}X^*(A_{p+1})\,,
\end{align}
where $\Sigma_{p+1}$ is the worldvolume of the $p$-brane and $X^*(A_{p+1})$ the pullback of $A_{p+1}$ to $\Sigma_{p+1}$.
The $p$-brane worldvolume is transverse to the compact $X_n$, in such a way that it acts as a codimension-$n$ defect (and $d=p+1$). Take a spherically symmetric ansatz,
\begin{equation}
    \hat\varphi=\hat\varphi(r)\,, \quad\dd s_{d+n}^2=e^{2\rho(r)}\eta_{\mu\nu}\dd x^\mu\dd x^\nu+e^{2\sigma(r)}(\dd r^2+r^2\dd\Omega_{n-1}^2)\quad \text{with }\int_{X_n}\dd r\,\dd\Omega_{n-1}\, r^{n-1}=1\,.
\end{equation}
In ten dimensions, this background is a BPS brane when $d \rho+(n-2)(\sigma-\log R)=0$ ($n>1$)~\cite{Horowitz:1991cd, Duff:1994an, Stelle:1996tz,Johnson:2003cvf} for some $R>0$, independent of $r$, that corresponds to a modulus controlling the size of the internal dimensions. Brane solutions are then given by (see, e.g., the aforementioned references)\footnote{\label{fn.flux sol}The $F_{p+2}$ flux is
\begin{equation}
    F_{p+2}=\dd H(r)^{-1}\wedge\dd x^0\wedge\dd x^1\wedge\dots\wedge\dd x^p\,,
\end{equation}
where $\dd x^0\wedge\dd x^1\wedge\dots\wedge\dd x^p$ is the volume element of the $p$-brane worldvolume.}
\begin{subequations}\label{eq. profiles}
    \begin{align}
        \rho(r)&=-\frac{4(n-2)}{\Delta(d+n-2)}\log H(r)\,,\\
        \sigma(r)&=\log R+\frac{4d}{\Delta(d+n-2)}\log H(r)\,,\\
        \hat\varphi(r)&=\hat{\varphi}_0+\frac{2\alpha}{\Delta}\log H(r)\,,
    \end{align}
\end{subequations}
where $H(r)$ is a harmonic function defined as\footnote{See Appendix \ref{app:compensators} for possible contributions from the compensator fields to the moduli space metric.}
\begin{equation}\label{eq. Hr}
    H(r)=\left\{
    \begin{array}{ll}
         1+h_pM_{{\rm Pl},d+2}^{-d}\mathcal{T}_p\log(R\, r)& \text{for }n=2\\
         1+h_pM_{{\rm Pl},d+n}^{-d}\mathcal{T}_p\, (R\,r)^{-(n-2)} &\text{for }n\geq 3
    \end{array}
    \right.\,,
\end{equation}
with
\begin{equation}\label{eq. general codim def}
    \Delta=\alpha^2+\frac{4d(n-2)}{d+n-2}\quad \text{and}\quad h_p=\left\{\begin{array}{ll}
         -\frac{N}{2\pi}\frac{\Delta}{8}&\text{for }n=2  \\
        \frac{2N}{(n-2)\Omega_{n-1}}\frac{\Delta}{8} &\text{for }n\geq 3 
    \end{array}\right.\,,\quad \text{with }\,\Omega_{n}=\tfrac{\pi^{n/2}}{\Gamma\left(\tfrac{n}{2}+1\right)}\,.
\end{equation}
We are considering branes of positive tension and charge. Here, $h_p$ is proportional to the electric charge of the brane and $N$ is the number of coincident $p$-branes. Note that $h_p<0$ for $n=2$ while $h_p>0$ for $n\geq 3$.  In the above solutions, the integration constants $\hat\varphi_0$ and $R$ are promoted to moduli that control the asymptotic behavior of the scalar $\hat{\varphi}$ and size of the internal dimensions. 

In ten-dimensional type II supergravities, $\hat{\varphi}$ is the canonically normalized dilaton and $\alpha=\frac{3-p}{\sqrt{2}}$ (and therefore $\Delta=8$). The tensions of the branes are then
\begin{equation}\label{eq.II tensions}
    \mathcal{T}_{{\rm D}p}\sim M_{\rm Pl,10}^{p+1}e^{\frac{p-3}{2\sqrt{2}}\hat{\varphi}}\,,\quad \mathcal{T}_{\rm NS5}\sim M_{\rm Pl,10}^6e^{-\frac{1}{\sqrt{2}}\hat{\varphi}}\,.
\end{equation}
Note that the exponential dependence of the D-brane tension, $\mathcal{T}_p\sim M_{{\rm Pl,}d+n}^{p+1}e^{\mu \hat{\varphi}}$, is such that 
\begin{equation}
    \mu=\frac{\epsilon}{2}\alpha\,,\quad\text{with }\ \epsilon=\left\{\begin{array}{ll}
         +1&\text{for electrically charged brane}  \\
         -1& \text{for magnetically charged brane}
    \end{array}\right. \,.
\end{equation}
NS5 branes are magnetically charged under the $B_2$ field, and $\mu_{\rm NS5}=-\frac{1}{\sqrt{2}}$, which is consistent with \eqref{eq.II tensions}.

Following the discussion in Section \ref{ssec:highly_warped_limit} for the general case with $n\geq 2$, we are interested in the asymptotic behavior of
\begin{equation}\label{eq. high warp large codim}
    \partial_r \log H(r)=\left\{\begin{array}{ll}
       \frac{\mathfrak{T}}{r}[1+\mathfrak{T}\log(R\, r)]^{-1}  &  \text{for }n=2\\
         \frac{\mathfrak{T}}{r}(2-n)(\mathfrak{T}+r^{n-2})^{-1}& \text{for } n\geq 3
    \end{array}\right.\,, \quad \text{with }\;\mathfrak{T}=h_pM_{{\rm Pl},d+n}^{-d}\mathcal{T}_p R^{-(n-2)}\,,
\end{equation}
where $\mathfrak{T}$ has dimensions of [energy]$^{2-n}$. The highly warped limits are those for which $\mathfrak{T}\not\to0$ along a given asymptotic trajectory, so that the gradients of the background fields in \eqref{eq. profiles} do not vanish asymptotically. Note that for $n\geq 3$ the gradient $\partial_r \log H(r)$ has only $\mathfrak{T}$ as a free parameter, and the only way of obtaining $\partial_r \log H(r)\to 0$ is to send $\mathfrak{T}$ to 0. The case $n=2$ is different: \textbf{(1)} the flux number $h_p$ (and as a result $\mathfrak{T}$) is negative, \textbf{(2)} $\mathfrak{T}$ and $R$ are independent quantities and $H(r)$ has an explicit (logarithmic) dependence on $R r$.
This has important implications for the highly warped limits that we discuss in Section \ref{S. codim2}, leaving the general discussion for higher codimensions to Section \ref{ss. gen exp}.\\

    \subsection{The issue with codimension-two\label{S. codim2}}
    
    The behavior of the warping and field profiles is quite different in codimension two, where $H(r)$ grows logarithmically with $r$. Taking $\mathcal{T}_p\sim M_{{\rm Pl},d+2}^{p+1}e^{\frac{\alpha}{2}\hat{\varphi}}$ and $\Delta=8$,
    \begin{equation}\label{gen codim2}
        \sigma(r)=\log R+\frac{1}{2}\log \left[1+h_pe^{\frac{\alpha}{2}\hat{\varphi}_0}\log(Rr)\right]\,,\quad \hat{\varphi}=\hat{\varphi}_0+\frac{\alpha}{2}\log \left[1+h_pe^{\frac{\alpha}{2}\hat{\varphi}_0}\log(Rr)\right]\,,
    \end{equation}
    and $\rho=0$. Using the above profiles, the metric in \eqref{eq:moduli_space_metric}, and the KK mass in \eqref{eq:KK_mass_estimate},  we obtain
    \begin{equation}
        |\vec\zeta_{\rm KK}|=\sqrt{\frac{2}{3}}=\sqrt{\frac{d}{2(d-2)}}
    \end{equation}
    in limits where $ e^{\frac{\alpha}{2}\hat{\varphi}_0}\log(Rr)\to 0$. This is the exponential rate that corresponds to the decompactification of two unwarped dimensions.
    The only way to obtain a different result is to have $e^{\frac{\alpha}{2}\hat{\varphi}_0}\log(Rr)$ non-vanishing in the decompactification limit. Note that since $h_p<0$, asking that the compact $X_2$ be of finite volume and that $\hat{\varphi}$ not diverge in the interior of $X_2$, we get condition 
   \begin{equation}\label{eq.pert cont FIRST}
       e^{\frac{\alpha}{2}\hat{\varphi}_0}\log(Rr_*)<|h_7|^{-1}\,,
   \end{equation}
   where $r_*$ is the diameter of $X_2$ in the metric $\mathsf{M}_{ij}$. Crucially, this obstructs the trajectory $R\to \infty$ with $\hat{\varphi}_0$ fixed. This presents a conundrum, since the limit of warped decompactification does not seem to be available in our EFT.
   
   \subsection*{An example: type IIB string theory with D7 branes and F-theory}

    How do we solve this puzzle for these types of backgrounds? We can gain some insight using a concrete example in which we have more control. Consider type IIB string theory compactified on a two-dimensional manifold in the presence of D7-branes. These are magnetic sources for the $F_1=\dd C_0$ flux and induce a non-trivial monodromy around them. In order to keep 8d $\mathcal{N}=1$ supersymmetry and cancel the tadpole, this must be a $\mathbb{P}^1$ manifold with the appropriate singularities, corresponding to F-theory on an elliptically fibered $\mathbb{E}\hookrightarrow{\rm K3}\to\mathbb{P}^1$; see, e.g., \cite{Weigand:2018rez}.
    
    Consider a stack of $N$ D7-branes, whose transverse space is locally parameterized by $z=y_1+iy_2\in\mathbb{C}$, with the branes located at $z=0$. The profile of the axio-dilaton $\tau=C_0+ie^{-\Phi}$ (see, for example, \cite[Chapter 18]{Blumenhagen:2013fgp}) is determined by
    \begin{equation}\label{eq. j tau}
        j(\tau)=e^{-2\pi i\tau}+744+196884e^{2\pi i\tau}+\, \dots\, =\left(\frac{z}{z_0}\right)^{-N}\,,
    \end{equation}
    where $j(\tau)$ is the modular-invariant $j$-function and $z_0$ a modulus which, as we will see, can be related to the value of the 10d dilaton $\Phi$ at some given point. The spacetime metric is given by
    \begin{equation}\label{eq. met IIB}
        \dd s^2_{10}=\eta_{\mu\nu}\dd x^\mu\dd x^\nu +({\rm Im}\,\tau)\frac{\left|\eta(\tau)\right|^4}{|z/z_0|^{\frac{N}{6}}}\dd z\dd\bar{z}\,,
    \end{equation}
    with $\eta(\tau)$ the Dirichlet $\eta$-function. For small $z$, we have $|\eta(\tau)|^4\sim e^{\frac{2\pi i}{6}\tau} \sim |z/z_0|^{N/6}$; therefore, close to the D7-brane stack, the metric and dilaton fields in the conventions of Section~\ref{ssec:warped} are
    \begin{subequations}\label{eq:7brane_IIB}
        \begin{align}
            \hat{\varphi}&=\frac{1}{\sqrt{2}}\Phi=\hat{\varphi}_0-\frac{1}{\sqrt{2}}\log\left[1+h_7e^{\sqrt{2}\hat\varphi_0}\log\left(R\sqrt{y_1^2+y_2^2}\right)\right] \,,\\
            \sigma&=-\frac{\hat\varphi_0}{\sqrt 2}+\log R+\frac{1}{2}\log\left[1+h_7e^{\sqrt{2}\hat\varphi_0}\log\left(R\sqrt{y_1^2+y_2^2}\right)\right]\,,
        \end{align}
    \end{subequations}
    and $\rho= 0$ (see~\eqref{gen codim2}), where we have rescaled the internal coordinates, $y_i\to y_i R$, so that $y_i\in[0,1]$. Note that we are working with the canonically normalized 10d dilaton, $\hat{\varphi}=\frac{1}{\sqrt{2}}\Phi$. As in \eqref{eq. profiles}, $R$ is the modulus that controls the overall size of the internal space in ten-dimensional Planck units, while $h_7=-\frac{N}{2\pi}<0$ controls the flux (see \eqref{eq. general codim def}), and $\hat{\varphi}_0$ and $z_0$ are related by
    \begin{equation}
        \hat\varphi_0=-\frac{1}{\sqrt{2}}\log\left(\frac{N}{2\pi}\log |z_0|\right)\,.
    \end{equation}

\begin{figure}[hbt!]
    \centering
    \includegraphics[width=\linewidth]{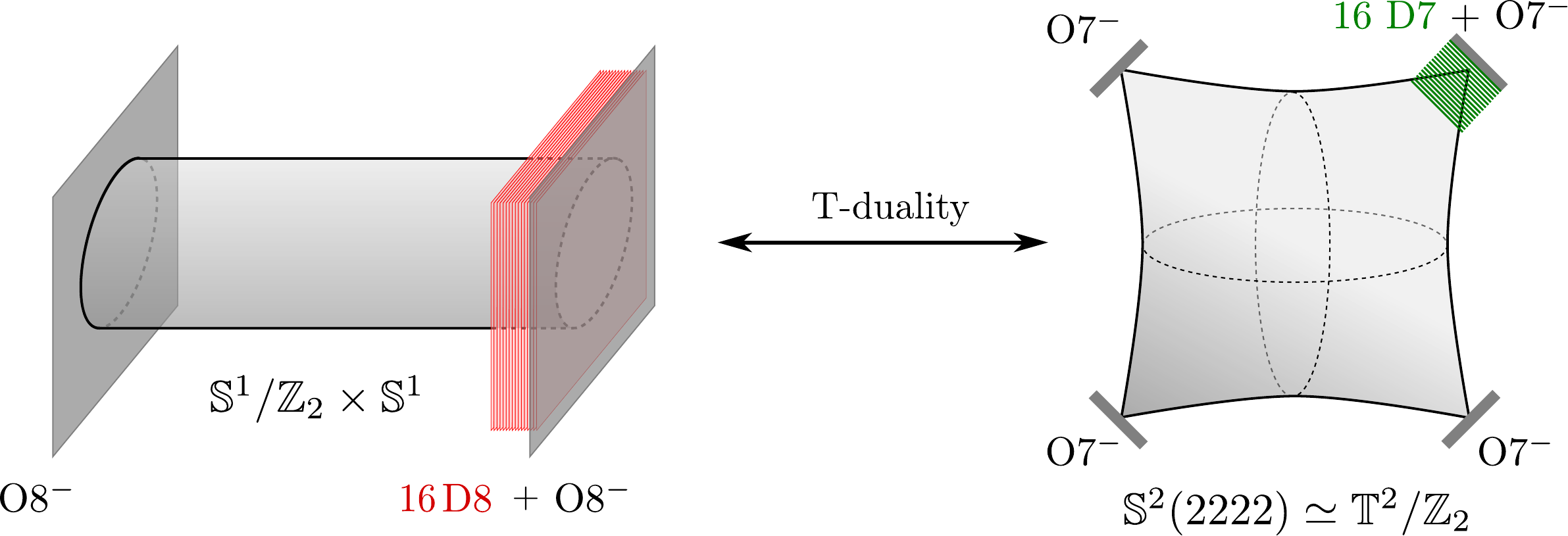}
    \caption{Type IIB string theory on the pillowcase manifold $\mathbb{S}^2(2222)\simeq \mathbb{T}^2/\mathbb{Z}_2$ with 16 D7-branes on one of the O7$^{-}$ singularities. This is T-dual to the massive type IIA configuration of Figure~\ref{fig.typeIprime} compactified on an additional $\mathbb{S}^1$.
    \label{fig.pillowcase}}
\end{figure}

    Let us consider a specific configuration of the D7 branes that can be related via dualities to a circle compactification of the Polchinski--Witten solution discussed in Section \ref{sec:warped_codim_1}. To see this, consider the Type I$'$ configuration in Figure~\ref{fig.typeIprime}. After compactifying on an additional $\mathbb{S}^1$, we can go to the small radius limit of this new circle and T-dualize to type IIB on $\mathbb{T}^2/\mathbb{Z}_2$, where the orientifold action identifies $(y_1,y_2)\sim (-y_1,-y_2)$. This gives the \emph{pillowcase} manifold $\mathbb{S}^2(2222)$, which is topologically $\mathbb{P}^1$ with four conical points of order two, corresponding to the four locations of the O7$^-$ planes. The 16 D8-branes are mapped to 16 D7-branes located on top of one of the fixed points/O7$^-$ planes, thus reproducing the same $SO(32)$ gauge group. This is represented in Figure~\ref{fig.pillowcase}.

    Parameterizing the internal coordinates as $(y_1,y_2)\in[0,1)\times\big[0,\tfrac{1}{2}\big]$,  we can apply \eqref{eq:moduli_space_metric} to obtain the moduli space metric $\mathsf{G}_{ab}$ in the $(R,\hat\varphi_0)$ space. $\mathsf{G}_{ab}$ is flat, but as in the codimension-one case, these coordinates are not orthogonal in general.\footnote{\label{eq. IIB metr}
    From \eqref{eq:moduli_space_metric}, one obtains 
    \begin{equation*}
        \mathsf{G}=\begin{pmatrix}
 \frac{14\big[1+e^{\sqrt{2}\hat{\varphi}_0}h_7(\log R+\mathsf{c}+\frac{1}{2})\big]^2-6\big[1+e^{\sqrt{2}\hat{\varphi}_0}h_7(\log R+\mathsf{c})\big]\big[1+e^{\sqrt{2}\hat{\varphi}_0}h_7(\log R+\mathsf{c}+1)\big]}{3R^2\big[1+e^{\sqrt{2}\hat{\varphi}_0}h_7(\log R+\mathsf{c})\big]^2} & -\frac{4\sqrt{2}\big[1+e^{\sqrt{2}\hat{\varphi}_0}h_7(\log R+\mathsf{c}+\frac{7}{8})\big]}{3R\big[1+e^{\sqrt{2}\hat{\varphi}_0}h_7(\log R+\mathsf{c})\big]^2} \\
 -\frac{4\sqrt{2}\big[1+e^{\sqrt{2}\hat{\varphi}_0}h_7(\log R+\mathsf{c}+\frac{7}{8})\big]}{3R\big[1+e^{\sqrt{2}\hat{\varphi}_0}h_7(\log R+\mathsf{c})\big]^2} & \frac{7}{3\big[1+e^{\sqrt{2}\hat{\varphi}_0}h_7(\log R+\mathsf{c})\big]^2} \\
\end{pmatrix}\,,
    \end{equation*}
where $\mathsf{c}=\frac{\pi}{4}-\frac{3}{2}+\frac{1}{2}\log\frac{5}{5}-\frac{1}{8}\arctan\frac{44}{117}\approx-0.647993$.

    The Christoffel symbols can then be computed to be
\begin{align*}    \mathsf{\Gamma}^{\hat{\varphi}_0}_{R\hat\varphi_0}&=\mathsf{\Gamma}^{\hat{\varphi}_0}_{\hat\varphi_0R}=-\frac{h_7e^{\sqrt{2}\hat{\varphi}_0}}{R[1+e^{\sqrt{2}\hat{\varphi}_0}h_7(\log R+\mathsf{c})]}\,,\quad \mathsf{\Gamma}^R_{RR}=-\frac{1}{R}\,,\\ \quad\mathsf{\Gamma}^{\hat{\varphi}_0}_{\hat\varphi_0\hat\varphi_0}&=-\sqrt{2}\left(1+\frac{e^{-\sqrt{2}\hat\varphi_0}}{h_7(\log R+\mathsf{c})}\right)^{-1}\,,\quad\mathsf{\Gamma}^{\hat{\varphi}_0}_{RR}=\frac{h_7^2e^{2\sqrt{2}\hat\varphi_0}}{\sqrt{2}R^2[1+e^{\sqrt{2}\hat{\varphi}_0}h_7(\log R+\mathsf{c})]}\,,
\end{align*}

from which one derives the result in the text.} We can then compute the scalar charge-to-mass ratio vector $\vec\zeta=-\vec\nabla\log\frac{m}{M_{{\rm Pl},d}}$ associated to the KK tower:
     \begin{equation}\label{eq. zeta IIB KK}
         \vec\zeta_{\rm KK}=\left(\frac{4}{3R}\Big(1+\frac{e^{\sqrt{2}\hat{\varphi}_0}h_7}{2\big[1+e^{\sqrt{2}\hat{\varphi}_0}h_7(\log R+\mathsf{c})\big]}\Big),-\frac{2\sqrt{2}}{3}\big[1+e^{\sqrt{2}\hat{\varphi}_0}h_7(\log R+\mathsf{c})\big]^{-1}\right)\,,
     \end{equation}
     
    in the (not orthonormal) basis $(R,\hat{\varphi}_0)$. Unlike in the codimension-one case, we can check using the metric given in Footnote \ref{eq. IIB metr} that its norm remains constant for any value of $R$ and $\varphi_0$: $|\vec\zeta_{\rm KK}|=\sqrt{\frac{2}{3}}$. In fact, this is the expected value for unwarped decompactifications! Naively, this would suggest that there is no sliding, but, as we will see momentarily, \eqref{eq. zeta IIB KK} is \emph{not} valid for the highly warped limits of interest.

    To analyze this in more detail, consider first the \emph{regime of validity} of the SUGRA description for our compactification. Note that the profiles in \eqref{eq:7brane_IIB} cease to make sense if \textbf{(a)} the dilaton diverges in the bulk of the $\mathbb{S}^2(2222)$ manifold (other than the conical singularities where the O7$^-$ planes are located) or \textbf{(b)} the compact volume\footnote{In ten-dimensional Planck units,
\begin{equation}\label{eq.res2}
    V_{X}=\int\dd^2y\, e^{2\sigma}=R^2e^{-\sqrt{2}\hat\varphi_0}\left[1+h_7 e^{\sqrt{2}\hat{\varphi}_0}\left(\log R+\mathsf{c}\right)\right]\,.
\end{equation}} becomes non-positive. Since $h_7<0$, in line with \eqref{eq.pert cont FIRST}, such pathologies will be avoided if
\begin{equation}\label{eq.pert cont}
    \log R< \frac{e^{-\sqrt{2}\hat\varphi_0}}{|h_7|}-\log\frac{\sqrt{5}}{2}\,\quad \text{and}\quad \log R <\frac{e^{-\sqrt{2}\hat\varphi_0}}{|h_7|}-\mathsf{c}\,.
\end{equation}
Note that the first inequality implies the second. In short, we need $e^{\sqrt{2}\hat{\varphi}_0}\log R\lesssim |h_7|^{-1}$ for the background to be under control.
Saturating the inequality leads to a dilaton divergence at the location of the O7$^-$ plane opposite to the D7-brane stack, see Figure~\ref{fig.pillowcase}.

In the language of highly warped limits from Section \ref{ssec:highly_warped_limit}, the above restrictions prevent us from taking a decompactification limit $R\to\infty$ where the gradients of the warped profiles do not vanish asymptotically. In fact, from
\begin{equation}
    \partial_r\hat\varphi, \,\partial_r\sigma\sim \frac{h_7e^{\sqrt{2}\hat\varphi_0}}{r}\big[1+h_7e^{\sqrt{2}\hat\varphi_0} \log(Rr)\big]^{-1}\,,
\end{equation}
we see that since $e^{\sqrt{2}\hat{\varphi}_0}\lesssim \big(|h_7|\log R\big)^{-1}\to 0$ along the perturbative trajectories (see \eqref{eq.pert cont}), the limits for which $e^{\sqrt{2}\hat{\varphi}_0}\log R\to0$ result in unwarped decompactifications. Borderline trajectories with $0<e^{\sqrt{2}\hat{\varphi}_0}\log R\lesssim |h_7|^{-1}$ can be shown to be non-geodesic (see Footnote~\ref{eq. IIB metr}), so that there is no well-defined asymptotic tangent vector $\hat{T}$ along them. Since the exponential change in the mass of the light towers is studied in terms of the geodesic distance, it makes no sense to compute the exponential mass decay rate in these cases.\\

Does this mean that the highly warped limit cannot be accessed in this moduli space? This would be surprising, due to the large amount of supersymmetry in the resulting 8d $\mathcal{N}=1$ theory. An important caveat is that we have been using supergravity profiles for $\sigma$ and $\hat{\varphi}$ to tackle these highly warped limits. It is clear that they are no longer valid in the regime where the dilaton diverges at the location of the O7$^-$-plane opposite to the brane stack. In this case, the complete description requires F-theory on $\mathbb{E}\hookrightarrow{\rm K3}\to\mathbb{S}^2(2222)\simeq\mathbb{P}^1$, with the fields as in \eqref{eq. j tau} and \eqref{eq. met IIB}. These equations are not invertible and do not lead to simple expressions in terms of $y_1$ and $y_2$. Numerical approaches may allow us to access the regime $e^{\sqrt{2}\hat{\varphi}_0}\log R\gtrsim |h_7|^{-1}$ and its effects on $\vec{\zeta}_{\rm KK}$, but we leave this for future work. \\

Instead, we are going to provide a different argument that provides further evidence for the absence of sliding of the $\zeta$-vectors in the highly warped limit discussed above. Type IIB string theory on $\mathbb{T}^2/\mathbb{Z}_2$ (more generally F-theory on $\mathbb{E}\hookrightarrow{\rm K3}\to\mathbb{P}^1$) is dual to heterotic string theory (for simplicity, we consider the $SO(32)$ case of Figure~\ref{fig.pillowcase}) on $\mathbb{T}^2$ \cite{Vafa:1996xn}. The moduli space is given by
\begin{equation}
    \mathcal{M}=O(2,18;\mathbb{Z})\backslash O(2,18;\mathbb{R})/(O(2;\mathbb{R})\times O(18;\mathbb{R}))\times \mathbb{R}^+\,,
\end{equation}
where the second factor corresponds to the 10d heterotic dilaton $\Phi_{\rm h,10}$. Among the $2(2+16)+1=37$ moduli that parameterize $\mathcal{M}$, 3 are non-compact: $\Phi_{\rm h,10}$ and the two radii of the heterotic torus, $R_a^{\rm h}$ and $R_b^{\rm h}$. Because of the amount of supersymmetry, all asymptotic directions in moduli space should be unobstructed. Additionally, there is a discrete subgroup of T-dualities, $D_4=\mathbb{Z}_4\rtimes\mathbb{Z}_2\simeq O(2,\mathbb{Z})$, which leaves invariant the \emph{self-dual line} parameterized by the 8d heterotic dilaton
\begin{equation}
    \Phi_{\rm h,8}=\Phi_{\rm h,10}-\frac{1}{2}\log\frac{R_{\rm h}^2\ell_{10}^2}{\ell_{\rm str,h}^2}\,.
\end{equation}
Since the mass scale of the heterotic string modes is given by $m_{\rm osc,h}=M_{\rm Pl,8}e^{\frac{1}{3}\Phi_{\rm h,8}}$, the vector $\vec\zeta_{\rm osc,h}$ is also invariant under these T-duality transformations.

For our purposes, we only move along the slice of $\mathcal{M}$ parameterized by $\Phi_{\rm h,8}$ and the overall volume of $\mathbb{T}^2$. Performing a series of dualities, we can cover the strongly coupled regions of the $SO(32)$ heterotic string theory on $\mathbb{T}^2$:
\begin{equation}
    \begin{array}{c}\label{eq.dualities}
        {\color{myGREEN}SO(32)\text{ heterotic on }\mathbb{T}^2}\\
        \phantom{\text{\small S-duality}}\Big\updownarrow{\text{\small S-duality}}\\
        {\color{myBLUE}\text{Type I on }\mathbb{T}^2\equiv \text{Type IIB on }\mathbb{T}^2+\text{O9}^{-}+32\text{ D9}}\\
        \phantom{\text{\small Double T-duality}}\Big\updownarrow{\text{\small Double T-duality}}\\
        {\color{myRED}\text{Type IIB on }\mathbb{S}^2(2222)+4\,\text{O7}^{-}+32\text{ D7}\equiv \text{F-theory on K3} }
    \end{array}
\end{equation}
The various 10d dilatons and radii in 10d Planck units, as well as the string scales, are related through the above dualities as follows:
\begin{equation*}
    \begin{array}{ccccc}
     {\color{myGREEN}SO(32)\text{ heterotic on }\mathbb{T}^2}&&   {\color{myBLUE}\text{Type I on }\mathbb{T}^2 } &&   {\color{myRED}\text{Type IIB on }\mathbb{S}^2(2222)} \\\hline
      \Phi_{\rm h,10} &&\Phi_{\rm I,10}=-\Phi_{\rm h,10} &&\Phi=\Phi_{\rm I,10}-\log\left(\frac{R_{\rm I}^2\ell_{10}^2}{\ell_{\rm str,I}^2}\right) =-\log\left(\frac{R_{\rm h}^2\ell_{10}^2}{\ell_{\rm str,h}^2}\right)\\
      R_{\rm h}&& R_{\rm I} =R_{\rm h} && R=\frac{\ell_{\rm str,IIB}\ell_{\rm str,I}}{\ell_{10}^2}R_{\rm I}^{-1}=e^{\frac{1}{2}\Phi_{\rm h,8}}\\
      \ell_{\rm str,h}&& \ell_{\rm str,I}=\ell_{\rm str,h}e^{\frac{1}{2}\Phi_{\rm h,10}} &&\ell_{\rm str,IIB}=\ell_{\rm str,I}
    \end{array}\,,
\end{equation*}

where we keep the previous notation and denote $\Phi\equiv\Phi_{\rm IIB,10}$ and $R\equiv R_{\rm IIB}$. 

Therefore, the strong coupling limit of the $SO(32)$ heterotic string on $\mathbb{T}^2$, with $\Phi_{\rm h,8}\to\infty$, corresponds to the type IIB/F-theory decompactification limit $R\to \infty$ with fixed $\Phi$, which, as we discussed, is a \emph{highly warped limit}. This highly warped direction is the same as that of the self-dual line, and therefore physical quantities must be invariant under the $D_4$ action when moving along it; for example, the type IIB KK scale $\frac{m_{\rm KK}}{M_{\rm Pl,8}}$. This implies that the type IIB $\vec\zeta_{\rm KK}$ vector points along the self-dual line, (anti)parallel to $\vec\zeta_{\rm osc,h}$. On the other hand, in the unwarped limit we have that ${m_{\rm KK}}\sim M_{\rm Pl,8}  R^{-\frac 43}$ (see \eqref{eq. KK unwarped}), and thus $\vec\zeta_{\rm KK}$ also points along the self-dual line in such regime. Since this happens for both warped and unwarped limits, we then conclude that $\vec\zeta_{\rm KK}$ stays fixed for any asymptotic trajectory in the type IIB on $\mathbb{S}^2(2222)$/F-theory on K3 duality frame, regardless of whether the type IIB SUGRA description is valid.\footnote{Here we have implicitly assumed that we are actually moving along the self-dual line, with zero impact parameter. With a finite impact parameter, $\vec\zeta_{\rm KK}$ could slide away from the self-dual line direction, though it would be strange that it would slide out and back from it, since as we argue $\vec\zeta_{\rm KK}$ is the same for 0 and infinite impact parameter. Lacking a more general understanding of the setup, this is the simplest behavior compatible with our argument.} \\

We conclude by stressing that this is just a qualitative argument for the non-sliding of the $\zeta$-vector, and a more thorough investigation would be needed to confirm these expectations.
    
    \subsection{General expectations for higher codimension\label{ss. gen exp}}

    We can try to generalize the above expectations to backgrounds of higher codimension. Assuming the type of dependence for the $\rho$, $\sigma$, and $\hat{\varphi}$ profiles as in \eqref{eq. profiles} and \eqref{eq. Hr}, the behavior of the harmonic function $H(r)$ in codimension $n\geq 3$ is radically different from that in codimension 1 or 2. Following \eqref{eq. high warp large codim}, we can write $H(r)= 1+\mathfrak{T}\,r^{2-n}$, where $\mathfrak{T}=h_pM_{{\rm Pl,}d+n}^{-d}\mathcal{T}_p R^{-(n-2)}$ is a decreasing function of $r$ and $R$, in such a way that in the decompactification limit $H(r)$ does not diverge. As in the codimension-two case, if the $r$-dependent part of $H(r)$ is negligible in the infinite-distance limit, one can expect to recover the unwarped results for $\vec\zeta_{\rm KK}$ and $\lambda_{\rm KK}$ since $\partial_r H(r)\to0$ and the warping is diluted. Thus, we would not be in a highly warped limit, as defined in Section \ref{ssec:highly_warped_limit}.
    
    We check this explicitly
    in Appendix \ref{App:BRANES}, where we obtain that trajectories in moduli space satisfying $\mathfrak{T}\to 0$ exhibit

    \begin{equation}
        \partial_\mu\rho\ll\partial_\mu\sigma\sim\partial_\mu \log R+\text{subl.}\,,\;\partial_\mu\hat{\varphi}\sim \partial_\mu\hat{\varphi}_0+\text{subl.}\,,\;\partial_\mu\theta=\frac{n}{d-2}\partial_\mu\log R+\text{subl.}\,,
    \end{equation}
    and the moduli space metric \eqref{eq:moduli_space_metric} asymptotically becomes that of the unwarped cases,
    \begin{equation}
        \mathsf{G}=\begin{pmatrix}
            \frac{n(d+n-2)}{d-2}R^{-2}&0\\0&1
        \end{pmatrix}\,,
    \end{equation}
    in the $\{R,\hat{\varphi}_0\}$ basis. Furthermore, the expression for the KK mass \eqref{eq:KK_mass_estimate} reduces to $m_{\rm KK}\sim M_{{\rm Pl},d}\,R^{-\frac{d+n-2}{d-2}}$ times a finite contribution, and therefore $|\vec\zeta_{\rm KK}|=\sqrt{\frac{d+n-2}{n(d-2)}}$, as in the unwarped case. On the other hand, when $\mathfrak{T}$ takes a finite asymptotic value, this results in subleading (but not negligible) contributions to both $\mathsf{G}$ and $\vec\zeta_{\rm KK}$. We show in Appendix \ref{App:BRANES} that, for phenomenologically interesting cases, this nevertheless reduces to a result that is similar to unwarped decompactifications. Finally, for trajectories along which $\mathfrak{T}$ blows up, we find that additional contributions to the KK mass and the moduli space metric may affect the decompactification rate (as in codimension-one backgrounds of Section \ref{sec:warped_codim_1}). \\

    Let us now check under which conditions $\mathfrak{T}$ does not blow up.
 The perturbative regime of the compactification associated with small coupling and large volume is given by  
    \begin{equation}\label{eq. pert reg}        \hat{\varphi}_0\leq0\quad\text{and}\quad R\geq\ell_{\rm str}M_{{\rm Pl},d+n}=\exp\left(-\frac{1}{\sqrt{d+n-2}}\hat{\varphi}\right)\,,
    \end{equation}
    where we take the higher-dimensional modulus $\hat{\varphi}$ to be the canonically normalized $(d+n)$-dimensional dilaton. By parametrizing the tension of the brane as $\mathcal{T}_p\sim M_{{\rm Pl,}d+n}^{d}e^{\mu \hat{\varphi}}$, we can check that $\mathfrak{T}\sim e^{\mu \hat{\varphi}}R^{-(n-2)}$ does not blow up within the perturbative regime \eqref{eq. pert reg} as long as 
    \begin{equation}\label{eq:general_requirement}
        \mu+\frac{n-2}{\sqrt{d+n-2}}\geq 0\,\quad\text{for }\ n>2\,.
    \end{equation}
    For ten-dimensional type II string theory with D$p$- or NS5-branes as sources, whose tensions appear in \eqref{eq.II tensions}, the condition in \eqref{eq:general_requirement} becomes 
    \begin{equation}
        \mu+\frac{n-2}{\sqrt{d+n-2}}=\left\{
        \begin{array}{ll}
           \frac{p-3}{2\sqrt{2}}+\frac{7-p}{2\sqrt{2}}=\sqrt{2}>0  & \text{ for D$p$-branes (with $p\leq 6$)} \\
            -\frac{1}{\sqrt{2}}+\frac{1}{\sqrt{2}}=0 & \text{ for NS5-branes}
        \end{array}
        \right.\,,
    \end{equation}
   where we have taken into account that $d+n=10$ and $d=p+1$.
    It is then clear that $\mathfrak{T}\to 0$ for all D-branes of high-enough codimension, while for NS5-branes $\mathfrak{T}$ takes a finite value for the trajectory $R\sim \exp(-\frac{1}{\sqrt{8}}\hat{\varphi})$. In Appendix \ref{App:BRANES}, we verify this explicitly with two examples of six-dimensional compactifications in which the sources are NS5 and D5-branes, and show how the exponential rate and $\zeta$-vector of the KK modes have the same behavior as in the unwarped case. 

    Note that in our derivation we have only considered the near-brane behavior of $H(r)$, where the most dramatic warping occurs. See however \cite{Andriot:2019hay,Andriot:2021gwv} for explicit examples where the full warping function is computed on $\mathbb{T}^n$. In the strict decompactification limit, their numerical results are compatible with our findings.

    \vspace{0.25cm}

    In summary, overall decompactification limits of warped backgrounds (recovering a ten-dimensional string theory) sourced by D$p$- or NS5-branes with codimension $\geq3$ exhibit the same asymptotic scaling of KK modes as in unwarped compactifications. Note that this does not necessarily mean that the background decompactifies to ten-dimensional Minwoski space. If $M_{{\rm Pl,}d+n}^{-d}\mathcal{T}_p R^{-(n-2)}$ approaches a non-vanishing constant $\mathfrak{T}>0$ in the decompactification limit, then $H(r)\to 1+\mathfrak{T} r^{2-n}\neq 1$, and the resulting higher-dimensional decompactified background feels the backreaction of a brane defect that might remain in the bulk.
    
    As shown in Appendix \ref{App:BRANES}, this is the case when NS5 branes are the sources of the warping and one takes the asymptotic trajectory $ \hat{\varphi}_0\sim-\sqrt{8}\log R\to-\infty$. In fact, such trajectory corresponds to an emergent string limit, so that the KK modes become light at the same rate as the string oscillator modes, $\lambda_{\rm KK}=\lambda_{\rm osc}=\frac{1}{2}$. This example, realized in F-theory compactifications on (elliptically fibered) K3-fibered three-folds with non-minimal degeneration points~\cite{Alvarez-Garcia:2023qqj,Alvarez-Garcia:2023gdd}, is analyzed in more detail in Appendix~\ref{App:BRANES}.

\section{Conclusions\label{sec. conc}}

In this work, we studied the asymptotic behavior of Kaluza–Klein towers in warped Minkowski compactifications in the overall decompactification limit. Focusing on scalar fluctuations, we derived the large-momentum scaling of KK masses in lower-dimensional Planck units. In codimension-one backgrounds---namely, for warped products of a Minkowski space times a one-dimensional compact space---we solved explicitly for the internal profiles and obtained a closed expression for the KK mass scaling whenever the warping is sourced by a higher-dimensional potential and suitable localized sources. 

We find that for a certain type of asymptotic trajectories, the warping effects do not get diluted in the decompactification limit and the KK scaling in lower-dimensional Planck units differs substantially from the unwarped result. Assuming a higher-dimensional potential of the form $V\sim V_0\exp(\gamma \hat\varphi)$ with $\hat\varphi$ a higher-dimensional (canonically normalized) scalar field, we derive the following result for the KK masses at large KK momentum, 
\begin{equation}\label{eq. lambda KK FINAL}
    \lambda_{\rm KK}=\sqrt{\frac{d-1}{d-2}}\left(1+\frac{4(d-2)}{(d-1)\gamma^2}\right)^{-1/2}\,.
\end{equation}
Therefore, the warping reduces $\lambda_{\rm KK}$ compared to the unwarped case, and sufficiently strong warping---arising from an extremely slowly-varying potential---could in principle lead to violations of the Sharpened Distance Conjecture bound $\lambda_{\rm KK}\geq \frac1{\sqrt{d-2}}$. Remarkably, the sharpened bound is satisfied precisely when $\gamma\geq \frac{2}{\sqrt{d-1}}$,
which coincides with the condition for the absence of asymptotic accelerated expansion in one higher dimension.

We also argue that this reduction of $\lambda_{\rm KK}$ is specific to codimension-one backgrounds, since for higher-codimension setups, the warping effects seem to be always diluted in the decompactification limit so that the KK scaling does not get modified to leading order.\\

We expect that our results could be used for several applications regarding warped compactifications. In the context of the Swampland program, perhaps the most surprising result is the relation between the sharpened bound for the KK tower and the condition forbidding asymptotic accelerated expansion in one dimension higher \cite{Bedroya:2019snp}---often referred to as the Strong de Sitter conjecture \cite{Rudelius:2021azq}. If one could establish the Sharpened Distance Conjecture on general quantum-gravitational grounds, independent of concrete string theory examples, this could provide a model-independent argument against realizing accelerated expansion at parametrically late-time in a consistent quantum gravity theory.\\

Our findings apply to overall decompactification limits starting from a warped Minkowski vacuum, suggesting two natural generalizations. The first is to consider compactifications with lower-dimensional spacetimes with non-vanishing vacuum energy.
This comprises both AdS vacua and runaway potentials, which would drastically increase the number of available examples given the vast literature on both, as well as the latter being of phenomenological interest. The second generalization is to refine our ansatz to address partial decompactification limits, thus including the dependence on other geometric moduli in addition to the overall volume. In such cases, however, one should be careful about mixing effects between KK modes of possibly different spin, since they may become relevant once the decompactification is not purely controlled by the overall volume modulus.

Although these mixing effects have been shown to be irrelevant for the asymptotic scaling of KK modes in codimension-one backgrounds \cite{Basile:2022ypo}, this remains an assumption in our analysis for higher-codimension setups. A more systematic treatment of such mixing would be important to confirm the robustness of our conclusions beyond codimension one.

Related to this, we argue that codimension $\geq 2$ backgrounds do not admit highly warped limits capable of modifying the asymptotic KK scaling. It would be valuable to confirm this explicitly. For instance, the relevant trajectory that could yield strong warping effects in the codimension-two example of Section \ref{S. codim2} takes us beyond the perturbative IIB regime. Although we argue that the KK scaling matches the unwarped result based on symmetry arguments of the moduli space, this could be confirmed explicitly using the full F-theory description.

To derive the moduli space metric, we have used the approach of directly dimensionally reducing the action. As we explained in this work, this has to be complemented by the addition of contributions from compensator fields in mixed terms of the metric ansatz. These do not affect the results of our work for the specific solutions that we consider, as shown in Appendix \ref{app:compensators}, but one should include the compensator for generic vacua, as in~\cite{Douglas:2008jx}. In particular, while we have used a Lagrangian approach, the Hamiltonian formulation seems better suited for more involved settings, such as cosmological solutions with non-vanishing vacuum energy in the lower-dimensional theory. We leave this analysis for future work.
\\

Another open question concerns the relationship between the KK tower and the species scale \cite{Han:2004wt,Dvali:2007wp, Dvali:2007hz,Anber:2011ut,vandeHeisteeg:2022btw,vandeHeisteeg:2023ubh,vandeHeisteeg:2023dlw,Castellano:2023aum,Calderon-Infante:2023ler,Caron-Huot:2024lbf,ValeixoBento:2025bmv}. In warped codimension-one decompactifications, the tower-species pattern found in \cite{Castellano:2023stg,Castellano:2023jjt} does not seem to hold in an obvious way. In other words, if one identifies the species scale with the string scale or the higher-dimensional Planck scale obtained in Section \ref{sec:warped_codim_1}, it is no longer true that $\vec\zeta_{\rm KK}\cdot \vec\zeta_{\rm species}=\frac1{d-2}$. However, the identification of the species scale in this warped compactification is highly subtle and it is not clear at all that it should coincide with the higher-dimensional Planck scale (or the string scale) derived here, as the density of states of a KK towers also changes for low KK momentum in the presence of warping. It would be interesting to determine the species scale in these setups directly from the scale suppressing higher-curvature corrections, so that we could test the pattern more explicitly.

An interesting related remark is that highly warped limits appear to yield always perturbative string limits. In particular, if we identify the scalar $\hat\varphi$ with a higher-dimensional volume modulus, the KK tower associated with $\hat\varphi$ becomes heavier than the higher-dimensional Planck scale in the highly warped limit, suggesting that further decompactification ceases to be meaningful. In contrast, everything seems consistent if $\hat\varphi$ is instead identified with a string dilaton, and we obtain that the string scale falls below the higher-dimensional Planck scale. This matches the fact that known warped Minkowski backgrounds in string theory decompactify to a time-dependent string theory solution (rather than to M-theory) when following the highly warped limit.\footnote{This would also suggest that the species scale should be somehow related to the string scale.} In this work, we also proposed a general expression for the asymptotic scaling of the string scale in these warped codimension-one setups, which is obtained from an educated guess based on the results of Type I$'$---unlike the KK scaling, which is instead obtained via explicit computation.  It would therefore be interesting to confirm our general proposal in other string theory examples.\\

As a side result, we also find a different family of solutions that resemble Randall--Sundrum-like models, which only appear when the exponential rate of the higher dimensional potential is $\gamma<\frac{2}{\sqrt{d-1}}$, violating all the proposed Swampland bounds; see the end of Section~\ref{ssec:scaling}. It would be interesting to explore these solutions further and check whether such correlation is specific to Minkowski warped compactifications or extends to more general warped backgrounds.\\

We also emphasize that our results apply to the asymptotic scaling of the KK tower at large KK momentum, rather than to the mass of the first KK excitation. It remains unclear whether Swampland bounds should be imposed on the asymptotic scaling of large-momentum modes or on the lightest state, and clarifying this point could be important to better understand the Swampland conjectures in warped backgrounds. In Appendix \ref{app:Full sol}, we compute the masses and profiles for the full KK spectrum of a scalar field in the scaling solutions of Section \ref{ssec:scaling}, and find that while the first modes are slightly heavier than their unwarped counterparts, they quickly asymptote to the expressions obtained through the WKB approximation.\\

Finally, our results have direct applications to 4d $\mathcal{N}=1$ theories coming from flux string compactifications. For instance, starting from a 4d $\mathcal{N}=1$ flux potential and compactifying one dimension in the presence of the BPS membranes sourcing the potential (see, e.g., \cite{Bandos:2018gjp,Herraez:2020tih,Lanza:2020qmt}) leads to warped Minkowski vacua where our analysis applies directly. In this context, nearly flat potentials associated with accelerated expansion \cite{Grimm:2019ixq,Calderon-Infante:2022nxb,Cremonini:2023suw,Cicoli:2023opf,Grimm:2025cpq} would translate into problematic KK towers in three dimensions violating the sharpened bound, further highlighting the impact of seemingly unrelated Swampland constraints in cosmology.\\

Much like Alice venturing into Wonderland, where new patterns emerge beneath apparent chaos, exploring KK spectra in warped compactifications leads into a “Warpland” where Swampland bounds manifest in surprising and interconnected ways.

\vspace{1cm}

\textbf{Acknowledgements}: 
We are grateful to David Andriot, José Calderón-Infante, Daniel Junghans, G. Bruno De Luca, Miguel Montero, Thomas Van Riet, Alessandro Tomasiello, Bruno Valeixo Bento, L. Germán Varona and Timo Weigand for very illuminating discussions and comments, as well as Muldrow Etheredge, Ben Heidenreich, Jacob McNamara and Tom Rudelius for collaboration in previous and related work. S.R. also wishes to acknowledge the hospitality of the Department of Theoretical Physics at CERN during different stages of this work, while I.R. thanks IFT UAM-CSIC for hospitality during a large part of the development of this paper. We also thank the Erwin Schr\"odinger International Institute for Mathematics and Physics of the University of Vienna for their hospitality during the program ``The Landscape vs. the Swampland'', where the idea behind this project was developped. This work was performed in part at the Aspen Center for Physics, which is supported by a grant from the Simons Foundation (1161654, Troyer). The authors thank the Spanish Agencia Estatal de Investigaci\'on through the grant ``IFT Centro de Excelencia Severo Ochoa'' CEX2020-001007-S and the grants PID2021-123017NB-I00 and PID2024-156043NB-I00, funded by MCIN/AEI/10.13039/ 501100011033 and by ERDF ``A way of making Europe''. S.R. is supported by the ERC Starting Grant QGuide101042568 - StG 2021. The work of I.R.  is
supported by the European Union through ERC Starting Grant SymQuaG-101163591 StG-2024, and he acknowledges the additional funding of the Spanish FPI grant No. PRE2020-094163 and the ERC Starting Grant QGuide101042568 - StG 2021. The work of I.V.  was supported by the grant RYC2019-028512-I from the MCI (Spain), the ERC Starting Grant QGuide101042568 - StG 2021, and the Project ATR2023-145703 funded by MCIN/AEI/10.13039/501100011033.

\appendix

\section{Consistency of lower-dimensional Minkowski}
\label{app:lower_potential}

In Section~\ref{ssec:warped}, the lower-dimensional spacetime has a Minkowski metric. This requires a balance of all the contributions to the $d$-dimensional on-shell scalar potential---internal curvature, higher-dimensional $\hat{V}(\hat{\varphi})$, and the tension of the extended defects---which must vanish. In the spirit of \cite{Maldacena:2000mw,Cribiori:2019clo}, in this appendix we study the general conditions for this to occur in codimension-one settings for the scaling solutions of Section~\ref{ssec:scaling}.

Similarly to the reduction of the higher $D$-dimensional action to obtain the moduli space metric in~\eqref{eq:moduli_space_metric}, we can obtain the lower-dimensional on-shell potential $V(\varphi)$, which must vanish for a $d$-dimensional Minkowski spacetime. To understand the conditions for this, consider the following terms that contribute to the lower-dimensional action:
\begin{align}
    S_D\supset&\int\dd^Dx\sqrt{-G_D}\left\{\frac{1}{2\kappa_D^2}\left[\mathcal{R}_D-\hat{\mathsf{G}}_{\hat{a}\hat{b}}\partial_M\hat{\varphi}^{\hat{a}}\partial^M\hat{\varphi}^{\hat{b}}\right]-\hat{V}(\hat{\varphi})\right\}+S_{\rm GHY}+S_{\rm sources}\notag\\
    \supset&\int\dd^dx\sqrt{-g}e^{-(d-2)\theta}\int_{X_n}\dd^n y\sqrt{\mathsf{M}}e^{(d-2)\rho+n\sigma}\left\{\frac{e^{2(\rho-\sigma-\theta)}}{2\kappa_D^2}\left[\mathcal{R}_{\mathsf{M}} - 2(n-1)\Delta_{\mathsf{M}}\sigma- 2d\Delta_{\mathsf{M}}\rho\right.\right.\notag \\
    &\left.  - (n-1)(n-2)\d_i\sigma\d^i\sigma - d(d+1)\d_i\rho\d^i\rho - 2d(n-2)\d_i\sigma\d^i\rho-\hat{\mathsf{G}}_{\hat{a}\hat{b}}\partial_i\hat{\varphi}^{\hat a}\partial^i\hat{\varphi}^{\hat{b}}\right]\notag\\
    &\left.-e^{2(\rho-\theta)}\hat{V}(\hat{\varphi})\right\}+S_{\rm GHY}+S_{\rm sources}\,,
\end{align}
where $S_{\rm GHY}$ is the Gibbons--Hawking--York boundary term and $S_{\rm sources}$ is the localized contribution; for example, the contribution of the brane sources. These source terms do not affect the (local) equations of motion, but must be taken into account when computing the lower-dimensional potential $V(\varphi)$, integrating over the internal space. From the first term in the above expression, we obtain the following contribution to the $d$-dimensional scalar potential:
\begin{align}
    V(\varphi)&=e^{-d\theta}\int_{X_n}\dd^ny\sqrt{\mathsf{M}}e^{d\rho+n\sigma}\left\{\hat{V}(\hat{\varphi})+\frac{e^{-2\sigma}}{2\kappa_{D}^2}\left[{2\Delta_{\mathsf{M}}[(n-1)\sigma+d\rho] + (n-1)(n-2)\d_i\sigma\d^i\sigma
   }\right.\right.\notag\\
    &~~~~~~~~~~~~~~~~~\left.\left.{ + d(d+1)\d_i\rho\d^i\rho+2d(n-2)\partial_i\sigma\partial^i\rho+\hat{\mathsf{G}}_{\hat{a}\hat{b}}\partial_i\hat{\varphi}^{\hat {a}}\partial^i\hat{\varphi}^{\hat{b}}-\mathcal{R}_{\mathsf{M}}}\right]\right\}\notag\\
    &=e^{-d\theta}\int_{X_n}\dd^ny\sqrt{\mathsf{M}}e^{d\rho+n\sigma}\left\{\hat{V}(\hat{\varphi})-\frac{e^{-2\sigma}}{2\kappa_{D}^2}\left[\mathcal{R}_{\mathsf{M}}+d(d-1)\partial_i\rho\partial^i\rho\right.\right.\notag\\
    &~~~~~~~~~~~~~~~~~~~~~~~~~~~~~~~~~~~~\left.\left.+(n-1)(n-2)\partial_i\sigma\partial^i\sigma+2d(n-1)\partial_i\rho\partial^i\sigma\right]\right\}\,,
\end{align}
where we have integrated by parts using the Gibbons--Hawking--York term. 

In the following, we focus on the case where the higher-dimensional scalars are canonically normalized, so that $\hat{\mathsf{G}}_{\hat a\hat b}=\delta_{\hat a\hat b}$. The above potential simplifies on shell, and using \eqref{e.eom1d} we get
\begin{equation}
    V^{\rm on-shell}(\varphi)\supset\frac{2\int\dd^n y\sqrt{\mathsf{M}}\,e^{d\rho+n\sigma}\left[\hat{V}(\hat{\varphi})-\tfrac{1}{2\kappa_{d+n}^2}e^{-2\sigma}\mathcal{R}_{\mathsf{M}}\right]}{\left(\int_{X_n}\dd^n y\sqrt{\mathsf{M}}e^{(d-2)\rho+n\sigma}\right)^{\frac{d}{d-2}}}\ell^{\frac{dn}{d-2}}\,,
\end{equation}
although for our purposes $\mathsf{M}_{ij}$ will be Ricci flat and will not contribute to the potential.

We now have to include the contribution of localized sources. Consider an arrangement of $(d-1)$-branes at $\{\vec y_i\}_{i=1}^k\subset X_n$, $\{\mathcal{T}_i\}_{i=1}^k$ and coupling $\{e^{\vec\mu_i\cdot\vec{\hat{\varphi}}}\}_{i=0}^k$ to the higher-dimensional scalars $\vec{\hat{\varphi}}$, so that
\begin{equation}
    S_{\rm sources}=-\sum_{i=0}^k\mathcal{T}_ie^{\frac{\gamma}{2}\hat{\varphi}(y_i)}\int_{y=y_i}\sqrt{-\tilde{G}_{d}^{(i)}}=-\sum_{i=0}^k\mathcal{T}_i\int\dd^{d+1}\sqrt{-G_D}e^{\frac{\gamma}{2}\hat{\varphi}}\frac{\delta(y-y_i)}{\sqrt{G_{yy}}}\,,
\end{equation}
where $\tilde{G}_{d}^{(i)}$ is the induced metric on the $(d-1)$-brane worldvolume. 

The equations of motion are then given by
\begin{subequations}\label{eq. gen EOM}
    \begin{align}
        \mathcal{R}_{MN}-\frac{1}{2}\mathcal{R}_DG_{MN}-\partial_M\hat{\varphi}\partial_N\hat{\varphi}+\frac{1}{2}\left[(\partial\hat{\varphi})^2+2\kappa_{d+n}^2\hat V(\hat{\varphi})\right]G_{MN}~~~~~~~~~~~~~~~&\notag\\+\kappa_{d+n}^2\sum_{i=0}^k\mathcal{T}_i\frac{\delta(\vec y-\vec y_i)}{\sqrt{G_{yy}}}e^{\vec{\mu}_i\cdot\vec{\hat\varphi}}\left(G_{MN}-\delta^m_M\delta_N^nG_{mn}\right)&=0 \,,\\
        \Delta_{G}\vec{\hat{\varphi}}-\kappa_{d+n}^2\vec\nabla \hat V(\hat\varphi)-\kappa_{d+n}^2\sum_{i=0}^k\mathcal{T}_i\vec{\mu}_ie^{\vec{\mu}_i\cdot\vec{\hat\varphi}}\frac{\delta(\vec y-\vec y_i)}{\sqrt{\tilde G^{(i)}_d}}&=0 \,, 
    \end{align}
\end{subequations}
where $\tilde{G}_{d}^{(i)}$ is the induced metric on the $(d-1)$-brane worldvolume. For our metric ansatz \eqref{eq:warped_cpt}, the above expressions simplify to
\begin{subequations}\label{e.eom}
    \begin{align}        \Delta_\mathsf{M}\vec{\hat{\varphi}}+\mathsf{M}^{ij}\partial_i\vec{\hat{\varphi}}\partial_ j\left[d\rho+(n-2)\sigma\right]&=\kappa_{d+n}^2e^{2\sigma}\bigg[\vec{\nabla}\hat{V}(\hat\varphi)+\sum_{i=0}^k\mathcal{T}_i\vec{\mu}_ie^{\vec{\mu}_i\cdot\vec{\hat\varphi}-n\sigma}\frac{\delta(\vec y-\vec y_i)}{\sqrt{\mathsf{M}}}\bigg]\,,\label{e.eom1}\\
    \Delta_\mathsf{M}\rho+\mathsf{M}^{ij}\partial_i\rho\partial_j\left[d\rho+(n-2)\sigma\right]&=-\frac{\kappa_{d+n}^2e^{2\sigma}}{d+n-2}\bigg[2\hat{V}(\hat{\varphi})+\sum_{i=0}^k\mathcal{T}_i e^{\vec{\mu}_i\cdot\vec{\hat\varphi}-n\sigma}\frac{\delta(\vec y-\vec y_i)}{\sqrt{\mathsf{M}}}\bigg]\label{e.eom2}\,,\\
    R_{ij}^{(\mathsf{M})}-\nabla_i\partial_j\left[d\rho+(n-2)\sigma\right]-&d\partial_i\rho\partial_j\rho+(n-2)\partial_i\sigma\partial_j\sigma+2d\partial_{(i}\rho\partial_{j)}\sigma\notag\\
    -\left\{\Delta_\mathsf{M}\sigma+\mathsf{M}^{lk}\partial_l\sigma\partial_k\left[d\rho+(n-2)\sigma\right]\right\}\mathsf{M}_{ij}&=\partial_i\vec{\hat{\varphi}}\cdot\partial_j\vec{\hat{\varphi}}+\frac{2\kappa_{d+n}^2}{d+n-2}e^{2\sigma}\mathsf{M}_{ij}\hat{V}(\hat{\varphi})\;,\label{e.eom3}
    \end{align}
\end{subequations}
where we have separated Einstein's equations into compact and non-compact parts.

Taking $n=1$ and $\hat{V}(\varphi)=\kappa_D^{-2}V_0e^{\gamma\varphi}$, together with an arrangement of $(d-1)$-branes at $0=y_0<y_1<\dots< y_k=1$ in the internal manifold $X_1$, parameterized by $y\in[0,1]$, with tensions $\{\mathcal{T}_i\}_{i=1}^k$ and coupling\footnote{Here we could have chosen arbitrary couplings $\{e^{\mu_i\hat{\varphi}}\}_{i=0}^k$; we will soon see that $\mu_i=\frac{\gamma}{2}$ must be imposed as a consistency constraint.} $e^{\frac{\gamma}{2}\hat{\varphi}}$ to the higher-dimensional scalar $\hat{\varphi}$, the equations of motion \eqref{e.eom} read:
\begin{subequations}\label{eq. gen warp APP}
    \begin{align}
        \mathcal{R}_{MN}-\frac{1}{2}\mathcal{R}_DG_{MN}-\partial_M\hat{\varphi}\partial_N\hat{\varphi}+\frac{1}{2}\left[(\partial\hat{\varphi})^2+2V_0e^{\gamma\hat{\varphi}}\right]G_{MN}&\notag\\+\kappa_{d+1}^2\sum_{i=0}^k\mathcal{T}_i\frac{\delta(y-y_i)}{\sqrt{G_{yy}}}e^{\frac{\gamma}{2}\hat{\varphi}}\left(G_{MN}-\delta^y_M\delta_N^yG_{yy}\right)&=0 \,,\\
        \Delta_{G}\hat{\varphi}-\gamma V_0 e^{\gamma \hat{\varphi}}-\kappa_{d+1}^2\frac{\gamma}{2}e^{\frac{\gamma}{2}\hat{\varphi}}\sum_{i=0}^k\mathcal{T}_i\frac{\delta(y-y_i)}{\sqrt{G_{yy}}}&=0 \,. 
        \end{align}
\end{subequations}
Taking the ansatz in \eqref{eq.met1d} and the scaling solutions of Section~\ref{ssec:scaling}, we obtain a generalization of \eqref{eq.rho equals sigma} including the sources:
\begin{subequations}\label{eq:appendix_rho_sigma_eq}
    \begin{align}        \hat{\varphi}''+(d-1)\rho'\hat{\varphi}'-\gamma V_0 e^{\gamma\hat{\varphi}+2\rho}-\kappa_{d+1}^2\frac{\gamma}{2}e^{\frac{\gamma}{2}\hat{\varphi}+\rho}\sum_{i=0}^k\mathcal{T}_i\delta(y-y_i)&=0 \,,\\
        \rho''+(d-1)(\rho')^2+\frac{2}{d-1}V_0 e^{\gamma\hat{\varphi}+2\rho}+\frac{\kappa_{d+1}^2}{d-1}e^{\frac{\gamma}{2}\hat{\varphi}+\rho}\sum_{i=0}^k\mathcal{T}_i\delta(y-y_i)&=0\,,\\
        d(d-1)(\rho')^2-(\hat{\varphi}')^2+2V_0e^{\gamma\hat{\varphi}+2\rho}&=0\,.
    \end{align}
\end{subequations}
We will take a parameterization of our solutions that differs from \eqref{eq.sols rho sigma}:
\begin{subequations}
    \begin{align}\label{eq.sols rho sigma NEW}
        \hat{\varphi}(y)&=-\frac{2 (d-1)\gamma}{ (d-1)\gamma ^2-4}\log[B(A\pm v_0y)] \,,\\
        \rho(y)&=\frac{4}{ (d-1)\gamma ^2-4}\log[B(A\pm v_0y)]+\log A+\frac{1}{2}\log \mathsf{C}_\gamma\,,
    \end{align}
\end{subequations}
which can be obtained from \eqref{eq.sols rho sigma} with
\begin{equation}
    v_0=\sqrt{|V_0|}\,,\quad \mathsf{C}_\gamma=\frac{2(d-1)[(d-1)\gamma^2-4d]}{[(d-1)\gamma^2-4]^2}\,.
\end{equation}
Following \cite{Polchinski:1995df} and \cite[Appendix A.1]{Etheredge:2023odp}, and using that $v_0$ can jump between the different $(d-1)$-branes while $\hat\varphi$ and $\rho$ must be continuous, we require $B$ to be constant and $A$ to jump at the locations of the branes,
\begin{equation}
    A(y_i^+)-A(y_i^-)=\mp y_i[v_0(y_i^+)-v_0(y_i^-)]\,.
\end{equation}
On the other hand, from the second order terms in~\eqref{eq:appendix_rho_sigma_eq},\footnote{It is here that the condition $\mu_i=\frac{\gamma}{2}\,\forall\, i=1,\dots,k$ must hold, otherwise the following equations would not be consistent with \eqref{eq.sols rho sigma}.}
\begin{subequations}
    \begin{align}\label{e. jump}
        \rho'(y_i^+)-\rho'(y_i^-)&=-\frac{\kappa_{d+1}^2}{d-1}\mathcal{T}_ie^{\frac{\gamma}{2}\hat{\varphi}(y_i)+\rho(y_i)}\,,\\
       \hat{\varphi}'(y_i^+)-\hat{\varphi}'(y_i^-)&=\frac{\gamma}{2} \kappa_{d+1}^2\mathcal{T}_ie^{\frac{\gamma}{2}\hat{\varphi}(y_i)+\rho(y_i)}\,,
    \end{align}
\end{subequations}
from which 
\begin{equation}
    v_0(y_i^+)-v_0(y_i^-)=\mp\frac{1}{2\sqrt2}\sqrt{\left|\gamma^2-\frac{4d}{d-1}\right|}\kappa_{d+1}^2\mathcal{T}_i\,.
\end{equation}
The upper and lower signs correspond to negative and positive tensions for the $(d-1)$-branes. We then obtain
\begin{subequations}
    \begin{align}
        v_0(y)&=v_\ast\mp\frac{\kappa_{d+1}^2}{2\sqrt{2}}\sqrt{\left|\gamma^2-\frac{4d}{d-1}\right|}\sum_{i=0}^k\mathcal{T}_i\Theta(y-y_i)\,,\\
        A(y)&=A_\ast+\frac{\kappa_{d+1}^2}{2\sqrt{2}}\sqrt{\left|\gamma^2-\frac{4d}{d-1}\right|}\sum_{i=0}^ky_i\mathcal{T}_i\Theta(y-y_i)  \,,      
    \end{align}
\end{subequations}
where $v_\ast$ is a constant, set by the boundary conditions. 

We can now obtain the lower-dimensional potential. We have
\begin{subequations}
    \begin{align}
        2\int_0^1\dd y\, e^{(d-1)\rho}\hat{V}  (\hat{\varphi})&=2\kappa_{d+1}^2\sum_{i=0}^{k-1}\int_{y_i}^{y_{i+1}}\dd y\,v_0^2e^{\gamma\hat{\varphi}+(d+1)\rho}\notag\\
        &=B^{\frac{(d-1)^2\gamma^2}{(d-1)\gamma^2-4}}\mathsf{C}_\gamma^{\frac{d}{2}}\left\{\pm\sum_{i=0}^{k-1}\mathcal{T}_i\left.(A^j+y_jv_0^j)^{\frac{4(d-1)}{(d-1)\gamma^2-4}-1}\right|_{j=i}^{k}\right.\notag\\
        &~~~\left.-2\sqrt{2}\left|\gamma^2-\frac{4d}{d-1}\right|^{-1/2}\kappa_{d-1}^{-2}v_\ast\left.(A^i+y_iv_0^i)^{\frac{4(d-1)}{(d-1)\gamma^2-4}-1}\right|_{i=0}^{k}\right\}\,,\\
        \sum_{i=0}^k\mathcal{T}_ie^{\frac{\gamma}{2}\hat{\varphi}+d\rho}&=c^{\frac{(d-1)^2\gamma^2}{(d-1)\gamma^2-4}}\mathsf{C}_\gamma^{\frac{d}{2}}\sum_{i=0}^k\mathcal{T}_i\left(a^i+y_iv_0^i\right)^{\frac{4(d-1)}{(d-1)\gamma^2-4}-1}\,,
  \end{align}
\end{subequations}
where $\left.f(x^i)\right|_{i=j}^k=f(x^k)-f(x^j)$ and
\begin{subequations}
    \begin{align}
        A^i=A_\ast+\frac{\kappa_{d+1}^2}{2\sqrt{2}}\sqrt{\left|\gamma^2-\frac{4d}{d-1}\right|}\sum_{j=0}^i\mathcal{T}_iy_i\,,\qquad
        v^i_0=v_\ast\mp\frac{\kappa_{d+1}^2}{2\sqrt{2}}\sqrt{\left|\gamma^2-\frac{4d}{d-1}\right|}\sum_{j=0}^i\mathcal{T}_i\,.
    \end{align}
\end{subequations}
For the two terms to cancel for all values of $B$ and $A_\ast$ (i.e., all points of the moduli space), we need
\begin{equation}\label{eq.cancellng}    \sum_{i=0}^k\mathcal{T}_i\left(A^i+y_iv_0^i\right)^{\#}\pm\sum_{i=0}^{k-1}\mathcal{T}_i\left.(A^j+y_jv_0^j)^{\#}\right|_{j=i}^{k}-\frac{2\sqrt{2}\kappa_{d+1}^{-2}v_\ast}{\sqrt{\left|\gamma^2-\frac{4d}{d-1}\right|}}\left.(A^i+y_iv_0^i)^{\#}\right|_{i=0}^{k}=0\,,
\end{equation}
where $\#=\frac{4(d-1)}{(d-1)\gamma^2-4}-1$. The numerical values of $\{A^i+y_iv_0^i\}_{i=0}^k$ depend on the positions and tensions of the $(d-1)$-branes.

Consider the example of ten-dimensional massive type IIA on an interval, with O8$^{-}$ planes at each boundary (see \cite{Polchinski:1995df,Etheredge:2023odp} and Figure~\ref{fig.typeIprime} for more details). For the setup with gauge group $SO(32)$, 16 D8-branes are located on top of one of the O8$^{-}$ planes. In this $d=9$ case, we have
\begin{equation}
    \gamma=\frac{5}{\sqrt{2}}\,,\quad \{0=y_0\leq y_1=1\}\,,\quad V_0=\frac{\kappa_{10}^2}{2}m_0^2\,,
\end{equation}
where $m_0$ is the Romans mass, and both 8-branes have positive tension, so that
\begin{equation}
    m_0(y)=m_\ast+\frac{1}{\sqrt{2\kappa_{10}^2}}(\mathcal{T}_i\Theta(y)+\mathcal{T}_1\Theta(y-1))\,.
\end{equation}
The value of $m_\ast$ is set by the orientifold projection,
\begin{equation}
    m_0(1)=m_\ast+\frac{\mathcal{T}_0+\mathcal{T}_1}{\sqrt{2\kappa_{10}^2}}=-m_0(0)=-m_\ast\Longrightarrow\left\{\begin{array}{l}
        m_\ast =0  \\
         \mathcal{T}_1=-\mathcal{T}_0
    \end{array}\right.\,.
\end{equation}
From \eqref{e. jump}, we obtain
\begin{equation}
    \mathcal{T}_1=-\mathcal{T}_0=\frac{2\sqrt{V_0}}{\kappa_{10}^2}\,,\quad A^0+y_0v_0^i=A^1+y_1v_0^1=A_\ast \,.
\end{equation}
Evaluating \eqref{eq.cancellng}, we get an exact cancellation, thus obtaining $V(\varphi)\equiv 0$ in the lower-dimensional theory, as required by the consistency with the Minkowski spacetime. A similar analysis can be performed for the $E_8\times E_8$ case, where each O8$^-$ plane has 7 D8-branes on top, while the remaining two branes are located symmetrically in the bulk of the interval, see \cite[Section 3.3]{Etheredge:2023odp} for more details.

\section{Moduli space metric: compensator analysis}\label{app:compensators}

In Section~\ref{ssec:warped}, we obtain the moduli space metric from dimensional reduction. It is well-known that this procedure must be refined in warped compactifications, with the introduction of compensator fields that generically change the moduli space metric~\cite{Giddings:2005ff,Shiu:2008ry,Douglas:2008jx}. In this appendix, we clarify the condition under which our metric in \eqref{eq:moduli_space_metric} remains the correct answer and compensator fields are not needed.

To understand the issue, consider our metric ansatz in~\eqref{eq:warped_cpt}. In Section~\ref{ssec:warped}, we are promoting the moduli, which we denote by $u^I$ in this appendix to avoid confusion with the $\varphi$ fields, to $x$-dependent fields.\footnote{As an illustrative example, in the context of Section \ref{ssec:scaling}, these are the $A$ and $B$ moduli.} This means that, for example, $\rho(y)\to \rho(y,u(x))$. Let $\d_I\rho\equiv \frac{\d \rho}{\d u^I}$, so that $\d_\mu \rho=\d_I\rho \d_\mu u^I$. Taking a solution to the equations of motion and promoting the moduli to $x$-dependent fields does not remain a solution to the equations of motion. In particular, since the metric ansatz of \eqref{eq:warped_cpt} has no mixed components, the $\mu i$ metric equations become
\begin{eqn}\label{eq:mixed_components_metric_eq}
    R_{\mu i}=\hat{G}_{\hat{a}\hat{b}}\d_\mu \hat{\varphi}^{\hat{a}}\d_i\hat{\varphi}^{\hat{b}}\,,
\end{eqn}
where $\{\hat{\varphi}^{\hat a}\}_{\hat a}$ are the higher-dimensional moduli fields, and there is no reason why this must hold in general. In practice, one cannot gauge-fix the background and the perturbations together by eliminating the mixed components from both.

In the ADM language, \eqref{eq:mixed_components_metric_eq} represents the momentum constraint~\cite{Douglas:2008jx}, which motivates the following resolution: one has to modify the metric ansatz by adding a mixed component,
\begin{eqn}\label{eq:compensator_metric}
    \dd s_D^2=e^{2\rho(y,u)}\eta_{\mu\nu}\dd x^\mu \dd x^\nu+2 B_{Ii}(y)\d_\mu u^I \dd x^\mu \dd y^i+e^{2\sigma(y,u)}\mathsf{M}_{ij}\dd y^i \dd y^j\,,
\end{eqn}
where $ B_{Ii}$ are non-propagating compensators, fixed by \eqref{eq:mixed_components_metric_eq}. 
The approach of~\cite{Douglas:2008jx} uses the Hamiltonian formalism to derive the kinetic matrix for the moduli starting from this modification. Here, we will use an equivalent Lagrangian approach that suffices to address the cases we are considering in this work.

The first key fact is that, at the two-derivative level in the moduli fields, $B_{Ii}\d_\mu u^I\equiv B_{\mu i}$ is the only ``correction'' to the naive metric ansatz that contributes to the action; see \cite{Douglas:2008jx}. In fact, it is clear that only $G_{\mu i}$ can be $\mathcal{O}(\d u)$ because of the index structure---any spacetime scalar or two-tensor would necessarily start at $\mathcal{O}(\d\d u)$. Then, let $E_{MN}$ be the Einstein field equations, so that the first order variation of the action is
\begin{eqn}
    \delta S=\int \dd^D x\sqrt{-G}E_{MN}\delta G^{MN}\,.
\end{eqn}
With the compensator field $B_{Ii}$ turned off, the equations fail to be satisfied, but this occurs at different orders: $E^{\mu i}|_{B=0}=\mathcal{O}(\d u)$, while $E^{\mu \nu}|_{B=0}$, $E^{ij}|_{B=0}$, or even the scalar field equation, are at least $\mathcal{O}(\d\d u)$, again based on the index structure. Thus, to build a term that is quadratic in the derivatives acting on $u^I$, the only option is to couple $E^{\mu i}$ with $\delta G_{\mu i}$.
Consider now the action as a functional of $B_{\mu i}$ (recall that this contains $\d u^I$). At the two-derivative order, and for simplicity setting $\kappa_D=1$, the action is the quadratic expansion
\begin{eqn}\label{eq:compensator_action}
    S[B]=S[0]+\int \dd^D x S^{\mu i}(x)B_{\mu i}(x)+\frac{1}{2}\int \dd^D x \dd^D x' B_{\mu i}(x) O^{\mu i,\nu j}(x,x')B_{\nu j}(x')\,,
\end{eqn}
where
\begin{eqn}
    \left. S^{\mu i}=\frac{\delta S}{\delta B_{\mu i}(x)}\right|_{B=0}\,,\qquad \left. O^{\mu i,\nu j}(x,x')=\frac{\delta^2 S}{\delta B_{\mu i}(x)\delta B_{\nu j}(x')}\right|_{B=0}\,.
\end{eqn}
Note that the source term is proportional to the $\mu i$ metric equation by definition,
\begin{eqn}\label{eq:compensator_source_terms}
    S^{\mu i}=\left. 2\sqrt{-G}E^{\mu i}\right|_{B=0}\,.
\end{eqn}

The second key fact is that the compensator is a non-propagating field, and hence it can be integrated out using its equation of motion. While the complete explanation for this can be found in~\cite{Douglas:2008jx}, one can see that letting $u^I=u^I(t)$, the only non-vanishing component of the metric with the compensator is $B_{0 i}$, which is the ADM shift vector on the constant-time slices. This is an auxiliary field, its equations being a constraint, so one can remove it using the equations inside the action. Therefore, extremizing the action in \eqref{eq:compensator_action}, we obtain the compensator $B_*$. Suppressing indices,
\begin{eqn}\label{eq:on-shell_compensator}
    O B_*=-S\,.
\end{eqn}
Using this in the action \eqref{eq:compensator_action}, we obtain that
\begin{eqn}
    \Delta S\equiv S[B_*]-S[0]=\frac{1}{2}\int \dd^D x S^{\mu i}{B_{*}}_{\mu i}\,.
\end{eqn}
The correction to the quadratic action is linear in the source term and, if $O$ is invertible, it can also be written as $-\frac{1}{2}\int  S O^{-1}S$, matching the expression in~\cite[eq.~(54)]{Douglas:2008jx}.

To see what this means for the moduli space metric, focus on $S^{\mu i}$ in \eqref{eq:compensator_source_terms}. The metric equation at $B=0$ is 
\begin{eqn}\label{eq:mixed_einst_eq}
    \left. E^{\mu i}\right|_{B=0}=G^{\mu \nu}G^{ij}(R_{\nu j}-T_{\nu j})\,.
\end{eqn}
While the gravitational term is universal and, for our metric ansatz, reads
\begin{eqn}
    R_{\nu j}=-(d-1)\d_\nu\d_j\rho-(n-1)\d_\nu\d_j\sigma+ (D-2)\d_j\rho\d_\mu\sigma\,,
\end{eqn}
the matter term depends on the setup that one is considering. Taking only scalar fields as in \eqref{eq:warped_action}, one obtains the combination
\begin{eqn}\label{eq:condition_no_compensator}
    \mathcal{C}_{iI}\d_\mu u^I\equiv\left[-(d-1)\d_i\d_I\rho-(n-1)\d_i\d_I\sigma+ (D-2)\d_i\rho\d_I\sigma-\hat{G}_{\hat{a}\hat{b}}\d_I\hat\varphi^{\hat{a}}\d_i\hat{\varphi}^{\hat{b}}\right]\d_\mu u^I\,.
\end{eqn}
Note that the potential does not enter the above combination. The source term is thus
\begin{eqn}
    S^{\mu i}=2\sqrt{-G}G^{\mu \nu}G^{ij}\mathcal{C}_{jI}\d_\nu u^I\,,
\end{eqn}
and $\Delta S$ can be written as
\begin{eqn}\label{eq.corr met}
    \Delta S=\int \dd^d x \d_\nu u^I\d_\mu u^J \int \dd^n y\sqrt{-G}G^{\mu \nu}G^{ij}\mathcal{C}_{jI}{B_*}_{Ji}\,,
\end{eqn}
in which one can clearly identify the correction to the moduli space metric. From this expression, we also get the key statement for our results: when $\mathcal{C}_{jI}=0$, the moduli space metric is not corrected by the compensators. Moreover, in this case one has $S=0$ and, if the operator $O$ in \eqref{eq:on-shell_compensator} is invertible, one can consistently set the compensator to zero. In practice, when $\mathcal{C}_{jI}=0$, one can obtain the metric directly from dimensional reduction, without carrying out the analysis with the compensator fields.\\

Indeed, $\mathcal{C}_{jI}=0$ for the backgrounds that we consider in this paper. For the scaling solutions \eqref{eq.sols rho sigma} with $\rho=\sigma$ from Section \ref{ssec:scaling}, where the moduli are denoted $A$ and $B$, we have
\begin{gather*}
    \partial_y\rho=\partial_A\rho=\frac{4(A+y)^{-1}}{(d-1)\gamma^2-4}\,,\;\;\partial_y\hat\varphi=\partial_A\hat\varphi=-\frac{2(d-1)\gamma(A+y)^{-1}}{(d-1)\gamma^2-4}\,,\;\;  \partial_y\partial_A\rho=-\frac{4(A+y)^{-2}}{(d-1)\gamma^2-4}\,,\\
    \partial_B\rho=\frac{(d-1)\gamma^2B^{-1}}{(d-1)\gamma^2-4}\,,\quad \partial_y\partial_B\rho=0\,,\quad\partial_B\hat\varphi=-\frac{2(d-1)\gamma B^{-1}}{(d-1)\gamma^2-4}\,,
\end{gather*}
from which $\mathcal{C}_{yA}=\mathcal{C}_{yB}=0$, and thus the moduli space metric is not affected by the inclusion of compensators. Similar results are recovered for the more general codimension-one solutions from Section \ref{ssec:codim_1_general}, since in the limiting behavior $V_0\to 0$ the expressions are analogous to the scaling ones above.\\

Let us now consider the higher-codimension solutions of \eqref{eq. profiles}, studied in Section \ref{sec:higher_cod}.
Here, there is a new ingredient: fluxes. Additional compensator fields may be present for the mixed gauge potential perturbations, and the condition that they do not change the moduli space metric is again that the mixed Einstein equations \eqref{eq:mixed_einst_eq} vanish, where now $T_{\nu i}$ contains the background flux contribution with $x$-dependent moduli (with the compensators set to zero). For the backgrounds of Section \ref{sec:higher_cod}, and assuming that fluxes induce no mixing in the regime of interest, as discussed at the beginning of Section \ref{ssec:warped}, there is actually a simple way to see that this imposes no further condition. In fact, the field strength would be $F_{p+2}=\dd H(r,u^I(x))^{-1}\wedge\dd x^0\wedge\dd x^1\wedge\dots\wedge\dd x^p$, and since this is already proportional to the external spacetime volume form, the contribution of $\d_\mu H(r,u^I(x))\dd x^\mu$ vanishes.
Therefore, fluxes add no further terms to \eqref{eq:condition_no_compensator}. This can also be seen explicitly from the flux contribution to $T_{\nu i}$, which is proportional to $F_{\nu M_2\dots M_{p+1}}{F_j}^{M_2\dots M_{p+1}}$. From our considerations above, the indices $M_2\dots M_{p+1}$ in ${F_j}^{M_2\dots M_{p+1}}$ must be all spacetime indices, but then $F_{\nu M_2\dots M_{p+1}}$ would have $(d+1)$ spacetime indices, which is not possible.
Hence, for our solutions of Section \ref{sec:higher_cod}, we find
\begin{equation}
    \mathcal{C}_{rR}=0\,,\quad \mathcal{C}_{r\hat\phi_0}=\left\{\begin{array}{ll}
         \frac{2 h_{d-1} (\alpha +2 \mu ) e^{\mu  \hat\phi_0}}{\alpha ^2 r \left[h_{d-1} e^{\mu  \hat\phi_0} \log (r R)+1\right]}& \text{for }n=2 \\
        -\frac{2 h_{d-1} (n-2) r R^2 (\alpha +2 \mu ) (d+n-2) e^{\mu  \hat\phi_0}}{\left[\alpha ^2 (d+n-2)+4 d (n-2)\right] \left[h_{d-1} r^2 R^2 e^{\mu   \hat\phi_0}+(r R)^n\right]} & \text{for }n>2
    \end{array}\right.\,.
\end{equation}
Since for our purposes, we are only interested in $\mathsf{G}_{RR}$ in order to study the decompactification limit, $ \mathcal{C}_{rR}=0$ implies that this component of the moduli space metric does not receive corrections from compensators. On the other hand, generically, $\mathcal{C}_{r\hat\phi_0}\neq0$ (though for $\alpha=-2\mu$, as in magnetically charged BPS branes studied in Section \ref{sec:higher_cod}, this also vanishes), which in principle would require the computation of the \ref{eq.corr met} correction to  $\mathsf{G}_{\hat\varphi_0\hat\varphi_0}$. 
%However, notice that this needs to be complemented by the inclusion of flux contributions to \eqref{eq:condition_no_compensator}, although we do not expect this to change the intuitive argument from Section \ref{ss. gen exp}. 
We leave a detailed investigation of this point for future work.

\section{Laplacian eigenvalues for codimension-one scaling solutions and WKB\label{app:Full sol}}

In this appendix, we obtain the general solution to the Laplacian eigenvalue equation \eqref{eq:eigenfunctions-laplacian} that is relevant for the codimension-one scaling solutions of Section \ref{ssec:scaling}. Note that this is the full Laplace operator for these types of solutions, see Footnote \ref{fn. exact lap}, and thus we are not suppressing subleading KK momentum corrections.

From the explicit expressions of $\rho$ and $\sigma$, 
\begin{equation}\label{eq.app lap}
    \frac{1}{\mu(y)}\partial_i\left[\mu(y)\partial^if(y)\right]+\lambda^2f(y)=f''(y)+\frac{4(d-1)}{(d-1)\gamma^2-4}\frac{f'(y)}{A+y}+\lambda^2f(y)=0\,,
\end{equation}
which has the following solution:
\begin{equation}
    f(y)=(A+y)^{-\Gamma}\left\{\mathsf{c}_JJ_\Gamma\left[\lambda\left(A+y\right)\right]+\mathsf{c}_YY_\Gamma\left[\lambda\left(A+y\right)\right]\right\}\, ,
\end{equation}
where
\begin{equation}
    \Gamma=\frac{4d-\gamma^2(d-1)}{2[\gamma^2(d-1)-4]}\left\{\begin{array}{ll}
        =0 &\text{for}\ \gamma=\frac{2}{\sqrt{d-1}},\, 2\sqrt{\frac{d}{d-1}}  \\>0&\text{for}\ \gamma\in\left(\frac{2}{\sqrt{d-1}},\, 2\sqrt{\frac{d}{d-1}}\right)\\
        <0         & \text{otherwise}
    \end{array}\right.\,,
\end{equation}
$J_\Gamma$ and $Y_\Gamma$ are Bessel functions of the first and second kind, and $\{\mathsf{c}_J,\mathsf{c}_Y\}$ are integration constants. Without loss of generality, we take $y\in[0,1]$. Requiring the presence of a zero mode in the highly warped limit with zero impact parameter, $A\to 0$, sets $\mathsf{c}_Y=0$, while $\mathsf{c}_J$ is set by the normalization in \eqref{eq.norm}. For a given boundary condition, $\lambda$ is quantized in terms of a function of the zeros of the $J_\Gamma$ function, $\{j_{\Gamma,k}\}_{k\in\mathbb{N}}$, with \cite[Table 10.21.vi]{NIST:DLMF} 
\begin{equation}
    j_{\Gamma,k} =\left(k+\frac{\Gamma}{2}-\frac{1}{4}\right)\pi-\frac{4\Gamma^2-1}{8\pi\left(k+\frac{\Gamma}{2}-\frac{1}{4}\right)}-\frac{4(4\Gamma^2-1)(28\Gamma^2-31)}{1536\pi^3\left(k+\frac{\Gamma}{2}-\frac{1}{4}\right)^3}+\mathcal{O}(k^{-5})\,,
\end{equation}
thus recovering the WKB approximation for $k\to\infty$. For example, setting the boundary condition $f(1)=0$, one obtains
\begin{equation}
    \lambda_{k}=j_{\Gamma,k}\left(A+1\right)^{-1}\,,\quad f_{k}(y)=\frac{\mathsf{c}_J}{(A+y)^{\Gamma}}J_{\Gamma}\left(\frac{A+y}{A+1}j_{\Gamma,k}\right)\,,
\end{equation}
with 
\begin{align}
    \mathsf{c}_J^{-2}=&\tfrac{(A+1)^2(\Gamma+1)}{(A+1)^{2(\Gamma+1)}-A^{2(\Gamma+1)}}\left\{J_{\Gamma-1}(j_{\Gamma,k})^2\right.\notag\\&\left.-(1+A^{-1})^{-2}\left[J_{\Gamma-1}\left(y_0\right)^2+J_{\Gamma}\left(y_0\right)^2-\tfrac{2\Gamma}{y_0}J_{\Gamma-1}\left(y_0\right)J_{\Gamma}\left(y_0\right)\right]_{y_0=\frac{Aj_{\Gamma,k}}{A+1}}\right\} \,,
\end{align}
where $\{f_k\}_{k\in\mathbb{N}}$ are normalized as in \eqref{eq.norm}. In the maximum warping limit with $A\to 0$, one recovers
\begin{equation}\label{eq.sols app full}
    \lambda_{k}=j_{\Gamma,k}\,,\quad f_{k}(y)=\frac{J_\Gamma(j_{\Gamma,k}y)}{\sqrt{\Gamma+1}J_{\Gamma-1}(j_{\Gamma,k})y^{\Gamma}}\,.
\end{equation}

\begin{figure}[!hbt]
    \centering
     \begin{subfigure}{0.8\textwidth}
    \includegraphics[width=\linewidth]{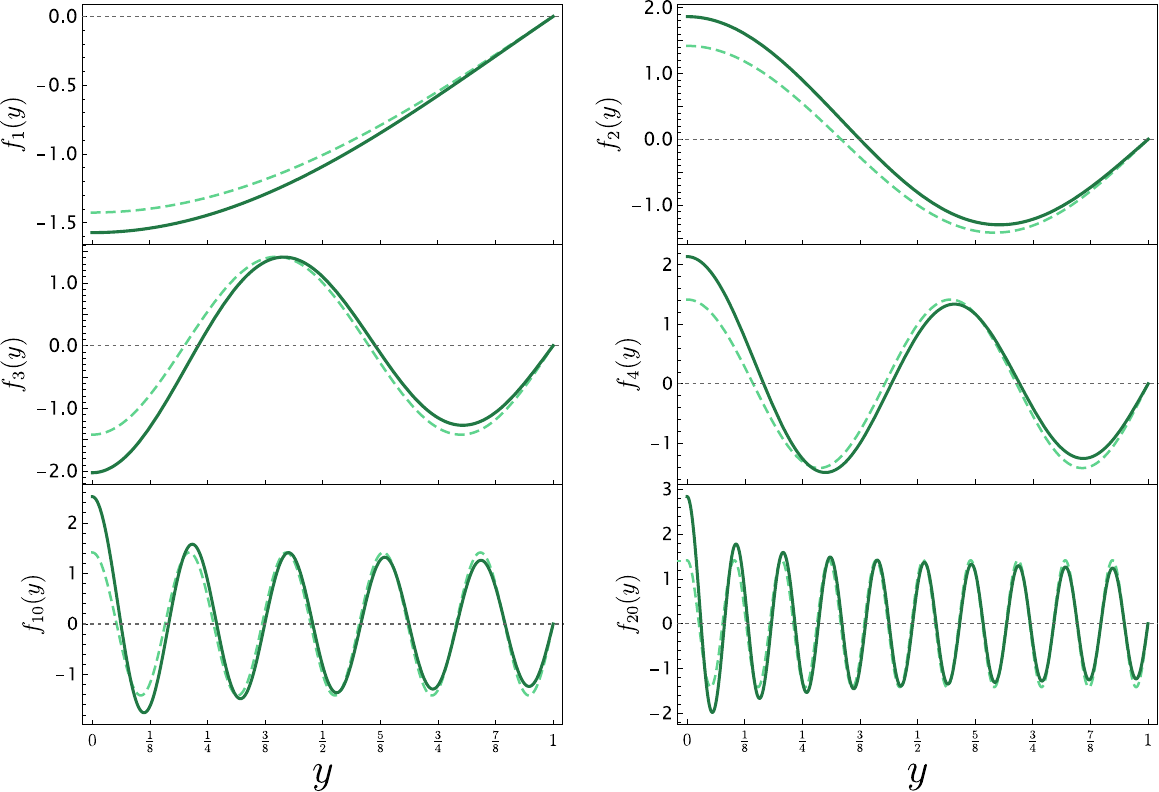}
    \caption{\hspace{-0.3em} Comparison between the scalar KK profiles obtained by solving the Laplacian equation (solid lines) and by using the WKB approximation (dashed) for $k\in\{1,2,3,4,10,20\}$.}
    \label{fig.KKmodes}
    \end{subfigure}
    \begin{subfigure}{0.85\textwidth}
    \includegraphics[width=\linewidth]{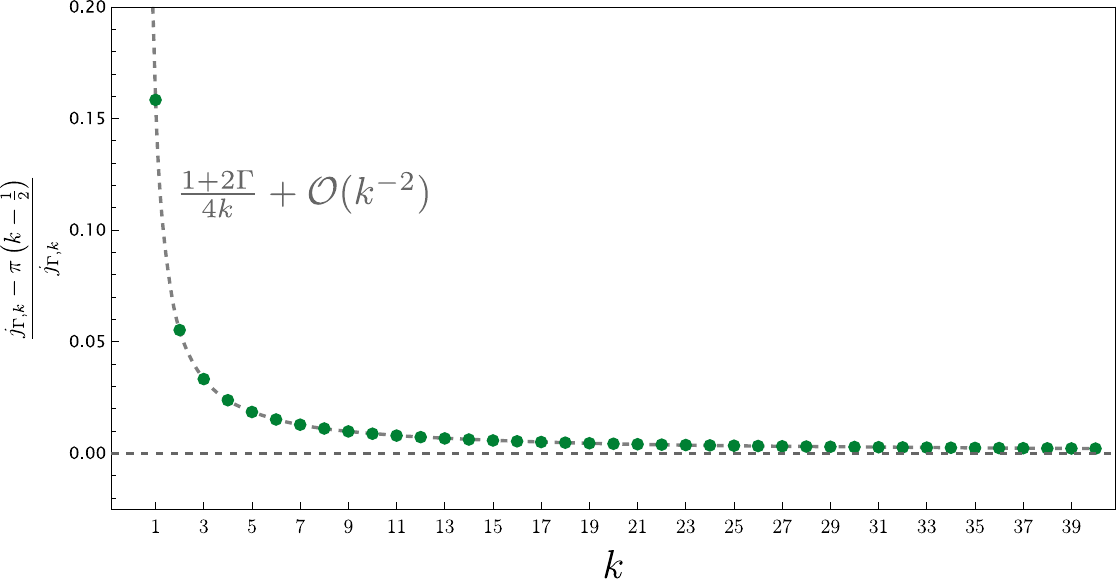}
    \caption{\hspace{-0.3em} Relative difference between the $k$-th eigenvalues of the Laplacian equation \eqref{eq.app lap} and the WKB approximation.}
    \label{fig.KKmasses}
    \end{subfigure}
    \caption{Comparison between eigenmodes and -values for the Laplacian equation \eqref{eq.app lap} for both the solution and the WKB approximation, in the $A\to 0$ maximal warping approximation.}
    \label{fig.KKcomp}
\end{figure}

Using the WKB approximation (i.e., suppressing the first order term in \eqref{eq.app lap}), and imposing the same boundary conditions and normalization,\footnote{The \emph{shape} of the modes in the WKB approximation is not affected by $A$ and $B$, but the \emph{normalization} through the measure $\mu(y)$ does. We depict the $A\to 0$ limit with maximal warping.} we obtain
\begin{equation}
    \lambda_{k}^{\text{\scriptsize WKB}}=\pi\left(k-\frac{1}{2}\right)\,,\quad f_{k}^{\text{\scriptsize WKB}}(y)=\sqrt{\frac{20}{3}}\frac{(-1)^{k}\cos\left[\pi\left(k-\tfrac{1}{2}\right)y\right]}{\pi(1+4k)\sqrt{{}_{1}F_{2}\left(\tfrac{5}{3};\tfrac{3}{2},\tfrac{8}{3};-\tfrac{\pi^2}{4}(1+4k)^2\right)}}
\end{equation}
for $k\geq 1$, with no dependence on $B$ or $A$ in $f_{k}^{\text{\scriptsize WKB}}$. 

In Figure~\ref{fig.KKcomp}, we compare the two with the exact solutions from \eqref{eq.sols app full}. One can see that the eigenmodes quickly converge at $\mathcal{O}(k^{-1})$:
\begin{equation}
    \lambda_{k}-\lambda_{k}^{\rm WKB}=\frac{\pi}{4}(2+\Gamma)+\frac{1-4\Gamma^2}{8\pi {k}^2}+\mathcal{O}(k^{-3})\,,\quad\frac{\lambda_{k}-\lambda_{k}^{\rm WKB}}{\lambda_{k}}=\frac{1+2\Gamma}{4k}+\mathcal{O}(k^{-2})\,,
\end{equation}
with an overall shift that becomes irrelevant for our purposes at large momentum $k\gg 1$. It is interesting to note that for $\Gamma >-\frac{1}{2}$ (equivalently, $\gamma>\frac{2}{\sqrt{d-1}}$), the KK modes are slightly heavier than in the unwarped case, 16\% heavier for $k=1$ for the example shown in Figure~\ref{fig.KKcomp}, with $d=9$ and $\gamma=\frac{5}{\sqrt{2}}$ (i.e. $\Gamma=-\frac{1}{3}$).
The Laplacian eigenfunctions are quite similar, with the main difference occurring at the position of maximal warping, $y=0$, where for $A\to 0$ the metric degenerates and $\hat{\varphi}$ diverges; still, $f_{k}(0)/f_{k}^{\rm WKB}(0)\sim 1$.

\section{Some comments on the \texorpdfstring{$A\to0$}{A->0} limit in scaling solutions with \texorpdfstring{$\gamma<\frac{2}{\sqrt{d-1}}$}{gamma<2/sqrt(d-1)}}\label{app. spooky solution}

In this appendix, we provide additional details on the $A\to 0$ limit for scaling solutions with $\gamma<\frac{2}{\sqrt{d-1}}$. As we discussed in Section \ref{ssec:scaling}, they have some unusual features that might signal something ``sick'' about these limits. From \eqref{eq. vol int}, the internal volume of the warped interval in $D$-dimensional Planck units is given by 
\begin{equation}
        \int_0^1\dd y \,e^\rho=\underbrace{\tfrac{\gamma ^2 (d-1)-4}{\gamma ^2 (d-1)}\left[(A+1)^{\frac{\gamma ^2 (d-1)}{\gamma ^2 (d-1)-4}}-A^{\frac{\gamma ^2 (d-1)}{\gamma ^2 (d-1)-4}}\right]}_{>0}B^{\frac{\gamma ^2 (d-1)}{\gamma^2 (d-1)-4}}\,.
    \end{equation}
This diverges along the trajectory $A\to 0$ with $B$ fixed only when $\gamma<\frac{2}{\sqrt{d-1}}$, which is therefore a decompactification limit. This limit is \emph{highly warped} since the gradients of the profiles do not vanish along the chosen trajectory, see \eqref{eq. grad codim1}. However, there are some distinctive features that characterize this limit.\\

First, consider the integral of the higher-dimensional potential on the internal interval:
    \begin{align}\label {eq. sick limits}
        \int_0^1{\rm d}y \,e^{\rho} \hat V(\hat\varphi)=&\tfrac{\sqrt{2|V_0|[4d-(d-1)\gamma^2]}}{\sqrt{d-1}\gamma^2{\rm sign}(V_0 (4-(d-1)\gamma^2))}\left((A+1)^{-\frac{\gamma ^2 (d-1)}{\gamma ^2 (d-1)-4}}-A^{-\frac{\gamma ^2 (d-1)}{\gamma ^2 (d-1)-4}}\right) B^{-\frac{\gamma ^2 (d-1)}{\gamma ^2 (d-1)-4}}\notag\\
        \to&\left\{\begin{array}{ll}
0&\text{if }B\text{ or }A\to\infty \text{ when }\gamma>\frac{2}{\sqrt{d-1}}\\
0&\text{if }B\to 0 \text{ and }A<\infty\text{ when }\gamma<\frac{2}{\sqrt{d-1}}\\
\propto B^{-\frac{(d-1)\gamma^2}{(d-1)\gamma^2-4}}\neq0&\text{if }A\to 0 \text{ and } B \text{ fixed when }\gamma<\frac{2}{\sqrt{d-1}}\\
\end{array}\right.\,.
    \end{align}
    This quantity is non-vanishing only in this special limit, $A\to 0$ with $B$ fixed and $\gamma<\frac{2}{\sqrt{d-1}}$. The reason is that along such trajectories, the higher-dimensional scalar remains finite throughout the bulk of the interval, $\hat{\varphi}(y)\to -\frac{2 (d-1)\gamma}{ (d-1)\gamma ^2-4}\log(By)$ for $y\neq 0$, and thus the higher-dimensional vacuum energy $\hat{V}(\hat{\varphi}(y))=V_0e^{\gamma\hat{\varphi}(y)}$ does not dilute at generic points. For the other decompactification limits, $\hat{\varphi}(y)\to-\infty$ at all points in the bulk of the warped interval.\\

    Consider then the moduli space metric for $A$ from \eqref{eq:moduli_space_metric}, $m_{\rm KK}$ from \eqref{eq:KK_mass_estimate}, and $M_{{\rm Pl},d+1}$ from \eqref{eq. warped MplD} in the same limit $A\to 0$, $B$ fixed, and $\gamma<\frac{2}{\sqrt{d-1}}$. We obtain
    \begin{subequations}
        \begin{equation}
            \mathsf{G}_{AA}=\left\{\begin{array}{ll}
              \frac{\gamma ^2 (d-1)^2 \left(-\gamma ^2+\left(\gamma ^2+4\right) d-8\right)}{(d-2) \left[\gamma ^2 (d-1)-4\right]^2}A^{-2} +\mathcal{O}(A^{-1})  & \text{for }\gamma\in\Big(0,\tfrac{2}{\sqrt{d-1}}\Big)  \\
               (d-2) (d-1) A^{d-4}+\mathcal{O}(A^{d-3})  & \text{for }\gamma=0
            \end{array}\right.\,,\label{eq. metric A0}
            \end{equation}
            \begin{equation}
            \frac{m_{\rm KK}}{M_{{\rm Pl},d}}\sim A^{-\frac{\gamma ^2(d-1)+4(d-2)}{(d-2) \left[\gamma ^2 (d-1)-4\right]}}\,,\qquad \frac{M_{{\rm Pl},d+1}}{M_{{\rm Pl},d}}\sim A^{-\frac{\gamma ^2 (d-1)}{(d-2) \left[\gamma ^2 (d-1)-4\right]}}\,,
            \end{equation}
    \end{subequations}
    where in $m_{\rm KK}$ and $M_{{\rm Pl},d+1}$ we are not displaying the $B$-dependence (since $B$ is kept fixed). From these we obtain the exponential rates for the trajectory under consideration, provided that $\gamma\neq 0$:
    \begin{equation}\label{eq. spooky kk}
        \lambda_{\rm KK}=\tfrac{1}{\sqrt{(d-1)(d-2)}}\left(1+\tfrac{4(d-2)}{(d-1)\gamma^2}\right)^{\frac12}\,,\quad \lambda_{{\rm Pl},d+1}=\tfrac{1}{\sqrt{(d-1)(d-2)}}\left(1+\tfrac{4(d-2)}{(d-1)\gamma^2}\right)^{-\frac12}\,.
    \end{equation}
    Note that $\lambda_{{\rm Pl},d+1}$ coincides with \eqref{eq.ex-Mpl}, which applies to the highly warped limits of \eqref{eq. highly warped limits scaling}. On the other hand, $\lambda_{\rm KK}$ differs from \eqref{eq:kk_rate_case_1}: it grows as $\gamma$ decreases, and eventually diverges for $\gamma\to0$.\footnote{We note that from \eqref{eq. spooky kk}, $\lambda_{\rm KK}>\sqrt{\frac{d-1}{d-2}}$ for $\gamma<\frac{2}{\sqrt{(d-1)(d-2)}}$. This range for $\gamma$ would imply a violation of the TCC bound \cite{Bedroya:2019snp}, and coincides with the slope at which the potential starts decreasing more slowly than the species scale, assuming the lower bound for the exponential rate of the species scale given in \cite{Calderon-Infante:2023ler,vandeHeisteeg:2023ubh}.} 
    This different behavior suggests that the KK mass behaves very differently when expressed in higher-dimensional Planck units, so that it becomes somehow sensitive to the warping, in apparent tension with the conclusions of \cite{DeLuca:2024fbc}.

    Moreover, it does not longer hold that the ratio between the exponential rates of $M_{{\rm Pl},d+1}$ and $m_{\rm KK}$ is a constant given by $\frac{\lambda_{{\rm Pl},d+1}}{\lambda_{\rm KK}}=\frac{1}{d+1}$ (see \eqref{eq. d-1 ratio}). This expectation came from the definition of the species scale (upon identifying $\Lambda_{\rm QG}\equiv M_{{\rm Pl},d+1}$) in terms of the number of light states and Weyl's law, as explained in Footnote \ref{fn. d-1 ratio}. In fact, one can check that, in this limit,
    \begin{equation}\label{eq:spooky_ratio_scaling}
        \frac{ \lambda_{{\rm Pl,d+1}}}{\lambda_{\rm KK}}=\underbrace{\left(1+\tfrac{4(d-2)}{(d-1)\gamma^2}\right)^{-1}}_{<\frac{1}{d-1}\text{ for }\gamma<\frac{2}{\sqrt{d-1}}}\to 0\quad \text{as }\;\gamma\to 0\,.
    \end{equation}
    As discussed around \eqref{eq. d-1 ratio}, $\frac{\lambda_{{\rm Pl},d+1}}{\lambda_{\rm KK}}=\frac{1}{d+1}$ holds for the other highly warped limits that we consider in this paper. 
    This suggests that the limit $A\to0$ exhibits some unusual features in terms of the KK spectrum; either the degeneracy of the spectrum for low KK momentum is drastically modified in comparison to Weyl's law expectations (so it becomes more dense),  or the higher-dimensional Planck scale cannot be identified with the species scale anymore.\\

    Note also that for finite values of $\gamma$ the limit $A\to0$ is at an infinite distance, while for $\gamma =0$ it is at a finite distance, with
    \begin{equation}\label{eq. pol scal app}
        \frac{m_{\rm KK}}{M_{{\rm Pl},d}}\sim A\sim (\Delta A)^{\frac{2}{d-2}}\,,\qquad \frac{M_{{\rm Pl},d+1}}{M_{{\rm Pl},d}}\to\mathcal{O}(1)|V_0|^{\frac{1}{2(d-2)}}\equiv\text{constant}\,,
    \end{equation}
    where $\Delta A$ is the moduli space distance computed through \eqref{eq. metric A0}. Hence, the KK scale vanishes at a finite distance, while $M_{{\rm Pl},d+1}$ does not become light in lower-dimensional Planck units. This matches the behavior of $\lambda_{\rm KK}$ and $\lambda_{{\rm Pl},d+1}$ in \eqref{eq. spooky kk} in the $\gamma\to 0$ limit.

    We finish by noting one last problem of this case: when $A=0$, the profiles of $\rho(y)$ and $\hat{\varphi}(y)$ diverge at $y=0$ regardless of the value of $\gamma$. As a consequence, higher-order corrections could spoil the solutions, the value of the moduli space metric, or our estimate for the KK scale. For the particular case of Type I$'$ from \cite{Polchinski:1995df}, we know that the theory remains under control even in the presence of singularities because these have a microscopic interpretation (O8$^-$ planes) and therefore the analogous $A\to 0$ limit (the vanishing of the impact parameter) is not problematic.
    However, this may not be the case for other setups with unclear UV completions, such as the solutions with $\gamma<\frac{2}{\sqrt{d-1}}$.
    
    In conclusion, the highly warped decompactification limit $A\to 0$ with fixed $B$ and $\gamma<\frac{2}{\sqrt{d-1}}$ leads to features that are quite different from natural expectations in decompactification regimes (warped or not). These may be indications that there is something ``sick'' with the asymptotic trajectory under consideration, and we do not consider it further in the paper. However, this is an interesting subject for future work.

\section{Codimension-one general solutions with exponential potentials}\label{app:codim_1_general}

In this appendix, we present the explicit codimension-one backgrounds of Section~\ref{ssec:codim_1_general}.

Consider first the case $V_0>0$. The parameterization in~\eqref{eq:param_1} reduces the equations of motion to~\eqref{eq:chi_eq_1}, which has the following solutions:
\begin{equation}\label{eq:chi_solution_1}
    \pm e^{\mp \chi} = \left\{\begin{array}{ll}
        \sqrt{\frac{\gamma+2\sqrt{\frac{d}{d-1}}}{\gamma-2\sqrt{\frac{d}{d-1}}}}\tanh\left[\zeta_{V_0,\gamma}^+(y)\right]&\qquad\text{if }\gamma>2\sqrt{\frac{d}{d-1}} \,,\\
         \sqrt{\frac{2d V_0}{d-1}} \, (y + C) &\qquad\text{if }\gamma=2\sqrt{\frac{d}{d-1}} \,, \\
         \sqrt{\frac{2\sqrt{\frac{d}{d-1}}+\gamma}{2\sqrt{\frac{d}{d-1}}-\gamma}}\tan\left[\zeta_{V_0,\gamma}^+(y)\right]&\qquad\text{if }\gamma<2\sqrt{\frac{d}{d-1}} \,,
    \end{array} \right.
\end{equation}
where 
\begin{equation}
    \zeta_{V_0,\gamma}^+(y)=\sqrt{\tfrac{V_0}{8}\left|\gamma^2-\tfrac{4d}{d-1}\right|} \, (y + C)\,.
\end{equation}
In~\eqref{eq:chi_solution_1}, the range of $y$ must be such that $e^\chi$ is always positive.
The sign ambiguity in~\eqref{eq:chi_solution_1} leads to two families of solutions for each value of $\gamma$. However, a family can be reached starting from the other by exchanging the $\sinh$ (or $\sin$) and the $\cosh$ (or $\cos$) functions in the following expressions---also taking into account the appropriate ranges of $y$. Therefore, we only write the explicit solution for the upper sign in~\eqref{eq:chi_solution_1}. This reads
\begin{subequations}\label{eq:codim_1_V_pos}
\begin{align}
    \hat{\varphi}(y)&=\left\{
    \begin{array}{ll}
        \hat{\varphi}_0+\frac{1}{2}\log\left[\left(\cosh\zeta_{V_0,\gamma}^+(y)\right)^{\frac{4\sqrt{d-1}}{\gamma\sqrt{d-1}-2\sqrt{d}}}\left(\sinh\zeta_{V_0,\gamma}^+(y)\right)^{\frac{4\sqrt{d-1}}{\gamma\sqrt{d-1}+2\sqrt{d}}}\right]&\quad \text{if }\gamma>2\sqrt{\frac{d}{d-1}}\,,\\
            \hat{\varphi}_0+\frac{1}{2}\sqrt{\frac{d-1}{d}}\left[\frac{1}{2}\left(\sqrt{\frac{2 d V_0}{d-1}}(y + C)\right)^2+\log\left(\sqrt{\frac{2 d V_0}{d-1}}(y + C)\right)\right]& \quad \text{if }\gamma=2\sqrt{\frac{d}{d-1}}\,,\\
                \hat{\varphi}_0+\frac{1}{2}\log\left[\left(\cos\zeta_{V_0,\gamma}^+(y)\right)^{\frac{4\sqrt{d-1}}{\gamma\sqrt{d-1}-2\sqrt{d}}}\left(\sin\zeta_{V_0,\gamma}^+(y)\right)^{\frac{4\sqrt{d-1}}{\gamma\sqrt{d-1}+2\sqrt{d}}}\right]&\quad \text{if }\gamma<2\sqrt{\frac{d}{d-1}}\,,
    \end{array}
    \right.\\
    \rho(y)&=\left\{
    \begin{array}{ll}
        \rho_0-\frac{1}{2d}\log\left[\left(\cosh\zeta_{V_0,\gamma}^+(y)\right)^{\frac{4\sqrt{d}}{\gamma\sqrt{d-1}-2\sqrt{d}}}\left(\sinh\zeta_{V_0,\gamma}^+(y)\right)^{-\frac{4\sqrt{d}}{\gamma\sqrt{d-1}+2\sqrt{d}}}\right]&  \text{ if }\gamma>2\sqrt{\frac{d}{d-1}}\,,\\
            \rho_0-\frac{1}{2d}\left[\frac{1}{2}\left(\sqrt{\frac{2 d V_0}{d-1}}(y + C)\right)^2-\log\left(\sqrt{\frac{2 d V_0}{d-1}}(y + C)\right)\right]& \text{ if }\gamma=2\sqrt{\frac{d}{d-1}}\,,\\
                \rho_0-\frac{1}{2d}\log\left[\left(\cos\zeta_{V_0,\gamma}^+(y)\right)^{\frac{4\sqrt{d}}{\gamma\sqrt{d-1}-2\sqrt{d}}}\left(\sin\zeta_{V_0,\gamma}^+(y)\right)^{-\frac{4\sqrt{d}}{\gamma\sqrt{d-1}+2\sqrt{d}}}\right]&\text{ if }\gamma<2\sqrt{\frac{d}{d-1}}\,.
    \end{array}
    \right.
\end{align}
\end{subequations}
In addition to these, there is a branch with $\chi'=0$ when $\gamma>2\sqrt{\frac{d}{d-1}}$. This corresponds, in the $\sigma=\rho$ coordinates, to the case that we have already studied in Section~\ref{ssec:scaling}.

The other case is $V_0<0$. Given $\hat{\varphi}'$ and $\rho'$ in~\eqref{eq:param_2}, one can solve the differential equation for $\chi$ in~\eqref{eq:chi_eq_2}, obtaining
\begin{equation} \label{eq:chi_solution_2}
    \pm e^{\mp \chi} = \left\{\begin{array}{ll}
         \sqrt{\frac{\gamma+2\sqrt{\frac{d}{d-1}}}{\gamma-2\sqrt{\frac{d}{d-1}}}}\tan\left[\zeta_{V_0,\gamma}^-(y)\right]&\qquad\text{if }\gamma>2\sqrt{\frac{d}{d-1}}\,, \\
         \sqrt{\frac{2d |V_0|}{d-1}} (y + C) &\qquad\text{if }\gamma=2\sqrt{\frac{d}{d-1}} \,, \\
         \sqrt{\frac{2\sqrt{\frac{d}{d-1}}+\gamma}{2\sqrt{\frac{d}{d-1}}-\gamma}}\tanh\left[\zeta_{V_0,\gamma}^-(y)\right]&\qquad\text{if }\gamma<2\sqrt{\frac{d}{d-1}}\,, 
    \end{array} \right.
\end{equation}
where
\begin{equation}
    \zeta_{V_0,\gamma}^-(y)=\sqrt{\frac{|V_0|}{8}\left|\gamma^2-\frac{4d}{d-1}\right|} \, (y + C) \,.
\end{equation}
Again, there are two families of solutions from~\eqref{eq:chi_solution_2}, which can be exchanged in the same way as in the $V_0>0$ case. Taking the upper sign in~\eqref{eq:chi_solution_2}, we find
\begin{subequations}\label{eq:codim_1_V_neg}
    \begin{align}
        \hat{\varphi}(y)&=\left\{
\begin{array}{ll}
     \hat{\varphi}_0+\frac{1}{2}\log\left[\left(\cos\zeta_{V_0,\gamma}^-(y)\right)^{\frac{4\sqrt{d-1}}{\gamma\sqrt{d-1}-2\sqrt{d}}}\left(\sin\zeta_{V_0,\gamma}^-(y)\right)^{\frac{4\sqrt{d-1}}{\gamma\sqrt{d-1}+2\sqrt{d}}}\right]& \quad \text{if }\gamma>2\sqrt{\frac{d}{d-1}}\,,\\
        \hat{\varphi}_0+\frac{1}{2}\sqrt{\frac{d-1}{d}}\left[-\frac{1}{2}\left(\sqrt{\frac{2 d |V_0|}{d-1}}(y + C)\right)^2+\log\left(\sqrt{\frac{2 d |V_0|}{d-1}}(y + C)\right)\right]& \quad \text{if }\gamma=2\sqrt{\frac{d}{d-1}}\,,\\
            \hat{\varphi}_0+\frac{1}{2}\log\left[\left(\cosh\zeta_{V_0,\gamma}^-(y)\right)^{\frac{4\sqrt{d-1}}{\gamma\sqrt{d-1}-2\sqrt{d}}}\left(\sinh\zeta_{V_0,\gamma}^-(y)\right)^{\frac{4\sqrt{d-1}}{\gamma\sqrt{d-1}+2\sqrt{d}}}\right]&\quad\text{if }\gamma<2\sqrt{\frac{d}{d-1}}\,,
    \end{array}
    \right.\\
    \rho(y)&=\left\{
\begin{array}{ll}
     \rho_0+\frac{1}{2d}\log\left[\left(\cos\zeta_{V_0,\gamma}^-(y)\right)^{\frac{4\sqrt{d}}{\gamma\sqrt{d-1}+2\sqrt{d}}}\left(\sin\zeta_{V_0,\gamma}^-(y)\right)^{-\frac{4\sqrt{d}}{\gamma\sqrt{d-1}-2\sqrt{d}}}\right]&  \text{ if }\gamma>2\sqrt{\frac{d}{d-1}}\,,\\
        \rho_0+\frac{1}{2d}\left[\frac{1}{2}\left(\sqrt{\frac{2 d |V_0|}{d-1}}(y + C)\right)^2+\log\left(\sqrt{\frac{2 d |V_0|}{d-1}}(y + C)\right)\right]& \text{ if }\gamma=2\sqrt{\frac{d}{d-1}}\,,\\
            \rho_0+\frac{1}{2d}\log\left[\left(\cosh\zeta_{V_0,\gamma}^-(y)\right)^{-\frac{4\sqrt{d}}{\gamma\sqrt{d-1}-2\sqrt{d}}}\left(\sinh\zeta_{V_0,\gamma}^-(y)\right)^{\frac{4\sqrt{d}}{\gamma\sqrt{d-1}+2\sqrt{d}}}\right]&\text{ if }\gamma<2\sqrt{\frac{d}{d-1}}\,.
    \end{array}
    \right.
    \end{align}
\end{subequations}
There is an additional solution with a constant value of $\chi$ when $\gamma<2\sqrt{\frac{d}{d-1}}$. When $\gamma\neq 0$, this reproduces a scaling codimension-one background of Section~\ref{ssec:scaling}, but when $\gamma=0$, it represents a new solution because one cannot shift $\hat\varphi$ to reabsorb $\sigma_0$ (see our comments after Footnote~\ref{ft:gauge_fixing_general}). In this special case, the gauge fixing is then $\sigma=\sigma_0$ and \eqref{eq:chi_eq_2} implies that $\chi=0$, thus leading to the solution
\begin{eqn}\label{eq:RS_as_codim_1}
    \varphi=\varphi_0\,,\qquad \rho=\pm e^{\sigma_0}\sqrt{\frac{|2V_0|}{d(d-1)}}y + \rho_0\,,\qquad \sigma=\sigma_0\,,
\end{eqn}
where we have explicitly shown the radion modulus $e^{\sigma_0}$, hidden within $V_0$ by our rescaling. This is a Randall--Sundrum-like solution, which matches the profile of \cite{Randall:1999ee} with $r_c\propto e^{\sigma_0}$.

Although our focus has been on exponential potentials, other codimension-one solutions associated with integrable systems are known analytically~\cite{Fre:2013vza,Pelliconi:2021eak}. It would be interesting to extend our analysis to those cases, computing the exponential rate $\lambda_{\rm KK}$ for more intricate codimension-one vacua, but we will not pursue this direction in the present work.

\subsection*{Moduli space metric for $\gamma=2\sqrt{\frac{d}{d-1}}$ and $V_0>0$}

We finish this appendix by explicitly computing the lower-dimensional moduli space metric for one of these warped compactifications. In general, the above expressions are quite involved due to the (hyperbolic) trigonometric functions used. Because of this, we focus on the simpler case with $\gamma=2\sqrt{\frac{d}{d-2}}$ and $V_0>0$. Evaluating \eqref{eq:moduli_space_metric} with \eqref{eq:codim_1_V_pos}, one finds:
\begin{subequations}
    \begin{align}
        \mathsf{G}_{V_0V_0}=&\frac{d-1}{8d(d-2)V_0^2}\big[d-1+8V_0(1+dV_0)\big]\notag\\&+\frac{d-1}{2dV_0}\tfrac{\Gamma\big(\tfrac{3d-1}{2d},C^2\big)-\Gamma\big(\tfrac{3d-1}{2d},(C+1)^2\big)}{\Gamma\big(\tfrac{d-1}{2d},C^2\big)-\Gamma\big(\tfrac{d-1}{2d},(C+1)^2\big)}-\frac{1}{2}\tfrac{\Gamma\big(\tfrac{5d-1}{2d},C^2\big)-\Gamma\big(\tfrac{5d-1}{2d},(C+1)^2\big)}{\Gamma\big(\tfrac{d-1}{2d},C^2\big)-\Gamma\big(\tfrac{d-1}{2d},(C+1)^2\big)}\,,\\
        \mathsf{G}_{CC}=&\frac{2(d-1)}{d-2}\left(\tfrac{C^{-\frac{1}{d}}e^{-C^2}-(1+C)^{-\frac{1}{d}}e^{-(1+C)^2}}{\Gamma\big(\tfrac{d-1}{2d},C^2\big)-\Gamma\big(\tfrac{d-1}{2d},(C+1)^2\big)}\right)^2+\frac{d^2-1}{2d}\tfrac{\Gamma\big(-\tfrac{d+1}{2d},C^2\big)-\Gamma\big(-\tfrac{d+1}{2d},(C+1)^2\big)}{\Gamma\big(\tfrac{d-1}{2d},C^2\big)-\Gamma\big(\tfrac{d-1}{2d},(C+1)^2\big)}\notag\\
        &+\frac{2(d-1)}{d}V_0-2V_0^2\tfrac{\Gamma\big(\tfrac{3d-1}{2d},C^2\big)-\Gamma\big(\tfrac{3d-1}{2d},(C+1)^2\big)}{\Gamma\big(\tfrac{d-1}{2d},C^2\big)-\Gamma\big(\tfrac{d-1}{2d},(C+1)^2\big)}\,,\\
        \mathsf{G}_{V_0C}=&\frac{d-1}{8dV_0}\tfrac{\Gamma\big(-\tfrac{1}{2d},C^2\big)-\Gamma\big(-\tfrac{1}{2d},(C+1)^2\big)}{\Gamma\big(\tfrac{d-1}{2d},C^2\big)-\Gamma\big(\tfrac{d-1}{2d},(C+1)^2\big)}-\frac{dV_0}{2(d-1)}\tfrac{\Gamma\big(\tfrac{2d-1}{2d},C^2\big)-\Gamma\big(\tfrac{2d-1}{2d},(C+1)^2\big)}{\Gamma\big(\tfrac{d-1}{2d},C^2\big)-\Gamma\big(\tfrac{d-1}{2d},(C+1)^2\big)}\,.
    \end{align}
\end{subequations}
There are two interesting aspects of the above metric. First, in the decompactification limit $V_0\to0$ with fixed impact parameter $C$, the dependence on $C$ is subleading, and thus all trajectories of this type have the same asymptotic unit tangent vector $\hat T=(\mathsf{G}_{V_0V_0})^{-1/2}\partial_{V_0}$. Additionally, one can see that $\mathsf{G}_{CC}=\mathcal{O}(C)^{-\frac{d+1}{d}}$ and all parallel trajectories are at finite distances from each other, including the one with zero impact parameter, $C=0$. 

\section{Further details on higher-codimension warped backgrounds\label{App:BRANES}}

In this appendix, we give more details on the arguments that we use in Section \ref{ss. gen exp} for the scaling of the KK modes in backgrounds with codimension-three or higher defects, where, following \eqref{eq. high warp large codim},
\begin{equation}\label{eq. def mathfrak T}
    \mathfrak{T}(\varphi)=h_pM_{{\rm Pl},d+n}^{-d}\mathcal{T}_p R^{-(n-2)}
\end{equation}
does not diverge along asymptotic trajectories in the perturbative regime of a given duality frame. This translates into having a non-highly warped limit, as defined in Section \ref{ssec:highly_warped_limit}. From \eqref{eq. profiles} and \eqref{eq. Hr}, one obtains
\begin{subequations}
    \begin{align}
        \partial_\mu\rho&=-\frac{4(n-2)}{\Delta(d+n-2)}\partial_\mathfrak{T}\log H(r)\partial_{\varphi^a}\mathfrak{T}\,\partial_\mu\varphi^a \,,\\
        \partial_\mu\sigma&=R^{-1}\partial_\mu R+\frac{4d}{\Delta(d+n-2)}\partial_\mathfrak{T}\log H(r)\partial_{\varphi^a}\mathfrak{T}\,\partial_\mu\varphi^a\,,\\
        \partial\hat\varphi&=\partial_\mu\hat{\varphi}_0+\frac{2\alpha}{\Delta}\partial_\mathfrak{T}\log H(r)\partial_{\varphi^a}\mathfrak{T}\,\partial_\mu\varphi^a\,,\\
        \partial_\mu\theta&=\frac{n}{d-2}R^{-1}\partial_\mu R+\partial_\mathfrak{T}\log\left(\int_{X_n}\dd^n y\sqrt{\mathsf{M}}H(r)^{\frac{8}{\Delta}}\right)\partial_{\varphi^a}\mathfrak{T}\,\partial_\mu\varphi^a\,.
    \end{align}
\end{subequations}
One can now compute the contributions of the warped profiles to the moduli space metric \eqref{eq:moduli_space_metric}, which read
\begin{align}\label{eq. metr brane H}
    \mathsf{G}_{ab}\partial_\mu\varphi^a\partial^\mu\varphi^b=&\tfrac{n(d+n-2)}{d-2}(\partial\log R)^2+(\partial\hat{\varphi}_0)^2\notag\\
    &+\tfrac{64}{\Delta^2}\left\{(d-1)(d-2)\langle\partial_\mathfrak{T}\log H\rangle^2-\tfrac{1}{4}\left(\tfrac{d(n-2)}{d+n-2}+\tfrac{\alpha^2}{4}-2\right)\langle(\partial_\mathfrak{T}\log H)^2\rangle\right\}(\partial\mathfrak{T})^2\notag\\
    &+\tfrac{16n}{\Delta} \left(\tfrac{d (n-1)}{d+n-2}+d-3\right)\langle\partial_\mathfrak{T}\log H\rangle\partial\mathfrak{T}\cdot\partial\log R+\tfrac{4\alpha}{\Delta}\langle\partial_\mathfrak{T}\log H\rangle\partial\mathfrak{T}\cdot\partial\hat{\varphi}_0\,,
\end{align}
where we define the following expectation value:
\begin{equation}
    \langle f\rangle=\frac{\int_{X_n}\dd^ny\sqrt{\mathsf{M}}H(r)^\frac{8}{\Delta}f(y)}{\int_{X_n}\dd^ny\sqrt{\mathsf{M}}H(r)^\frac{8}{\Delta}}\in\left[\inf_{X_n}f,\sup_{X_n}f\right]\,,
\end{equation}
and we denote $\partial\mathfrak{T}=\partial_{\varphi^a}\mathfrak{T}\partial_\mu\varphi^a$. One can see that in the limit $\mathfrak{T}\to 0$ we have $H(r)\to 1$, and thus $\langle\partial_\mathfrak{T}\log H\rangle=\langle(\partial_\mathfrak{T}\log H)^2\rangle=0$, thus recovering the unwarped metric
\begin{equation}\label{eq.metr as}
    \mathsf{G}=\begin{pmatrix}
        \frac{n(d+n-2)}{d-2}R^{-2}&0\\0&1
    \end{pmatrix}
\end{equation}
in $(R,\hat{\varphi}_0)$ coordinates.

On the other hand, for the KK mass \eqref   {eq:KK_mass_estimate} we have
\begin{equation}\label{eq.KK as} 
        \frac{m_{\rm KK}}{M_{{\rm Pl},d}}\sim \left[\int_{X_n} \dd^n y\sqrt{\mathsf{M}}\,  e^{d(\rho-\theta)+(n-2)\sigma} \right]^{\frac{1}{2}}= R^{-\frac{d+n-2}{d-2}}{\underbrace{\left(\int_{X_n}\dd^ny\sqrt{\mathsf{M}}\,H(r)^\frac{8}{\Delta}\right)}_{\text{finite contribution}}}^{-\frac{d}{2(d-2)}}\,,
\end{equation}
where the finite contribution amounts to a numeric prefactor bounded by $(1+\mathfrak{T}(\varphi)r_{\rm max}^{n-2})^{-\frac{4d}{\Delta(d-2)}}$ from below and by 1 from above, so that
\begin{align}
   (\zeta_{\rm KK})_a&= -\partial_{\varphi^a}\log\frac{m_{\rm KK}}{M_{{\rm Pl},d}}=\frac{d+n-2}{d-2}\partial_{\varphi^a}\log R+\frac{4d}{\Delta(d-2)}\langle\partial_\mathfrak{T}\log H\rangle\partial_{\varphi^a}\mathfrak{T}\,.
\end{align}
Once again, in the limit in which $\mathfrak{T}\to 0$, we see that the additional contributions are not present, and we recover the unwarped case with $|\vec\zeta_{\rm KK}|=\sqrt{\frac{d+n-2}{d-2}}$. 

As for the case when $\mathfrak{T}$ takes a finite value along some asymptotic limit, from \eqref{eq. def mathfrak T} we have $\mathfrak{T}=h_pM_{{\rm Pl},d+n}^{-d}\mathcal{T}_p R^{-(n-2)}$. This means that $\mathfrak{T}$ takes a finite value only along the direction(s) for which $\mathcal{T}_p\sim M_{{\rm Pl},d+n}^dR^{n-2}$. From \eqref{eq. metr brane H} and \eqref{eq.KK as}, we see that when $\langle\partial_\mathfrak{T}\log H\rangle,\,\langle(\partial_\mathfrak{T}\log H)^2\rangle\neq 0$ one has additional contributions to both the $\vec\zeta_{\rm KK}$ vector and the moduli space metric, which in general depend on the codimension, $n$, of the defect and on the $\hat{\varphi}$-dependence of the defect tension $\mathcal{T}_p$. We cannot show in full generality that one always recovers $|\vec\zeta_{\rm KK}|$ along this type of limit, but we show how this is the case for a physically relevant case.

We can illustrate this with two different examples in $d=6$ and codimension $n=4$. We first consider D5-branes, with $\mu=\frac{1}{\sqrt
2}$, see \eqref{eq.II tensions}, so that $\mathfrak{T}\sim\exp\left(\frac{1}{\sqrt{2}}\hat{\varphi}_0\right)R^{-2}
$. Using \eqref{eq. profiles} and \eqref{eq. Hr}, we can compute the moduli space metric \eqref{eq:moduli_space_metric} and the $\zeta$-vector associated with the KK mass \eqref{eq:KK_mass_estimate} in terms of the moduli $R$ and $\hat\varphi_0$:
\begin{subequations}
    \begin{align}
    \mathsf{G}&= \begin{pmatrix}
        \frac{1}{R^2}\left(1+\frac{7+4h_5R^{-2}e^{\frac{1}{\sqrt{2}}\hat\varphi_0}}{\Big(1+2h_5R^{-2}e^{\frac{1}{\sqrt{2}}\hat\varphi_0}\Big)^2}\right)&\frac{\sqrt{2}h_5R^{-2}e^{\frac{1}{\sqrt{2}}\hat\varphi_0}\Big(4+3h_5R^{-2}e^{\frac{1}{\sqrt{2}}\hat\varphi_0}\Big)}{R^{-5}\Big(1+2h_5R^{-2}e^{\frac{1}{\sqrt{2}}\hat\varphi_0}\Big)^2}\\
        \frac{\sqrt{2}h_5R^{-2}e^{\frac{1}{\sqrt{2}}\hat\varphi_0}\Big(4+3h_5R^{-2}e^{\frac{1}{\sqrt{2}}\hat\varphi_0}\Big)}{R^{-5}\Big(1+2h_5R^{-2}e^{\frac{1}{\sqrt{2}}\hat\varphi_0}\Big)^2}&\frac{9}{8}\left(1-\frac{1+12h_5R^{-2}e^{\frac{1}{\sqrt{2}}\hat\varphi_0}}{9\Big(1+2h_5R^{-2}e^{\frac{1}{\sqrt{2}}\hat\varphi_0}\Big)^2}\right)
    \end{pmatrix}    \,,   \\
    \vec\zeta_{\rm KK}&=\left(\frac{1}{2R}\Bigg(1+\frac{3}{1+2h_5R^{-2}e^{\frac{1}{\sqrt{2}}\hat\varphi_0}}\Bigg),\frac{3h_5R^{-2}e^{\frac{1}{\sqrt{2}}\hat\varphi_0}}{2\sqrt{2}\big(1+2h_5R^{-2}e^{\frac{1}{\sqrt{2}}\hat\varphi_0}\big)}\right) \,.       
    \end{align}
\end{subequations}
Using these equations, one can check that $|\vec\zeta_{\rm KK}|=\frac{1}{\sqrt{2}}$, thus recovering the value of unwarped compactifications with $d=6$ and $n=4$ in \eqref{eq. KK unwarped}. More intuitively, it is also clear that for any perturbative limit, 
\begin{equation}\label{eq.pert reg d6}
    R\to\infty\,,\quad\hat{\varphi}_0\leq 0\qquad\text{with}\quad R\geq e^{-\frac{1}{2\sqrt{2}}\hat\varphi_0} \,,
\end{equation}
we have $\mathfrak{T}\sim R^{-2}e^{\frac{1}{\sqrt{2}}\hat{\varphi}_0}=R^{-4}(R e^{-\frac{1}{2\sqrt{2}}\hat{\varphi}_0})^2\leq R^{-4}\to 0$, so that we recover 
\begin{equation}\label{eq.asympt exp}
    \mathsf{G}=\begin{pmatrix}
        \frac{8}{R^2}&0\\0&1
    \end{pmatrix}\;,\quad \vec\zeta_{\rm KK}=\left(\frac{2}{R},0\right)\;,
\end{equation}
as expected from \eqref{eq.metr as} and \eqref{eq.KK as}, thus obtaining the same results as in the unwarped limit.

\vspace{0.5cm}

The case with NS5 branes, where $\mu=-\frac{1}{\sqrt{2}}$ and $\mathfrak{T}\sim\exp\left(-\frac{1}{\sqrt{2}}\hat{\varphi}_0\right)R^{-2}
$, is more interesting. We obtain the following moduli space metric and $\zeta$-vector associated with the KK mass:
    \begin{subequations}
        \begin{align}
            \mathsf{G}&=\begin{pmatrix}
        \frac{1}{R^2}\left(1+\frac{7+4h_5R^{-2}e^{-\frac{1}{\sqrt{2}}\hat\varphi_0}}{\Big(1+2h_5R^{-2}e^{-\frac{1}{\sqrt{2}}\hat\varphi_0}\Big)^2}\right)&\frac{\sqrt{2}h_5R^{-2}e^{-\frac{1}{\sqrt{2}}\hat\varphi_0}\Big(4+3h_5R^{-2}e^{-\frac{1}{\sqrt{2}}\hat\varphi_0}\Big)}{R^{-5}\Big(1+2h_5R^{-2}e^{-\frac{1}{\sqrt{2}}\hat\varphi_0}\Big)^2}\\\frac{\sqrt{2}h_5R^{-2}e^{-\frac{1}{\sqrt{2}}\hat\varphi_0}\Big(4+3h_5R^{-2}e^{-\frac{1}{\sqrt{2}}\hat\varphi_0}\Big)}{R^{-5}\Big(1+2h_5R^{-2}e^{-\frac{1}{\sqrt{2}}\hat\varphi_0}\Big)^2}&1+\frac{h_5R^{-2}e^{-\frac{1}{\sqrt{2}}\hat\varphi_0}\Big(2+9h_5R^{-2}e^{-\frac{1}{\sqrt{2}}\hat\varphi_0}\Big)}{2\Big(h_5R^{-2}e^{-\frac{1}{\sqrt{2}}\hat\varphi_0}\Big)^2}    \end{pmatrix}\,,\\
            \vec\zeta_{\rm KK}&=\left(\frac{2}{R}-\frac{3h_5R^{-2}e^{-\frac{1}{\sqrt{2}}\hat\varphi_0}}{R\big(1+2 h_5R^{-2}e^{-\frac{1}{\sqrt{2}}\hat\varphi_0}\big)},-\frac{3 h_5R^{-2}e^{-\frac{1}{\sqrt{2}}\hat\varphi_0}}{2\sqrt{2} \big(1+2 h_5R^{-2}e^{-\frac{1}{\sqrt{2}}\hat\varphi_0}\big)}\right)\,.
        \end{align}
    \end{subequations}
    Again, $ |\vec\zeta_{\rm KK}|=\frac{1}{\sqrt2}$, but the situation is not as straightforward as in the D5 case. First of all, given that the perturbative regime is \eqref{eq.pert reg d6}, we have that
    \begin{equation}
        \mathfrak{T}\sim h_p R^{-2}e^{-\frac{1}{\sqrt{2}}\hat\varphi_0}\lesssim \mathcal{O}(1)\,,
    \end{equation}
    and thus the moduli space metric and the KK $\zeta$-vector along asymptotic trajectories are not \eqref{eq.asympt exp} but rather 
    \begin{equation}\label{eq. G zeta NS5}
            \mathsf{G}=\begin{pmatrix}
        \frac{1}{R^2}\left(1+\frac{7+4\mathfrak{T}}{(1+2\mathfrak{T})^2}\right)&\frac{\sqrt{2}\mathfrak{T}(4+3\mathfrak{T})}{R^{-5}(1+2\mathfrak{T})^2}\\\frac{\sqrt{2}\mathfrak{T}(4+3\mathfrak{T})}{R^{-5}(1+2\mathfrak{T})^2}&1+\frac{\mathfrak{T}(2+9\mathfrak{T})}{2\mathfrak{T}^2}    \end{pmatrix}\;,\quad
            \vec\zeta_{\rm KK}=\left(\tfrac{1}{R}\big(2-\tfrac{3\mathfrak{T}}{1+2 \mathfrak{T}}\big),-\tfrac{3 \mathfrak{T}}{2\sqrt{2} (1+2 \mathfrak{T})}\right)\,.
        \end{equation}
        If $\mathfrak{T}$ is not asymptotically 0, the above expressions do not automatically match the unwarped case. Let us consider the different limits in more detail. Let $\hat T_1$ be the unit tangent vector along the decompactification direction (i.e., $R\to\infty$ with $\hat{\varphi}_0$ fixed). We have 
    \begin{align}
        \lambda_{\rm KK}=\hat{T}_1\cdot\vec\zeta_{\rm KK}&=
        \frac{2+\mathfrak{T}}{2\sqrt{(1+\mathfrak{T})^2+1}}
        \notag\\&= \frac{2+h_5R^{-2}e^{-\frac{1}{\sqrt{2}}\hat{\varphi}_0}}{2\sqrt{(1+h_5R^{-2}e^{-\frac{1}{\sqrt{2}}\hat{\varphi}_0})^2+1}}\to\frac{1}{\sqrt{2}}\quad\text{as }\, R\to\infty\,,
        \end{align}
    so that $\vec\zeta_{\rm KK}\propto \hat{T}_1$ and there is no sliding in this direction. For more general trajectories in the perturbative regime, we can consider
    \begin{equation}
        \hat{\varphi}_0=-(1-\alpha)\sqrt{8}\log R\,,\quad\text{with }\alpha\in[0,1]\ \text{ and }\, R\to\infty\,,
    \end{equation}
    with unit tangent vector $\hat{T}_\alpha$, and where the asymptotic $\mathfrak{T}\neq0$ limit corresponds to $\alpha=0$. For this, we find 
    \begin{equation}\label{eq. lambda KK NS5}
        \lambda_{\rm KK}=(\hat{T}_\alpha\cdot \vec\zeta_{\rm KK})|_{R\to\infty}=\frac{1}{\sqrt{2}}\frac{1}{\sqrt{2+\alpha(\alpha-2)}}=|\vec\zeta_{\rm KK}|\cos(\widehat{\hat{T}_\alpha,\vec\zeta_{\rm KK}})\,,
    \end{equation}
    where $\widehat{\hat{T}_\alpha,\vec\zeta_{\rm KK}}$ is the angle between the two directions that can be computed using \eqref{eq. G zeta NS5}. We therefore conclude that $\vec\zeta_{\rm KK}$ does not slide for any asymptotic trajectory in the perturbative regime, including the limiting $ \hat{\varphi}_0\sim-\sqrt{8}\log R\to\infty$ (i.e., $\alpha=0$ and $\mathfrak{T}>0$) where $H(r)=1+\frac{\mathfrak{T}}{r^2}$ rather than $H(r)\to 1$, which indeed corresponds to a highly warped limit as in Section \ref{ssec:highly_warped_limit}. 

    To understand the type of spacetime to which we are decompactifying, consider a coordinate transformation $\mathtt{r}=R r\in[0,R]$,\footnote{Since the internal space $X_4$ has a fixed volume in the internal metric $\mathsf{M}_{ij}$, which we approximate as $\dd s_4^2=\mathsf{M}_{ij}\dd y^i\dd y^j\approx\dd r^2+r^2\dd \Omega_2^2$ around the warping defect, we can effectively take $r\in[0,1]$ without loss of generality.}, where the new radial direction has a growing range as we decompactify. The warped profiles \eqref{eq. profiles} can be rewritten as
    \begin{subequations}
        \begin{align}
            \dd s_{10}^2&=\left[1+\mathfrak{T}\bigg(\frac{R}{\mathtt{r}}\bigg)^2\right]^{-\frac{1}{4}}\dd\vec{x}_{\parallel}+\left[1+\mathfrak{T}\bigg(\frac{R}{\mathtt{r}}\bigg)^2\right]^{\frac{3}{4}}(\dd\mathtt{r}^2+\mathtt{r}^2\dd\Omega_3^2)\,,\label{eq. decomp ds10}\\
            \hat{\varphi}&=\hat{\varphi}_0-\frac{1}{2\sqrt{2}}\log\left[1+\mathfrak{T}\bigg(\frac{R}{\mathtt{r}}\bigg)^2\right]\sim-\sqrt{8}\log R-\frac{1}{2\sqrt{2}}\log\left[1+\mathfrak{T}\bigg(\frac{R}{\mathtt{r}}\bigg)^2\right]\,,\\
            H_3=\dd B_2&=-\frac{\mathfrak{T}R^2}{\mathtt{r}^3}\star(\dd x^0\wedge\dots\wedge\dd x^5\wedge\dd \mathtt{r})\,,
        \end{align}
    \end{subequations}
    where $\dd\vec{x}_{\parallel}=-\dd t^2+\sum_{i=1}^5(\dd x^i)^2$ is the NS5 worldvolume element. The above expressions are reminiscent of the 10d NS5 black brane solution; see, e.g., \cite[Chapter 18.5]{Blumenhagen:2013fgp}. In this way, we are decompactifying to a non-empty spacetime, threaded by non-vanishing $H_3$ field strength and Ricci scalar
    \begin{equation}
        \mathcal{R}_{10}=\frac{3}{2}\mathfrak{T}\left(\frac{R}{\mathtt{r}}\right)^4 \left[1+\mathfrak{T}\bigg(\frac{R}{\mathtt{r}}\bigg)^2\right]^{-\frac{11}{4}}\mathtt{r}^{-2}\,,
    \end{equation}
    which is non-vanishing at finite distance from the NS5-brane.
    
    Note that, as $R\to \infty$, the space decompactifies,\footnote{To see this, note that from \eqref{eq. decomp ds10}, the transverse distance to the brane grows as
    \begin{equation}
        \int_0^{R}\dd \mathtt{r} \left[1+\mathfrak{T}\bigg(\frac{R}{\mathtt{r}}\bigg)^2\right]^{\frac{3}{8}}=R-\frac{8 \sqrt{\pi } \, \Gamma \left(\frac{9}{8}\right)}{\Gamma \left(-\frac{3}{8}\right)}\mathfrak{T}^{\frac{1}{2}}+\mathcal{O}(R^{-1})\to\infty\quad\text{as}\quad R\to\infty\,.
    \end{equation}    
    } so that additional (anti)NS5-branes are pushed away towards infinity (or the non-trivial gauge bundle that cancel the $H_3$ charge gets diluted away) and we can have a single NS5-brane in the bulk. Clearly, far from the NS5-brane, spacetime looks locally like a Minkowski one; however, when performing the decompactification in this limit, we remain ``close'' to one of the warping defects while the rest is carried away.

    This limit has the caveat that we are sending the ten-dimensional dilaton to weak coupling, $\hat{\varphi}\to -\infty$, which would lower the quantum gravity cutoff. As a matter of fact, from \eqref{eq. lambda KK NS5} we can see that for this limit with finite $\mathfrak{T}$ and $\hat{\varphi}_0\sim -\sqrt{8}\log R\to-\infty$, the exponential rate of the KK tower is $\lambda_{\rm KK}=\frac{1}{2}=\frac{1}{\sqrt{6-2}}$, thus saturating the bound in \eqref{eq. sharpened DC}. We then expect to have string oscillating modes with $\lambda_{\rm osc}=\lambda_{\rm KK}=\frac{1}{2}$, so that this is actually an \emph{emergent string limit} in which, as we discussed at the end of Section \ref{ss. results and bounds}, the string modes are accompanied by a KK tower.
    
    Any other limit with $\hat{\varphi}_0\ll-\sqrt{8}\log R\to\infty$ has effectively $\mathfrak{T}\to 0$, decompactifying to empty ten-dimensional spacetime where there is a parametric separation between the KK and string scales.

    The setting discussed above is interesting on its own, since it realizes the example discussed in \cite{Alvarez-Garcia:2023gdd, Alvarez-Garcia:2023qqj}, where the authors considered 6d compactifications of F-theory on (elliptically fibered) K3-fibered three-folds $Y_3$, dual to heterotic string theory on a (different) elliptically fibered K3:

    \begin{equation*}
    \centering
        \begin{array}{cccc}
           \textbf{F-theory}  && \textbf{Heterotic}&\phantom{XXXXXX}  \\
            \begin{tikzcd}[
  column sep={2cm,between origins},
  column 1/.style={column sep=1.5cm},
  column 2/.style={column sep=1.5cm}
]
                \mathbb{E}\arrow[r,hook, myRED]\arrow[rd, hook,orange]&\text{K3}_{\rm F-th}\arrow[r,myRED]\arrow[d,hook,myGREEN]&\mathbb{P}^1\arrow[d,hook,myBLUE]\\                &Y_3\arrow[r,orange]\arrow[rd,myGREEN]&\mathcal{B}_2\simeq\mathbb{F}_n\arrow[d,myBLUE]\\
                & & \mathcal{B}_1\simeq\mathbb{P}^1_{\rm b}
            \end{tikzcd} &\longleftrightarrow&  \begin{tikzcd}
\mathbb{T}^2_{\rm het}\arrow[d, hook,myBLUE]\\ \text{K3}_{\rm het}\arrow[d,myBLUE]\\ \mathcal{B}_1\simeq\mathbb{P}^1_{\rm b}
\end{tikzcd}
        \end{array}
    \end{equation*}
    Considering $Y_3$ as a K3-fibration, the base $\mathcal{B}_1$ is given by $\mathbb{P}^1_{\rm b}$, while seeing $Y_3$ as elliptically fibered, the base is a Hirzebruch surface $\mathcal{B}_2\simeq \mathbb{F}_n$. If K3 has non-minimal degeneration points, upon adiabatic decompactification limits where both $\mathcal{B}_1$ and $\mathcal{B}_2$ become large, these are understood as 7-branes wrapping the $\mathbb{P}^1$ fiber of $\mathcal{B}_1$. In the dual heterotic frame, these defects are NS5 branes (M5 branes in the bulk of the Ho\v{r}ava--Witten interval in the heterotic M-theory picture) located at the degeneration points of the $\mathbb{T}^2_{\rm het}$ over $\mathbb{P}^1_{\rm b}$. From our analysis (at least in the frame where the perturbative description is the heterotic one), the scaling of the associated KK modes should be the same as in homogeneous compactifications to Minkowski spacetime, regardless of the scaling between the internal volume and the heterotic coupling.

\bibliographystyle{JHEP}
\bibliography{ref}

\end{document}